\documentclass{aa}  

\usepackage{graphicx}
\usepackage{subfigure}
\usepackage[varg]{txfonts}
\usepackage{natbib}
\bibpunct{(}{)}{;}{a}{}{,} % to follow the A&A style

\newcommand{\forloop}[5][1]%
{%
\setcounter{#2}{#3}%
\ifthenelse{#4}%
	{%
	#5%
	\addtocounter{#2}{#1}%
	\forloop[#1]{#2}{\value{#2}}{#4}{#5}%
	}%
	{%
	}%
}% 

\newcommand{\ctbd}[1]{}

\newcommand{\lc}{light curve}
\newcommand{\lcs}{light curves}
\newcommand{\Lc}{Light curve}

\newcommand{\band}[1]{\ensuremath{#1}~band}

\newcommand{\chisq}{\ensuremath{\chi^2}}

\newcommand{\kms}{\ensuremath{\rm km\,s^{-1}}}
\newcommand{\ms}{\ensuremath{\rm m\,s^{-1}}}

\newcommand{\gcmc}{\ensuremath{\rm g\,cm^{-3}}}
\newcommand{\ergscmsq}{\ensuremath{\rm erg\,s^{-1}\,cm^{-2}}}

\newcommand{\masyr}{\ensuremath{\rm mas\,yr^{-1}}}
\providecommand{\arcdeg}{\hbox{$^\circ$}}

\newcommand{\vsini}{\ensuremath{v \sin{i}}}
\newcommand{\feh}{\ensuremath{\rm [Fe/H]}}

\newcommand{\rsun}{\ensuremath{R_\sun}}
\newcommand{\msun}{\ensuremath{M_\sun}}
\newcommand{\lsun}{\ensuremath{L_\sun}}

\newcommand{\rstar}{\ensuremath{R_\star}}
\newcommand{\mstar}{\ensuremath{M_\star}}
\newcommand{\lstar}{\ensuremath{L_\star}}

\newcommand{\teffstar}{\ensuremath{T_{\rm eff\star}}}

\newcommand{\loggstar}{\ensuremath{\log{g_{\star}}}}

\newcommand{\rpl}{\ensuremath{R_{p}}}
\newcommand{\mpl}{\ensuremath{M_{p}}}

\newcommand{\rhopl}{\ensuremath{\rho_{p}}}

\newcommand{\arstar}{\ensuremath{a/\rstar}}
\newcommand{\zrstar}{\ensuremath{\zeta/\rstar}}

\newcommand{\rjup}{\ensuremath{R_{\rm J}}}
\newcommand{\mjup}{\ensuremath{M_{\rm J}}}

\newcommand{\reffigl}[1]{Figure~\ref{fig:#1}}

\newcommand{\reftabl}[1]{Table~\ref{tab:#1}}

\newcommand{\hatcurhtrxxxxA}{HTR366-001}                               % Original HTR name of target
\newcommand{\hatcurfieldxxxxA}{366}                                    % Original HTR field
\newcommand{\hatcurCCraxxxxA}{\ensuremath{09^{\mathrm h}01^{\mathrm m}22.66{\mathrm s}}}                             % Right Ascension
\newcommand{\hatcurCCdecxxxxA}{\ensuremath{+06{\arcdeg}05{\arcmin}50.0{\arcsec}}}                            % Declination
\newcommand{\hatcurCCmagxxxxA}{12.168}                                 % apparent V-band magnitude
\newcommand{\hatcurCCtwomassxxxxA}{2MASS~09012265+0605500}             % 2MASS identifier
\newcommand{\hatcurCCgscxxxxA}{GSC~0232-01451}                         % GSC(1.2) identifier
\newcommand{\hatcurCCtassmvxxxxA}{\ensuremath{12.168\pm0.030}}                              % note, APASS V-band magnitude
\newcommand{\hatcurCCtassmbxxxxA}{\ensuremath{12.827\pm0.020}}                              % note, APASS B-band magnitude
\newcommand{\hatcurCCtwomassJmagxxxxA}{\ensuremath{10.960\pm0.023}}    % 2MASS ORIG MAG
\newcommand{\hatcurCCtwomassHmagxxxxA}{\ensuremath{10.677\pm0.024}}    % 2MASS ORIG MAG
\newcommand{\hatcurCCtwomassKmagxxxxA}{\ensuremath{10.626\pm0.026}}    % 2MASS ORIG MAG
\newcommand{\hatcurCCcitJmagxxxxA}{\ensuremath{10.979\pm0.023}}        % 2MASS CIT MAG
\newcommand{\hatcurCCcitHmagxxxxA}{\ensuremath{10.672\pm0.024}}        % 2MASS CIT MAG
\newcommand{\hatcurCCcitKmagxxxxA}{\ensuremath{10.650\pm0.026}}        % 2MASS CIT MAG
\newcommand{\hatcurCCbbJmagxxxxA}{\ensuremath{11.025\pm0.025}}         % 2MASS BB MAG
\newcommand{\hatcurCCbbHmagxxxxA}{\ensuremath{10.693\pm0.025}}         % 2MASS BB MAG
\newcommand{\hatcurCCbbKmagxxxxA}{\ensuremath{10.670\pm0.026}}         % 2MASS BB MAG
\newcommand{\hatcurCCesoJmagxxxxA}{\ensuremath{11.027\pm0.026}}        % 2MASS ESO MAG
\newcommand{\hatcurCCesoHmagxxxxA}{\ensuremath{10.688\pm0.028}}        % 2MASS ESO MAG
\newcommand{\hatcurCCesoKmagxxxxA}{\ensuremath{10.669\pm0.027}}        % 2MASS ESO MAG
\newcommand{\hatcurCCesoJHmagxxxxA}{\ensuremath{0.339\pm0.036}}        % 2MASS ESO JH COLOR
\newcommand{\hatcurCCesoJKmagxxxxA}{\ensuremath{0.359\pm0.037}}        % 2MASS ESO JK COLOR
\newcommand{\hatcurCCesoHKmagxxxxA}{\ensuremath{0.019\pm0.039}}        % 2MASS ESO HK COLOR
\newcommand{\hatcurLCdipxxxxA}{\ensuremath{7.6}}                       % BLS detected dip (mmag)
\newcommand{\hatcurLCrprstarxxxxA}{\ensuremath{0.0858\pm0.0032}}       % Rp/R*
\newcommand{\hatcurLCbsqxxxxA}{\ensuremath{0.332_{-0.136}^{+0.084}}}   % impact parameter square
\newcommand{\hatcurLCimpxxxxA}{\ensuremath{0.577_{-0.155}^{+0.067}}}   % impact parameter
\newcommand{\hatcurLCzetaxxxxA}{\ensuremath{13.40\pm0.14}}             % zeta/R*
\newcommand{\hatcurLCdurxxxxA}{\ensuremath{0.1680\pm0.0038}}           % transit duration (days)
\newcommand{\hatcurLCdurshortxxxxA}{\ensuremath{0.1680}}               % transit duration (days)
\newcommand{\hatcurLCdurhrxxxxA}{\ensuremath{4.033\pm0.091}}           % transit duration (hours)
\newcommand{\hatcurLCdurhrshortxxxxA}{\ensuremath{4.033}}              % transit duration (hours)
\newcommand{\hatcurLCqxxxxA}{\ensuremath{0.0362\pm0.0008}}             % fractional transit duration (days)
\newcommand{\hatcurLCqshortxxxxA}{\ensuremath{0.036}}                  % fractional transit duration (days)
\newcommand{\hatcurLCingdurxxxxA}{\ensuremath{0.0193\pm0.0036}}        % ingress/egress duration (days)
\newcommand{\hatcurLCPxxxxA}{\ensuremath{4.641876\pm0.000032}}         % period (days)
\newcommand{\hatcurLCPprecxxxxA}{\ensuremath{4.6418761}}               % period (days)
\newcommand{\hatcurLCPshortxxxxA}{\ensuremath{4.6419}}                 % period (days)
\newcommand{\hatcurLCTxxxxA}{\ensuremath{2455952.52606\pm0.00077}}     % epoch (BJD)
\newcommand{\hatcurLCTAxxxxA}{\ensuremath{2455506.90591\pm0.00295}}    % TA (BJD)
\newcommand{\hatcurLCTBxxxxA}{\ensuremath{2455989.66106\pm0.00087}}    % TB (BJD)
\newcommand{\hatcurLChatnetmxxxxA}{\ensuremath{11.9999\pm0.0001}}      % HATNet OOT level
\newcommand{\hatcurLCiblendxxxxA}{\ensuremath{0.56\pm0.04}}            % HATNet iblend factor
\newcommand{\hatcurSMEiteffxxxxA}{\ensuremath{5838\pm50}}              % Ini SME, stellar effective temperature
\newcommand{\hatcurSMEizfehxxxxA}{\ensuremath{0.33\pm0.08}}            % Ini SME, stellar metallicity
\newcommand{\hatcurSMEizfehshortxxxxA}{\ensuremath{0.33}}              % Ini SME, stellar metallicity
\newcommand{\hatcurSMEiloggxxxxA}{\ensuremath{4.33\pm0.10}}            % Ini SME, stellar surface gravity
\newcommand{\hatcurSMEivsinxxxxA}{\ensuremath{3.4\pm0.5}}              % Ini SME, stellar rotational velocity
\newcommand{\hatcurSMEivmacxxxxA}{\ensuremath{NULL}}                   % Ini SME, stellar macroturbulence
\newcommand{\hatcurSMEivmicxxxxA}{\ensuremath{NULL}}                   % Ini SME, stellar microturbulence
\newcommand{\hatcurSMEiiteffxxxxA}{\ensuremath{5743\pm50}}             % Final SME, stellar effective temperature
\newcommand{\hatcurSMEiizfehxxxxA}{\ensuremath{0.27\pm0.08}}           % Final SME, stellar metallicity
\newcommand{\hatcurSMEiizfehshortxxxxA}{\ensuremath{0.27}}             % Final SME, stellar metallicity
\newcommand{\hatcurSMEiiloggxxxxA}{\ensuremath{4.15\pm0.07}}           % Final SME, stellar surface gravity
\newcommand{\hatcurSMEiivsinxxxxA}{\ensuremath{3.5\pm0.5}}             % Final SME, stellar rotational velocity
\newcommand{\hatcurSMEiivmacxxxxA}{\ensuremath{NULL}}                  % Final SME, stellar macroturbulence
\newcommand{\hatcurSMEiivmicxxxxA}{\ensuremath{NULL}}                  % Final SME, stellar microturbulence
\newcommand{\hatcurDSteffxxxxA}{\ensuremath{NULL\pmNULL}}              % DS stellar effective temperature
\newcommand{\hatcurDSzfehxxxxA}{\ensuremath{NULL\pmNULL}}              % DS stellar metallicity
\newcommand{\hatcurDSloggxxxxA}{\ensuremath{NULL\pmNULL}}              % DS stellar surface gravity
\newcommand{\hatcurDSvsinixxxxA}{\ensuremath{NULL\pmNULL}}             % DS stellar rotational velocity
\newcommand{\hatcurDSgammaxxxxA}{\ensuremath{NULL\pmNULL}}             % DS absolute gamma velocity
\newcommand{\hatcurDSnumspecxxxxA}{\ensuremath{0}}                     % DS number of spectra
\newcommand{\hatcurDSspanxxxxA}{\ensuremath{0}}                        % DS stellar surface gravity
\newcommand{\hatcurDSrvrmsxxxxA}{\ensuremath{0.00}}                    % DS rms of RV values [km/s]
\newcommand{\hatcurTRESteffxxxxA}{\ensuremath{5838\pm50}}              % TRES stellar effective temperature
\newcommand{\hatcurTRESzfehxxxxA}{\ensuremath{0.33\pm0.08}}            % TRES stellar metallicity
\newcommand{\hatcurTRESloggxxxxA}{\ensuremath{4.33\pm0.10}}            % TRES stellar surface gravity
\newcommand{\hatcurTRESvsinixxxxA}{\ensuremath{3.4\pm0.5}}             % TRES stellar rotational velocity
\newcommand{\hatcurTRESgammaxxxxA}{\ensuremath{21.258\pm0.286}}        % TRES absolute gamma velocity
\newcommand{\hatcurTRESnumspecxxxxA}{\ensuremath{13}}                  % TRES number of spectra
\newcommand{\hatcurTRESspanxxxxA}{\ensuremath{0}}                      % TRES stellar surface gravity
\newcommand{\hatcurTRESrvrmsxxxxA}{\ensuremath{0.00}}                  % TRES rms of RV values [km/s]
\newcommand{\hatcurFIESteffxxxxA}{\ensuremath{NULL\pmNULL}}            % FIES stellar effective temperature
\newcommand{\hatcurFIESzfehxxxxA}{\ensuremath{NULL\pmNULL}}            % FIES stellar metallicity
\newcommand{\hatcurFIESloggxxxxA}{\ensuremath{NULL\pmNULL}}            % FIES stellar surface gravity
\newcommand{\hatcurFIESvsinixxxxA}{\ensuremath{NULL\pmNULL}}           % FIES stellar rotational velocity
\newcommand{\hatcurFIESgammaxxxxA}{\ensuremath{NULL\pmNULL}}           % FIES absolute gamma velocity
\newcommand{\hatcurFIESnumspecxxxxA}{\ensuremath{0}}                   % FIES number of spectra
\newcommand{\hatcurFIESspanxxxxA}{\ensuremath{0}}                      % FIES stellar surface gravity
\newcommand{\hatcurFIESrvrmsxxxxA}{\ensuremath{0.00}}                  % FIES rms of RV values [km/s]
\newcommand{\hatcurLBizxxxxA}{\ensuremath{0.2191}}                     % Limb darkening parameters, Gamma1, z-band
\newcommand{\hatcurLBiizxxxxA}{\ensuremath{0.3294}}                    % Limb darkening parameters, Gamma2, z-band
\newcommand{\hatcurLBiixxxxA}{\ensuremath{0.2863}}                     % Limb darkening parameters, Gamma1, i-band
\newcommand{\hatcurLBiiixxxxA}{\ensuremath{0.3274}}                    % Limb darkening parameters, Gamma2, i-band
\newcommand{\hatcurLBiIxxxxA}{\ensuremath{0.2637}}                     % Limb darkening parameters, Gamma1, I-band
\newcommand{\hatcurLBiiIxxxxA}{\ensuremath{0.3288}}                    % Limb darkening parameters, Gamma2, I-band
\newcommand{\hatcurLBigxxxxA}{\ensuremath{0.5885}}                     % Limb darkening parameters, Gamma1, g-band
\newcommand{\hatcurLBiigxxxxA}{\ensuremath{0.2118}}                    % Limb darkening parameters, Gamma2, g-band
\newcommand{\hatcurLBirxxxxA}{\ensuremath{0.3823}}                     % Limb darkening parameters, Gamma1, r-band
\newcommand{\hatcurLBiirxxxxA}{\ensuremath{0.3149}}                    % Limb darkening parameters, Gamma2, r-band
\newcommand{\hatcurLBiRxxxxA}{\ensuremath{0.3556}}                     % Limb darkening parameters, Gamma1, R-band
\newcommand{\hatcurLBiiRxxxxA}{\ensuremath{0.3194}}                    % Limb darkening parameters, Gamma2, R-band
\newcommand{\hatcurLBikepxxxxA}{\ensuremath{}}                 % darkening parameters, Gamma1, Kep-band
\newcommand{\hatcurLBiikepxxxxA}{\ensuremath{}}                % darkening parameters, Gamma2, Kep-band
\newcommand{\hatcurISOmxxxxA}{\ensuremath{1.18\pm0.07}}                % stellar mass
\newcommand{\hatcurISOmshortxxxxA}{\ensuremath{1.18}}                  % stellar mass
\newcommand{\hatcurISOmlongxxxxA}{\ensuremath{1.179\pm0.067}}          % stellar mass
\newcommand{\hatcurISOrxxxxA}{\ensuremath{1.53\pm0.14}}                % stellar radius
\newcommand{\hatcurISOrshortxxxxA}{\ensuremath{1.53}}                  % stellar radius
\newcommand{\hatcurISOrlongxxxxA}{\ensuremath{1.528\pm0.136}}          % stellar radius
\newcommand{\hatcurISOrhoxxxxA}{\ensuremath{0.46_{-0.08}^{+0.16}}}     % stellar density (cgs)
\newcommand{\hatcurISOloggxxxxA}{\ensuremath{4.14\pm0.07}}             % stellar surface gravity from isochrones
\newcommand{\hatcurISOlumxxxxA}{\ensuremath{2.27\pm0.42}}              % stellar luminosity
\newcommand{\hatcurISOlumshortxxxxA}{\ensuremath{2.27}}                % stellar luminosity
\newcommand{\hatcurISOmvxxxxA}{\ensuremath{3.94\pm0.21}}               % stellar absolute magnitude
\newcommand{\hatcurISOvixxxxA}{\ensuremath{0.713\pm0.016}}             % stellar V-I index
\newcommand{\hatcurISOagexxxxA}{\ensuremath{5.1_{-0.7}^{+1.8}}}        % stellar age
\newcommand{\hatcurISOsigmaxxxxA}{\ensuremath{0.00060\pm0.00012}}      % system mass-correction sigma parameter
\newcommand{\hatcurISOMJxxxxA}{\ensuremath{2.78\pm0.20}}               % stellar absolute J magnitude
\newcommand{\hatcurISOMHxxxxA}{\ensuremath{2.44\pm0.20}}               % stellar absolute H magnitude
\newcommand{\hatcurISOMKxxxxA}{\ensuremath{2.38\pm0.20}}               % stellar absolute K magnitude
\newcommand{\hatcurISOJKxxxxA}{\ensuremath{0.39\pm0.01}}               % J-K color index from isochrones.
\newcommand{\hatcurISOspecxxxxA}{G}                                    % stellar spectral type
\newcommand{\hatcurRVKxxxxA}{\ensuremath{105.8\pm13.0}}                % RV semi-amplitude [m/s]
\newcommand{\hatcurRVrkxxxxA}{\ensuremath{0.000\pm0.000}}              % sqrt(e)*cos(omega)
\newcommand{\hatcurRVrhxxxxA}{\ensuremath{0.000\pm0.000}}              % sqrt(e)*sin(omega)
\newcommand{\hatcurRVkxxxxA}{\ensuremath{0.000\pm0.000}}               % e*cos(omega)
\newcommand{\hatcurRVhxxxxA}{\ensuremath{0.000\pm0.000}}               % e*sin(omega)
\newcommand{\hatcurRVtronexxxxA}{\ensuremath{0.0000\pm0.0000}}         % RV linear trend tr1 factor
\newcommand{\hatcurRVtrtwoxxxxA}{\ensuremath{0.0000\pm0.0000}}         % RV linear trend tr2 factor
\newcommand{\hatcurRVgammaAxxxxA}{\ensuremath{20561.1\pm13.9}}         % RV gamma velocity, relative scale
\newcommand{\hatcurRVjitterAxxxxA}{\ensuremath{0.0}}                   % RV jitter (m/s)
\newcommand{\hatcurRVfitrmsAxxxxA}{\ensuremath{.1fym}}                 % 
\newcommand{\hatcurRVgammaBxxxxA}{\ensuremath{14.2\pm15.2}}            % RV gamma velocity, relative scale
\newcommand{\hatcurRVjitterBxxxxA}{\ensuremath{35.0}}                  % RV jitter (m/s)
\newcommand{\hatcurRVfitrmsBxxxxA}{\ensuremath{.1fym}}                 % 
\newcommand{\hatcurRVeccenxxxxA}{\ensuremath{0.000\pm0.000}}           % eccentricity
\newcommand{\hatcurRVomegaxxxxA}{\ensuremath{0\pm0}}                   % argument of pericenter
\newcommand{\hatcurPPixxxxA}{\ensuremath{85.9_{-0.8}^{+1.3}}}          % orbital inclination
\newcommand{\hatcurPPgxxxxA}{\ensuremath{14.8_{-2.8}^{+4.9}}}          % planetary surface gravity (m/s^2)
\newcommand{\hatcurPPloggxxxxA}{\ensuremath{3.17\pm0.11}}              % planetary surface gravity (log cgs)
\newcommand{\hatcurPParxxxxA}{\ensuremath{8.09_{-0.54}^{+0.79}}}       % relative orbital radius (a/R*)
\newcommand{\hatcurPParelxxxxA}{\ensuremath{0.0575\pm0.0011}}          % semimajor axis (AU)
\newcommand{\hatcurPPrhoxxxxA}{\ensuremath{0.58_{-0.15}^{+0.31}}}      % planetary density (cgs)
\newcommand{\hatcurPPmxxxxA}{\ensuremath{0.97\pm0.13}}                 % planetary mass (M_jup)
\newcommand{\hatcurPPmshortxxxxA}{\ensuremath{0.97}}                   % planetary mass (M_jup)
\newcommand{\hatcurPPmlongxxxxA}{\ensuremath{0.975\pm0.126}}           % planetary mass (M_jup)
\newcommand{\hatcurPPmexxxxA}{\ensuremath{309.8\pm40.0}}               % planetary mass (M_earth)
\newcommand{\hatcurPPmeshortxxxxA}{\ensuremath{309.8}}                 % planetary mass (M_earth)
\newcommand{\hatcurPPmelongxxxxA}{\ensuremath{309.76\pm40.03}}         % planetary mass (M_earth)
\newcommand{\hatcurPPrxxxxA}{\ensuremath{1.28\pm0.15}}                 % planetary radius (R_jup)
\newcommand{\hatcurPPrshortxxxxA}{\ensuremath{1.28}}                   % planetary radius (R_jup)
\newcommand{\hatcurPPrlongxxxxA}{\ensuremath{1.277\pm0.149}}           % planetary radius (R_jup)
\newcommand{\hatcurPPrexxxxA}{\ensuremath{14.3\pm1.7}}                 % planetary radius (R_earth)
\newcommand{\hatcurPPreshortxxxxA}{\ensuremath{14.3}}                  % planetary radius (R_earth)
\newcommand{\hatcurPPrelongxxxxA}{\ensuremath{14.32\pm1.67}}           % planetary radius (R_earth)
\newcommand{\hatcurPPmrcorrxxxxA}{\ensuremath{0.18}}                   % mass/radius correlation
\newcommand{\hatcurPPteffxxxxA}{\ensuremath{1427\pm58}}                % planetary temperature (K)
\newcommand{\hatcurPPthetaxxxxA}{\ensuremath{0.074_{-0.011}^{+0.015}}} % Safranov number
\newcommand{\hatcurPPfluxperixxxxA}{\ensuremath{9.36\pm1.52}}          % flux @ periastron (CGS)
\newcommand{\hatcurPPfluxperidimxxxxA}{\ensuremath{8}}                 % flux @ periastron (CGS) units.
\newcommand{\hatcurPPfluxapxxxxA}{\ensuremath{9.36\pm1.52}}            % flux @ apastron (CGS)
\newcommand{\hatcurPPfluxapdimxxxxA}{\ensuremath{8}}                   % flux @ apastron (CGS) units.
\newcommand{\hatcurPPfluxavgxxxxA}{\ensuremath{9.36\pm1.52}}           % flux on average (CGS)
\newcommand{\hatcurPPfluxavgdimxxxxA}{\ensuremath{8}}                  % flux average (CGS) units.
\newcommand{\hatcurXsecphasexxxxA}{\ensuremath{0.5000\pm0.0000}}       % Phase of secondary eclipse
\newcommand{\hatcurXsecondaryxxxxA}{\ensuremath{2455954.847\pm0.001}}  % Secondary eclipse epoch
\newcommand{\hatcurXsecdurxxxxA}{\ensuremath{0.1680\pm0.0038}}         % sec eclipse duration (days)
\newcommand{\hatcurXsecingdurxxxxA}{\ensuremath{0.0193\pm0.0036}}      % sec I/E duration (days)
\newcommand{\hatcurPPphiconjxxxxA}{\ensuremath{0.2500\pm0.0000}}       % phase diff between conjunction and periastron
\newcommand{\hatcurPPperixxxxA}{\ensuremath{2455951.37\pm0.00}}        % time of periastron passage.
\newcommand{\hatcurPPaequivxxxxA}{\ensuremath{0.0382_{-0.0026}^{+0.0038}}} % equivalent semi-major axis
\newcommand{\hatcurPPtcircxxxxA}{\ensuremath{747.7_{-262.7}^{+825.0}}} % circularization timescale
\newcommand{\hatcurPPtinfallxxxxA}{\ensuremath{1876.2_{-537.4}^{+1271.9}}} % infall timescale
\newcommand{\hatcurXdistxxxxA}{\ensuremath{453\pm41}}                  % distance (pc), no reddenning correction
\newcommand{\hatcurXAvxxxxA}{\ensuremath{0.000\pm0.020}}               % Av (mag)
\newcommand{\hatcurXdistredxxxxA}{\ensuremath{447\pm40}}               % distance with Av correction (pc)
\newcommand{\hatcurXEBVxxxxA}{\ensuremath{0.000\pm0.007}}              % E(B-V) (mag)
\newcommand{\hatcurXmvisoredxxxxA}{\ensuremath{12.195\pm0.026}}        % Expected m_v with reddening (mag)
\newcommand{\hatcurXmiisoredxxxxA}{\ensuremath{11.479\pm0.016}}        % Expected m_i with reddening (mag)
\newcommand{\hatcurXmjisoredxxxxA}{\ensuremath{11.030\pm0.014}}        % Expected m_j with reddening (mag)
\newcommand{\hatcurXmhisoredxxxxA}{\ensuremath{10.694\pm0.016}}        % Expected m_h with reddening (mag)
\newcommand{\hatcurXmkisoredxxxxA}{\ensuremath{10.638\pm0.017}}        % Expected m_k with reddening (mag)
\newcommand{\hatcurXviisoredxxxxA}{\ensuremath{0.716\pm0.015}}         % Expected V-I with reddening (mag)
\newcommand{\hatcurXvkisoredxxxxA}{\ensuremath{1.556\pm0.031}}         % Expected V-K with reddening (mag)
\newcommand{\hatcurXjhisoredxxxxA}{\ensuremath{0.336\pm0.009}}         % Expected J-H with reddening (mag)
\newcommand{\hatcurXjkisoredxxxxA}{\ensuremath{0.392\pm0.010}}         % Expected J-K with reddening (mag)
\newcommand{\hatcurCCpmraxxxxA}{\ensuremath{-6.2\pm1.9}}               % proper motion, in RA
\newcommand{\hatcurCCpmdecxxxxA}{\ensuremath{-29.3\pm2.0}}             % proper motion, in DEC
\newcommand{\hatcurCCpmxxxxA}{\ensuremath{29.9488\pm2.75862}}          % proper motion

\newcommand{\hatcurhtrxxxxB}{HTR365-006}                               % Original HTR name of target
\newcommand{\hatcurfieldxxxxB}{317}                                    % Original HTR field
\newcommand{\hatcurCCraxxxxB}{\ensuremath{08^{\mathrm h}35^{\mathrm m}42.18{\mathrm s}}}                             % Right Ascension
\newcommand{\hatcurCCdecxxxxB}{\ensuremath{+10{\arcdeg}12{\arcmin}24.0{\arcsec}}}                            % Declination
\newcommand{\hatcurCCmagxxxxB}{13.356}                                 % apparent V-band magnitude
\newcommand{\hatcurCCtwomassxxxxB}{2MASS~08354217+1012239}             % 2MASS identifier
\newcommand{\hatcurCCgscxxxxB}{GSC~0801-00608}                         % GSC(1.2) identifier
\newcommand{\hatcurCCtassmvxxxxB}{\ensuremath{13.356\pm0.030}}                              % note, APASS V-band magnitude
\newcommand{\hatcurCCtassmbxxxxB}{\ensuremath{14.120\pm0.060}}                              % note, APASS B-band magnitude
\newcommand{\hatcurCCtwomassJmagxxxxB}{\ensuremath{12.146\pm0.021}}    % 2MASS ORIG MAG
\newcommand{\hatcurCCtwomassHmagxxxxB}{\ensuremath{11.809\pm0.029}}    % 2MASS ORIG MAG
\newcommand{\hatcurCCtwomassKmagxxxxB}{\ensuremath{11.764\pm0.023}}    % 2MASS ORIG MAG
\newcommand{\hatcurCCcitJmagxxxxB}{\ensuremath{12.162\pm0.021}}        % 2MASS CIT MAG
\newcommand{\hatcurCCcitHmagxxxxB}{\ensuremath{11.804\pm0.029}}        % 2MASS CIT MAG
\newcommand{\hatcurCCcitKmagxxxxB}{\ensuremath{11.788\pm0.023}}        % 2MASS CIT MAG
\newcommand{\hatcurCCbbJmagxxxxB}{\ensuremath{12.213\pm0.023}}         % 2MASS BB MAG
\newcommand{\hatcurCCbbHmagxxxxB}{\ensuremath{11.825\pm0.030}}         % 2MASS BB MAG
\newcommand{\hatcurCCbbKmagxxxxB}{\ensuremath{11.808\pm0.023}}         % 2MASS BB MAG
\newcommand{\hatcurCCesoJmagxxxxB}{\ensuremath{12.215\pm0.025}}        % 2MASS ESO MAG
\newcommand{\hatcurCCesoHmagxxxxB}{\ensuremath{11.819\pm0.033}}        % 2MASS ESO MAG
\newcommand{\hatcurCCesoKmagxxxxB}{\ensuremath{11.807\pm0.024}}        % 2MASS ESO MAG
\newcommand{\hatcurCCesoJHmagxxxxB}{\ensuremath{0.395\pm0.039}}        % 2MASS ESO JH COLOR
\newcommand{\hatcurCCesoJKmagxxxxB}{\ensuremath{0.409\pm0.034}}        % 2MASS ESO JK COLOR
\newcommand{\hatcurCCesoHKmagxxxxB}{\ensuremath{0.013\pm0.041}}        % 2MASS ESO HK COLOR
\newcommand{\hatcurLCdipxxxxB}{\ensuremath{16.5}}                      % BLS detected dip (mmag)
\newcommand{\hatcurLCrprstarxxxxB}{\ensuremath{0.1194\pm0.0018}}       % Rp/R*
\newcommand{\hatcurLCbsqxxxxB}{\ensuremath{0.039_{-0.026}^{+0.053}}}   % impact parameter square
\newcommand{\hatcurLCimpxxxxB}{\ensuremath{0.197_{-0.101}^{+0.098}}}   % impact parameter
\newcommand{\hatcurLCzetaxxxxB}{\ensuremath{16.60\pm0.09}}             % zeta/R*
\newcommand{\hatcurLCdurxxxxB}{\ensuremath{0.1355\pm0.0011}}           % transit duration (days)
\newcommand{\hatcurLCdurshortxxxxB}{\ensuremath{0.1355}}               % transit duration (days)
\newcommand{\hatcurLCdurhrxxxxB}{\ensuremath{3.251\pm0.026}}           % transit duration (hours)
\newcommand{\hatcurLCdurhrshortxxxxB}{\ensuremath{3.251}}              % transit duration (hours)
\newcommand{\hatcurLCqxxxxB}{\ensuremath{0.0407\pm0.0003}}             % fractional transit duration (days)
\newcommand{\hatcurLCqshortxxxxB}{\ensuremath{0.041}}                  % fractional transit duration (days)
\newcommand{\hatcurLCingdurxxxxB}{\ensuremath{0.0150\pm0.0008}}        % ingress/egress duration (days)
\newcommand{\hatcurLCPxxxxB}{\ensuremath{3.332688\pm0.000016}}         % period (days)
\newcommand{\hatcurLCPprecxxxxB}{\ensuremath{3.3326877}}               % period (days)
\newcommand{\hatcurLCPshortxxxxB}{\ensuremath{3.3327}}                 % period (days)
\newcommand{\hatcurLCTxxxxB}{\ensuremath{2455997.37105\pm0.00033}}     % epoch (BJD)
\newcommand{\hatcurLCTAxxxxB}{\ensuremath{2455504.13328\pm0.00230}}    % TA (BJD)
\newcommand{\hatcurLCTBxxxxB}{\ensuremath{2456010.70181\pm0.00036}}    % TB (BJD)
\newcommand{\hatcurLChatnetmAxxxxB}{\ensuremath{13.2562\pm0.0002}}     % HATNet OOT level
\newcommand{\hatcurLCiblendAxxxxB}{\ensuremath{0.78\pm0.07}}           % HATNet iblend factor
\newcommand{\hatcurLChatnetmBxxxxB}{\ensuremath{13.2559\pm0.0002}}     % HATNet OOT level
\newcommand{\hatcurLCiblendBxxxxB}{\ensuremath{0.86\pm0.07}}           % HATNet iblend factor
\newcommand{\hatcurSMEiteffxxxxB}{\ensuremath{5738\pm74}}              % Ini SME, stellar effective temperature
\newcommand{\hatcurSMEizfehxxxxB}{\ensuremath{0.28\pm0.08}}            % Ini SME, stellar metallicity
\newcommand{\hatcurSMEizfehshortxxxxB}{\ensuremath{0.28}}              % Ini SME, stellar metallicity
\newcommand{\hatcurSMEiloggxxxxB}{\ensuremath{4.47\pm0.13}}            % Ini SME, stellar surface gravity
\newcommand{\hatcurSMEivsinxxxxB}{\ensuremath{2.4\pm0.5}}              % Ini SME, stellar rotational velocity
\newcommand{\hatcurSMEivmacxxxxB}{\ensuremath{NULL}}                   % Ini SME, stellar macroturbulence
\newcommand{\hatcurSMEivmicxxxxB}{\ensuremath{NULL}}                   % Ini SME, stellar microturbulence
\newcommand{\hatcurSMEiiteffxxxxB}{\ensuremath{5645\pm74}}             % Final SME, stellar effective temperature
\newcommand{\hatcurSMEiizfehxxxxB}{\ensuremath{0.23\pm0.08}}           % Final SME, stellar metallicity
\newcommand{\hatcurSMEiizfehshortxxxxB}{\ensuremath{0.23}}             % Final SME, stellar metallicity
\newcommand{\hatcurSMEiiloggxxxxB}{\ensuremath{4.38\pm0.02}}           % Final SME, stellar surface gravity
\newcommand{\hatcurSMEiivsinxxxxB}{\ensuremath{2.4\pm0.5}}             % Final SME, stellar rotational velocity
\newcommand{\hatcurSMEiivmacxxxxB}{\ensuremath{NULL}}                  % Final SME, stellar macroturbulence
\newcommand{\hatcurSMEiivmicxxxxB}{\ensuremath{NULL}}                  % Final SME, stellar microturbulence
\newcommand{\hatcurDSteffxxxxB}{\ensuremath{NULL\pmNULL}}              % DS stellar effective temperature
\newcommand{\hatcurDSzfehxxxxB}{\ensuremath{NULL\pmNULL}}              % DS stellar metallicity
\newcommand{\hatcurDSloggxxxxB}{\ensuremath{NULL\pmNULL}}              % DS stellar surface gravity
\newcommand{\hatcurDSvsinixxxxB}{\ensuremath{NULL\pmNULL}}             % DS stellar rotational velocity
\newcommand{\hatcurDSgammaxxxxB}{\ensuremath{NULL\pmNULL}}             % DS absolute gamma velocity
\newcommand{\hatcurDSnumspecxxxxB}{\ensuremath{0}}                     % DS number of spectra
\newcommand{\hatcurDSspanxxxxB}{\ensuremath{0}}                        % DS stellar surface gravity
\newcommand{\hatcurDSrvrmsxxxxB}{\ensuremath{0.00}}                    % DS rms of RV values [km/s]
\newcommand{\hatcurTRESteffxxxxB}{\ensuremath{5738\pm74}}              % TRES stellar effective temperature
\newcommand{\hatcurTRESzfehxxxxB}{\ensuremath{0.28\pm0.1}}             % TRES stellar metallicity
\newcommand{\hatcurTRESloggxxxxB}{\ensuremath{4.47\pm0.13}}            % TRES stellar surface gravity
\newcommand{\hatcurTRESvsinixxxxB}{\ensuremath{2.4\pm0.5}}             % TRES stellar rotational velocity
\newcommand{\hatcurTRESgammaxxxxB}{\ensuremath{-4.42\pm0.16}}          % TRES absolute gamma velocity
\newcommand{\hatcurTRESnumspecxxxxB}{\ensuremath{2}}                   % TRES number of spectra
\newcommand{\hatcurTRESspanxxxxB}{\ensuremath{0}}                      % TRES stellar surface gravity
\newcommand{\hatcurTRESrvrmsxxxxB}{\ensuremath{0.00}}                  % TRES rms of RV values [km/s]
\newcommand{\hatcurFIESteffxxxxB}{\ensuremath{5738\pm74}}              % FIES stellar effective temperature
\newcommand{\hatcurFIESzfehxxxxB}{\ensuremath{0.28\pm0.08}}            % FIES stellar metallicity
\newcommand{\hatcurFIESloggxxxxB}{\ensuremath{4.47\pm0.13}}            % FIES stellar surface gravity
\newcommand{\hatcurFIESvsinixxxxB}{\ensuremath{2.4\pm0.5}}             % FIES stellar rotational velocity
\newcommand{\hatcurFIESgammaxxxxB}{\ensuremath{-4.42\pm0.16}}          % FIES absolute gamma velocity
\newcommand{\hatcurFIESnumspecxxxxB}{\ensuremath{2}}                   % FIES number of spectra
\newcommand{\hatcurFIESspanxxxxB}{\ensuremath{0}}                      % FIES stellar surface gravity
\newcommand{\hatcurFIESrvrmsxxxxB}{\ensuremath{0.00}}                  % FIES rms of RV values [km/s]
\newcommand{\hatcurLBizxxxxB}{\ensuremath{0.2310}}                     % Limb darkening parameters, Gamma1, z-band
\newcommand{\hatcurLBiizxxxxB}{\ensuremath{0.3221}}                    % Limb darkening parameters, Gamma2, z-band
\newcommand{\hatcurLBiixxxxB}{\ensuremath{0.3003}}                     % Limb darkening parameters, Gamma1, i-band
\newcommand{\hatcurLBiiixxxxB}{\ensuremath{0.3180}}                    % Limb darkening parameters, Gamma2, i-band
\newcommand{\hatcurLBiIxxxxB}{\ensuremath{0.2772}}                     % Limb darkening parameters, Gamma1, I-band
\newcommand{\hatcurLBiiIxxxxB}{\ensuremath{0.3200}}                    % Limb darkening parameters, Gamma2, I-band
\newcommand{\hatcurLBigxxxxB}{\ensuremath{0.6101}}                     % Limb darkening parameters, Gamma1, g-band
\newcommand{\hatcurLBiigxxxxB}{\ensuremath{0.1944}}                    % Limb darkening parameters, Gamma2, g-band
\newcommand{\hatcurLBirxxxxB}{\ensuremath{0.3995}}                     % Limb darkening parameters, Gamma1, r-band
\newcommand{\hatcurLBiirxxxxB}{\ensuremath{0.3031}}                    % Limb darkening parameters, Gamma2, r-band
\newcommand{\hatcurLBiRxxxxB}{\ensuremath{0.3719}}                     % Limb darkening parameters, Gamma1, R-band
\newcommand{\hatcurLBiiRxxxxB}{\ensuremath{0.3082}}                    % Limb darkening parameters, Gamma2, R-band
\newcommand{\hatcurLBikepxxxxB}{\ensuremath{}}                 % darkening parameters, Gamma1, Kep-band
\newcommand{\hatcurLBiikepxxxxB}{\ensuremath{}}                % darkening parameters, Gamma2, Kep-band
\newcommand{\hatcurISOmxxxxB}{\ensuremath{1.05_{-0.04}^{+0.03}}}       % stellar mass
\newcommand{\hatcurISOmshortxxxxB}{\ensuremath{1.05}}                  % stellar mass
\newcommand{\hatcurISOmlongxxxxB}{\ensuremath{1.048_{-0.042}^{+0.031}}} % stellar mass
\newcommand{\hatcurISOrxxxxB}{\ensuremath{1.10_{-0.02}^{+0.04}}}       % stellar radius
\newcommand{\hatcurISOrshortxxxxB}{\ensuremath{1.10}}                  % stellar radius
\newcommand{\hatcurISOrlongxxxxB}{\ensuremath{1.104_{-0.022}^{+0.036}}} % stellar radius
\newcommand{\hatcurISOrhoxxxxB}{\ensuremath{1.09_{-0.09}^{+0.05}}}     % stellar density (cgs)
\newcommand{\hatcurISOloggxxxxB}{\ensuremath{4.37\pm0.02}}             % stellar surface gravity from isochrones
\newcommand{\hatcurISOlumxxxxB}{\ensuremath{1.12\pm0.10}}              % stellar luminosity
\newcommand{\hatcurISOlumshortxxxxB}{\ensuremath{1.12}}                % stellar luminosity
\newcommand{\hatcurISOmvxxxxB}{\ensuremath{4.73\pm0.11}}               % stellar absolute magnitude
\newcommand{\hatcurISOvixxxxB}{\ensuremath{0.744\pm0.024}}             % stellar V-I index
\newcommand{\hatcurISOagexxxxB}{\ensuremath{5.7_{-1.1}^{+1.9}}}        % stellar age
\newcommand{\hatcurISOsigmaxxxxB}{\ensuremath{0.00070\pm0.00010}}      % system mass-correction sigma parameter
\newcommand{\hatcurISOMJxxxxB}{\ensuremath{3.52\pm0.08}}               % stellar absolute J magnitude
\newcommand{\hatcurISOMHxxxxB}{\ensuremath{3.16\pm0.07}}               % stellar absolute H magnitude
\newcommand{\hatcurISOMKxxxxB}{\ensuremath{3.11\pm0.06}}               % stellar absolute K magnitude
\newcommand{\hatcurISOJKxxxxB}{\ensuremath{0.42\pm0.02}}               % J-K color index from isochrones.
\newcommand{\hatcurISOspecxxxxB}{G}                                    % stellar spectral type
\newcommand{\hatcurRVKxxxxB}{\ensuremath{87.5\pm10.8}}                 % RV semi-amplitude [m/s]
\newcommand{\hatcurRVrkxxxxB}{\ensuremath{0.000\pm0.000}}              % sqrt(e)*cos(omega)
\newcommand{\hatcurRVrhxxxxB}{\ensuremath{0.000\pm0.000}}              % sqrt(e)*sin(omega)
\newcommand{\hatcurRVkxxxxB}{\ensuremath{0.000\pm0.000}}               % e*cos(omega)
\newcommand{\hatcurRVhxxxxB}{\ensuremath{0.000\pm0.000}}               % e*sin(omega)
\newcommand{\hatcurRVtronexxxxB}{\ensuremath{0.0000\pm0.0000}}         % RV linear trend tr1 factor
\newcommand{\hatcurRVtrtwoxxxxB}{\ensuremath{0.0000\pm0.0000}}         % RV linear trend tr2 factor
\newcommand{\hatcurRVgammaxxxxB}{\ensuremath{-5081.6\pm8.6}}           % RV gamma velocity, relative scale
\newcommand{\hatcurRVjitterxxxxB}{\ensuremath{0.0}}                    % RV jitter (m/s)
\newcommand{\hatcurRVfitrmsxxxxB}{\ensuremath{10.0}}                   % RVfitrms
\newcommand{\hatcurRVeccenxxxxB}{\ensuremath{0.000\pm0.000}}           % eccentricity
\newcommand{\hatcurRVomegaxxxxB}{\ensuremath{0\pm0}}                   % argument of pericenter
\newcommand{\hatcurPPixxxxB}{\ensuremath{88.7\pm0.7}}                  % orbital inclination
\newcommand{\hatcurPPgxxxxB}{\ensuremath{9.9\pm1.4}}                   % planetary surface gravity (m/s^2)
\newcommand{\hatcurPPloggxxxxB}{\ensuremath{2.99\pm0.06}}              % planetary surface gravity (log cgs)
\newcommand{\hatcurPParxxxxB}{\ensuremath{8.63_{-0.24}^{+0.12}}}       % relative orbital radius (a/R*)
\newcommand{\hatcurPParelxxxxB}{\ensuremath{0.0443_{-0.0006}^{+0.0004}}} % semimajor axis (AU)
\newcommand{\hatcurPPrhoxxxxB}{\ensuremath{0.38\pm0.06}}               % planetary density (cgs)
\newcommand{\hatcurPPmxxxxB}{\ensuremath{0.66\pm0.08}}                 % planetary mass (M_jup)
\newcommand{\hatcurPPmshortxxxxB}{\ensuremath{0.66}}                   % planetary mass (M_jup)
\newcommand{\hatcurPPmlongxxxxB}{\ensuremath{0.660\pm0.083}}           % planetary mass (M_jup)
\newcommand{\hatcurPPmexxxxB}{\ensuremath{209.6\pm26.4}}               % planetary mass (M_earth)
\newcommand{\hatcurPPmeshortxxxxB}{\ensuremath{209.6}}                 % planetary mass (M_earth)
\newcommand{\hatcurPPmelongxxxxB}{\ensuremath{209.62\pm26.36}}         % planetary mass (M_earth)
\newcommand{\hatcurPPrxxxxB}{\ensuremath{1.28_{-0.03}^{+0.06}}}        % planetary radius (R_jup)
\newcommand{\hatcurPPrshortxxxxB}{\ensuremath{1.28}}                   % planetary radius (R_jup)
\newcommand{\hatcurPPrlongxxxxB}{\ensuremath{1.283_{-0.034}^{+0.057}}} % planetary radius (R_jup)
\newcommand{\hatcurPPrexxxxB}{\ensuremath{14.4_{-0.4}^{+0.6}}}         % planetary radius (R_earth)
\newcommand{\hatcurPPreshortxxxxB}{\ensuremath{14.4}}                  % planetary radius (R_earth)
\newcommand{\hatcurPPrelongxxxxB}{\ensuremath{14.38_{-0.38}^{+0.64}}}  % planetary radius (R_earth)
\newcommand{\hatcurPPmrcorrxxxxB}{\ensuremath{0.07}}                   % mass/radius correlation
\newcommand{\hatcurPPteffxxxxB}{\ensuremath{1361\pm24}}                % planetary temperature (K)
\newcommand{\hatcurPPthetaxxxxB}{\ensuremath{0.043\pm0.006}}           % Safranov number
\newcommand{\hatcurPPfluxperixxxxB}{\ensuremath{7.75\pm0.54}}          % flux @ periastron (CGS)
\newcommand{\hatcurPPfluxperidimxxxxB}{\ensuremath{8}}                 % flux @ periastron (CGS) units.
\newcommand{\hatcurPPfluxapxxxxB}{\ensuremath{7.75\pm0.54}}            % flux @ apastron (CGS)
\newcommand{\hatcurPPfluxapdimxxxxB}{\ensuremath{8}}                   % flux @ apastron (CGS) units.
\newcommand{\hatcurPPfluxavgxxxxB}{\ensuremath{7.75\pm0.54}}           % flux on average (CGS)
\newcommand{\hatcurPPfluxavgdimxxxxB}{\ensuremath{8}}                  % flux average (CGS) units.
\newcommand{\hatcurXsecphasexxxxB}{\ensuremath{0.5000\pm0.0000}}       % Phase of secondary eclipse
\newcommand{\hatcurXsecondaryxxxxB}{\ensuremath{2455999.037\pm0.000}}  % Secondary eclipse epoch
\newcommand{\hatcurXsecdurxxxxB}{\ensuremath{0.1355\pm0.0011}}         % sec eclipse duration (days)
\newcommand{\hatcurXsecingdurxxxxB}{\ensuremath{0.0150\pm0.0008}}      % sec I/E duration (days)
\newcommand{\hatcurPPphiconjxxxxB}{\ensuremath{0.2500\pm0.0000}}       % phase diff between conjunction and periastron
\newcommand{\hatcurPPperixxxxB}{\ensuremath{2455996.54\pm0.00}}        % time of periastron passage.
\newcommand{\hatcurPPaequivxxxxB}{\ensuremath{0.0420\pm0.0014}}        % equivalent semi-major axis
\newcommand{\hatcurPPtcircxxxxB}{\ensuremath{107.9\pm21.5}}            % circularization timescale
\newcommand{\hatcurPPtinfallxxxxB}{\ensuremath{2372.4\pm394.5}}        % infall timescale
\newcommand{\hatcurXdistxxxxB}{\ensuremath{549_{-14}^{+19}}}           % distance (pc), no reddenning correction
\newcommand{\hatcurXAvxxxxB}{\ensuremath{0.000\pm0.019}}               % Av (mag)
\newcommand{\hatcurXdistredxxxxB}{\ensuremath{543\pm18}}               % distance with Av correction (pc)
\newcommand{\hatcurXEBVxxxxB}{\ensuremath{0.000\pm0.006}}              % E(B-V) (mag)
\newcommand{\hatcurXmvisoredxxxxB}{\ensuremath{13.407_{-0.034}^{+0.044}}} % Expected m_v with reddening (mag)
\newcommand{\hatcurXmiisoredxxxxB}{\ensuremath{12.662\pm0.021}}        % Expected m_i with reddening (mag)
\newcommand{\hatcurXmjisoredxxxxB}{\ensuremath{12.197\pm0.014}}        % Expected m_j with reddening (mag)
\newcommand{\hatcurXmhisoredxxxxB}{\ensuremath{11.839\pm0.019}}        % Expected m_h with reddening (mag)
\newcommand{\hatcurXmkisoredxxxxB}{\ensuremath{11.780\pm0.022}}        % Expected m_k with reddening (mag)
\newcommand{\hatcurXviisoredxxxxB}{\ensuremath{0.746\pm0.021}}         % Expected V-I with reddening (mag)
\newcommand{\hatcurXvkisoredxxxxB}{\ensuremath{1.626_{-0.044}^{+0.061}}} % Expected V-K with reddening (mag)
\newcommand{\hatcurXjhisoredxxxxB}{\ensuremath{0.358_{-0.012}^{+0.017}}} % Expected J-H with reddening (mag)
\newcommand{\hatcurXjkisoredxxxxB}{\ensuremath{0.416_{-0.014}^{+0.020}}} % Expected J-K with reddening (mag)
\newcommand{\hatcurCCpmraxxxxB}{\ensuremath{-10.3\pm2.6}}              % proper motion, in RA
\newcommand{\hatcurCCpmdecxxxxB}{\ensuremath{-16.0\pm3.2}}             % proper motion, in DEC
\newcommand{\hatcurCCpmxxxxB}{\ensuremath{19.0287\pm4.12311}}          % proper motion
\newcommand{\hatcurCCbbHmag}[1]{\ifnum#1=42 %
\hatcurCCbbHmagxxxxA
\else
\ifnum#1=43 %
\hatcurCCbbHmagxxxxB
\else
??????\fi
\fi
}
\newcommand{\hatcurCCbbJmag}[1]{\ifnum#1=42 %
\hatcurCCbbJmagxxxxA
\else
\ifnum#1=43 %
\hatcurCCbbJmagxxxxB
\else
??????\fi
\fi
}
\newcommand{\hatcurCCbbKmag}[1]{\ifnum#1=42 %
\hatcurCCbbKmagxxxxA
\else
\ifnum#1=43 %
\hatcurCCbbKmagxxxxB
\else
??????\fi
\fi
}
\newcommand{\hatcurCCcitHmag}[1]{\ifnum#1=42 %
\hatcurCCcitHmagxxxxA
\else
\ifnum#1=43 %
\hatcurCCcitHmagxxxxB
\else
??????\fi
\fi
}
\newcommand{\hatcurCCcitJmag}[1]{\ifnum#1=42 %
\hatcurCCcitJmagxxxxA
\else
\ifnum#1=43 %
\hatcurCCcitJmagxxxxB
\else
??????\fi
\fi
}
\newcommand{\hatcurCCcitKmag}[1]{\ifnum#1=42 %
\hatcurCCcitKmagxxxxA
\else
\ifnum#1=43 %
\hatcurCCcitKmagxxxxB
\else
??????\fi
\fi
}
\newcommand{\hatcurCCdec}[1]{\ifnum#1=42 %
\hatcurCCdecxxxxA
\else
\ifnum#1=43 %
\hatcurCCdecxxxxB
\else
??????\fi
\fi
}
\newcommand{\hatcurCCesoHKmag}[1]{\ifnum#1=42 %
\hatcurCCesoHKmagxxxxA
\else
\ifnum#1=43 %
\hatcurCCesoHKmagxxxxB
\else
??????\fi
\fi
}
\newcommand{\hatcurCCesoHmag}[1]{\ifnum#1=42 %
\hatcurCCesoHmagxxxxA
\else
\ifnum#1=43 %
\hatcurCCesoHmagxxxxB
\else
??????\fi
\fi
}
\newcommand{\hatcurCCesoJHmag}[1]{\ifnum#1=42 %
\hatcurCCesoJHmagxxxxA
\else
\ifnum#1=43 %
\hatcurCCesoJHmagxxxxB
\else
??????\fi
\fi
}
\newcommand{\hatcurCCesoJKmag}[1]{\ifnum#1=42 %
\hatcurCCesoJKmagxxxxA
\else
\ifnum#1=43 %
\hatcurCCesoJKmagxxxxB
\else
??????\fi
\fi
}
\newcommand{\hatcurCCesoJmag}[1]{\ifnum#1=42 %
\hatcurCCesoJmagxxxxA
\else
\ifnum#1=43 %
\hatcurCCesoJmagxxxxB
\else
??????\fi
\fi
}
\newcommand{\hatcurCCesoKmag}[1]{\ifnum#1=42 %
\hatcurCCesoKmagxxxxA
\else
\ifnum#1=43 %
\hatcurCCesoKmagxxxxB
\else
??????\fi
\fi
}
\newcommand{\hatcurCCgsc}[1]{\ifnum#1=42 %
\hatcurCCgscxxxxA
\else
\ifnum#1=43 %
\hatcurCCgscxxxxB
\else
??????\fi
\fi
}
\newcommand{\hatcurCCmag}[1]{\ifnum#1=42 %
\hatcurCCmagxxxxA
\else
\ifnum#1=43 %
\hatcurCCmagxxxxB
\else
??????\fi
\fi
}
\newcommand{\hatcurCCpm}[1]{\ifnum#1=42 %
\hatcurCCpmxxxxA
\else
\ifnum#1=43 %
\hatcurCCpmxxxxB
\else
??????\fi
\fi
}
\newcommand{\hatcurCCpmdec}[1]{\ifnum#1=42 %
\hatcurCCpmdecxxxxA
\else
\ifnum#1=43 %
\hatcurCCpmdecxxxxB
\else
??????\fi
\fi
}
\newcommand{\hatcurCCpmra}[1]{\ifnum#1=42 %
\hatcurCCpmraxxxxA
\else
\ifnum#1=43 %
\hatcurCCpmraxxxxB
\else
??????\fi
\fi
}
\newcommand{\hatcurCCra}[1]{\ifnum#1=42 %
\hatcurCCraxxxxA
\else
\ifnum#1=43 %
\hatcurCCraxxxxB
\else
??????\fi
\fi
}
\newcommand{\hatcurCCtassmv}[1]{\ifnum#1=42 %
\hatcurCCtassmvxxxxA
\else
\ifnum#1=43 %
\hatcurCCtassmvxxxxB
\else
??????\fi
\fi
}
\newcommand{\hatcurCCtassmb}[1]{\ifnum#1=42 %
\hatcurCCtassmbxxxxA
\else
\ifnum#1=43 %
\hatcurCCtassmbxxxxB
\else
??????\fi
\fi
}
\newcommand{\hatcurCCtwomass}[1]{\ifnum#1=42 %
\hatcurCCtwomassxxxxA
\else
\ifnum#1=43 %
\hatcurCCtwomassxxxxB
\else
??????\fi
\fi
}
\newcommand{\hatcurCCtwomassHmag}[1]{\ifnum#1=42 %
\hatcurCCtwomassHmagxxxxA
\else
\ifnum#1=43 %
\hatcurCCtwomassHmagxxxxB
\else
??????\fi
\fi
}
\newcommand{\hatcurCCtwomassJmag}[1]{\ifnum#1=42 %
\hatcurCCtwomassJmagxxxxA
\else
\ifnum#1=43 %
\hatcurCCtwomassJmagxxxxB
\else
??????\fi
\fi
}
\newcommand{\hatcurCCtwomassKmag}[1]{\ifnum#1=42 %
\hatcurCCtwomassKmagxxxxA
\else
\ifnum#1=43 %
\hatcurCCtwomassKmagxxxxB
\else
??????\fi
\fi
}
\newcommand{\hatcurDSgamma}[1]{\ifnum#1=42 %
\hatcurDSgammaxxxxA
\else
\ifnum#1=43 %
\hatcurDSgammaxxxxB
\else
??????\fi
\fi
}
\newcommand{\hatcurDSlogg}[1]{\ifnum#1=42 %
\hatcurDSloggxxxxA
\else
\ifnum#1=43 %
\hatcurDSloggxxxxB
\else
??????\fi
\fi
}
\newcommand{\hatcurDSnumspec}[1]{\ifnum#1=42 %
\hatcurDSnumspecxxxxA
\else
\ifnum#1=43 %
\hatcurDSnumspecxxxxB
\else
??????\fi
\fi
}
\newcommand{\hatcurDSrvrms}[1]{\ifnum#1=42 %
\hatcurDSrvrmsxxxxA
\else
\ifnum#1=43 %
\hatcurDSrvrmsxxxxB
\else
??????\fi
\fi
}
\newcommand{\hatcurDSspan}[1]{\ifnum#1=42 %
\hatcurDSspanxxxxA
\else
\ifnum#1=43 %
\hatcurDSspanxxxxB
\else
??????\fi
\fi
}
\newcommand{\hatcurDSteff}[1]{\ifnum#1=42 %
\hatcurDSteffxxxxA
\else
\ifnum#1=43 %
\hatcurDSteffxxxxB
\else
??????\fi
\fi
}
\newcommand{\hatcurDSvsini}[1]{\ifnum#1=42 %
\hatcurDSvsinixxxxA
\else
\ifnum#1=43 %
\hatcurDSvsinixxxxB
\else
??????\fi
\fi
}
\newcommand{\hatcurDSzfeh}[1]{\ifnum#1=42 %
\hatcurDSzfehxxxxA
\else
\ifnum#1=43 %
\hatcurDSzfehxxxxB
\else
??????\fi
\fi
}
\newcommand{\hatcurfield}[1]{\ifnum#1=42 %
\hatcurfieldxxxxA
\else
\ifnum#1=43 %
\hatcurfieldxxxxB
\else
??????\fi
\fi
}
\newcommand{\hatcurFIESgamma}[1]{\ifnum#1=42 %
\hatcurFIESgammaxxxxA
\else
\ifnum#1=43 %
\hatcurFIESgammaxxxxB
\else
??????\fi
\fi
}
\newcommand{\hatcurFIESlogg}[1]{\ifnum#1=42 %
\hatcurFIESloggxxxxA
\else
\ifnum#1=43 %
\hatcurFIESloggxxxxB
\else
??????\fi
\fi
}
\newcommand{\hatcurFIESnumspec}[1]{\ifnum#1=42 %
\hatcurFIESnumspecxxxxA
\else
\ifnum#1=43 %
\hatcurFIESnumspecxxxxB
\else
??????\fi
\fi
}
\newcommand{\hatcurFIESrvrms}[1]{\ifnum#1=42 %
\hatcurFIESrvrmsxxxxA
\else
\ifnum#1=43 %
\hatcurFIESrvrmsxxxxB
\else
??????\fi
\fi
}
\newcommand{\hatcurFIESspan}[1]{\ifnum#1=42 %
\hatcurFIESspanxxxxA
\else
\ifnum#1=43 %
\hatcurFIESspanxxxxB
\else
??????\fi
\fi
}
\newcommand{\hatcurFIESteff}[1]{\ifnum#1=42 %
\hatcurFIESteffxxxxA
\else
\ifnum#1=43 %
\hatcurFIESteffxxxxB
\else
??????\fi
\fi
}
\newcommand{\hatcurFIESvsini}[1]{\ifnum#1=42 %
\hatcurFIESvsinixxxxA
\else
\ifnum#1=43 %
\hatcurFIESvsinixxxxB
\else
??????\fi
\fi
}
\newcommand{\hatcurFIESzfeh}[1]{\ifnum#1=42 %
\hatcurFIESzfehxxxxA
\else
\ifnum#1=43 %
\hatcurFIESzfehxxxxB
\else
??????\fi
\fi
}
\newcommand{\hatcurhtr}[1]{\ifnum#1=42 %
\hatcurhtrxxxxA
\else
\ifnum#1=43 %
\hatcurhtrxxxxB
\else
??????\fi
\fi
}
\newcommand{\hatcurISOage}[1]{\ifnum#1=42 %
\hatcurISOagexxxxA
\else
\ifnum#1=43 %
\hatcurISOagexxxxB
\else
??????\fi
\fi
}
\newcommand{\hatcurISOJK}[1]{\ifnum#1=42 %
\hatcurISOJKxxxxA
\else
\ifnum#1=43 %
\hatcurISOJKxxxxB
\else
??????\fi
\fi
}
\newcommand{\hatcurISOlogg}[1]{\ifnum#1=42 %
\hatcurISOloggxxxxA
\else
\ifnum#1=43 %
\hatcurISOloggxxxxB
\else
??????\fi
\fi
}
\newcommand{\hatcurISOlum}[1]{\ifnum#1=42 %
\hatcurISOlumxxxxA
\else
\ifnum#1=43 %
\hatcurISOlumxxxxB
\else
??????\fi
\fi
}
\newcommand{\hatcurISOlumshort}[1]{\ifnum#1=42 %
\hatcurISOlumshortxxxxA
\else
\ifnum#1=43 %
\hatcurISOlumshortxxxxB
\else
??????\fi
\fi
}
\newcommand{\hatcurISOm}[1]{\ifnum#1=42 %
\hatcurISOmxxxxA
\else
\ifnum#1=43 %
\hatcurISOmxxxxB
\else
??????\fi
\fi
}
\newcommand{\hatcurISOMH}[1]{\ifnum#1=42 %
\hatcurISOMHxxxxA
\else
\ifnum#1=43 %
\hatcurISOMHxxxxB
\else
??????\fi
\fi
}
\newcommand{\hatcurISOMJ}[1]{\ifnum#1=42 %
\hatcurISOMJxxxxA
\else
\ifnum#1=43 %
\hatcurISOMJxxxxB
\else
??????\fi
\fi
}
\newcommand{\hatcurISOMK}[1]{\ifnum#1=42 %
\hatcurISOMKxxxxA
\else
\ifnum#1=43 %
\hatcurISOMKxxxxB
\else
??????\fi
\fi
}
\newcommand{\hatcurISOmlong}[1]{\ifnum#1=42 %
\hatcurISOmlongxxxxA
\else
\ifnum#1=43 %
\hatcurISOmlongxxxxB
\else
??????\fi
\fi
}
\newcommand{\hatcurISOmshort}[1]{\ifnum#1=42 %
\hatcurISOmshortxxxxA
\else
\ifnum#1=43 %
\hatcurISOmshortxxxxB
\else
??????\fi
\fi
}
\newcommand{\hatcurISOmv}[1]{\ifnum#1=42 %
\hatcurISOmvxxxxA
\else
\ifnum#1=43 %
\hatcurISOmvxxxxB
\else
??????\fi
\fi
}
\newcommand{\hatcurISOr}[1]{\ifnum#1=42 %
\hatcurISOrxxxxA
\else
\ifnum#1=43 %
\hatcurISOrxxxxB
\else
??????\fi
\fi
}
\newcommand{\hatcurISOrho}[1]{\ifnum#1=42 %
\hatcurISOrhoxxxxA
\else
\ifnum#1=43 %
\hatcurISOrhoxxxxB
\else
??????\fi
\fi
}
\newcommand{\hatcurISOrlong}[1]{\ifnum#1=42 %
\hatcurISOrlongxxxxA
\else
\ifnum#1=43 %
\hatcurISOrlongxxxxB
\else
??????\fi
\fi
}
\newcommand{\hatcurISOrshort}[1]{\ifnum#1=42 %
\hatcurISOrshortxxxxA
\else
\ifnum#1=43 %
\hatcurISOrshortxxxxB
\else
??????\fi
\fi
}
\newcommand{\hatcurISOsigma}[1]{\ifnum#1=42 %
\hatcurISOsigmaxxxxA
\else
\ifnum#1=43 %
\hatcurISOsigmaxxxxB
\else
??????\fi
\fi
}
\newcommand{\hatcurISOspec}[1]{\ifnum#1=42 %
\hatcurISOspecxxxxA
\else
\ifnum#1=43 %
\hatcurISOspecxxxxB
\else
??????\fi
\fi
}
\newcommand{\hatcurISOvi}[1]{\ifnum#1=42 %
\hatcurISOvixxxxA
\else
\ifnum#1=43 %
\hatcurISOvixxxxB
\else
??????\fi
\fi
}
\newcommand{\hatcurLBig}[1]{\ifnum#1=42 %
\hatcurLBigxxxxA
\else
\ifnum#1=43 %
\hatcurLBigxxxxB
\else
??????\fi
\fi
}
\newcommand{\hatcurLBii}[1]{\ifnum#1=42 %
\hatcurLBiixxxxA
\else
\ifnum#1=43 %
\hatcurLBiixxxxB
\else
??????\fi
\fi
}
\newcommand{\hatcurLBiI}[1]{\ifnum#1=42 %
\hatcurLBiIxxxxA
\else
\ifnum#1=43 %
\hatcurLBiIxxxxB
\else
??????\fi
\fi
}
\newcommand{\hatcurLBiig}[1]{\ifnum#1=42 %
\hatcurLBiigxxxxA
\else
\ifnum#1=43 %
\hatcurLBiigxxxxB
\else
??????\fi
\fi
}
\newcommand{\hatcurLBiii}[1]{\ifnum#1=42 %
\hatcurLBiiixxxxA
\else
\ifnum#1=43 %
\hatcurLBiiixxxxB
\else
??????\fi
\fi
}
\newcommand{\hatcurLBiiI}[1]{\ifnum#1=42 %
\hatcurLBiiIxxxxA
\else
\ifnum#1=43 %
\hatcurLBiiIxxxxB
\else
??????\fi
\fi
}
\newcommand{\hatcurLBiikep}[1]{\ifnum#1=42 %
\hatcurLBiikepxxxxA
\else
\ifnum#1=43 %
\hatcurLBiikepxxxxB
\else
??????\fi
\fi
}
\newcommand{\hatcurLBiir}[1]{\ifnum#1=42 %
\hatcurLBiirxxxxA
\else
\ifnum#1=43 %
\hatcurLBiirxxxxB
\else
??????\fi
\fi
}
\newcommand{\hatcurLBiiR}[1]{\ifnum#1=42 %
\hatcurLBiiRxxxxA
\else
\ifnum#1=43 %
\hatcurLBiiRxxxxB
\else
??????\fi
\fi
}
\newcommand{\hatcurLBiiz}[1]{\ifnum#1=42 %
\hatcurLBiizxxxxA
\else
\ifnum#1=43 %
\hatcurLBiizxxxxB
\else
??????\fi
\fi
}
\newcommand{\hatcurLBikep}[1]{\ifnum#1=42 %
\hatcurLBikepxxxxA
\else
\ifnum#1=43 %
\hatcurLBikepxxxxB
\else
??????\fi
\fi
}
\newcommand{\hatcurLBir}[1]{\ifnum#1=42 %
\hatcurLBirxxxxA
\else
\ifnum#1=43 %
\hatcurLBirxxxxB
\else
??????\fi
\fi
}
\newcommand{\hatcurLBiR}[1]{\ifnum#1=42 %
\hatcurLBiRxxxxA
\else
\ifnum#1=43 %
\hatcurLBiRxxxxB
\else
??????\fi
\fi
}
\newcommand{\hatcurLBiz}[1]{\ifnum#1=42 %
\hatcurLBizxxxxA
\else
\ifnum#1=43 %
\hatcurLBizxxxxB
\else
??????\fi
\fi
}
\newcommand{\hatcurLCbsq}[1]{\ifnum#1=42 %
\hatcurLCbsqxxxxA
\else
\ifnum#1=43 %
\hatcurLCbsqxxxxB
\else
??????\fi
\fi
}
\newcommand{\hatcurLCdip}[1]{\ifnum#1=42 %
\hatcurLCdipxxxxA
\else
\ifnum#1=43 %
\hatcurLCdipxxxxB
\else
??????\fi
\fi
}
\newcommand{\hatcurLCdur}[1]{\ifnum#1=42 %
\hatcurLCdurxxxxA
\else
\ifnum#1=43 %
\hatcurLCdurxxxxB
\else
??????\fi
\fi
}
\newcommand{\hatcurLCdurhr}[1]{\ifnum#1=42 %
\hatcurLCdurhrxxxxA
\else
\ifnum#1=43 %
\hatcurLCdurhrxxxxB
\else
??????\fi
\fi
}
\newcommand{\hatcurLCdurhrshort}[1]{\ifnum#1=42 %
\hatcurLCdurhrshortxxxxA
\else
\ifnum#1=43 %
\hatcurLCdurhrshortxxxxB
\else
??????\fi
\fi
}
\newcommand{\hatcurLCdurshort}[1]{\ifnum#1=42 %
\hatcurLCdurshortxxxxA
\else
\ifnum#1=43 %
\hatcurLCdurshortxxxxB
\else
??????\fi
\fi
}
\newcommand{\hatcurLChatnetm}[1]{\ifnum#1=42 %
\hatcurLChatnetmxxxxA
\else
??????\fi
}
\newcommand{\hatcurLChatnetmA}[1]{\ifnum#1=43 %
\hatcurLChatnetmAxxxxB
\else
??????\fi
}
\newcommand{\hatcurLChatnetmB}[1]{\ifnum#1=43 %
\hatcurLChatnetmBxxxxB
\else
??????\fi
}
\newcommand{\hatcurLCiblend}[1]{\ifnum#1=42 %
\hatcurLCiblendxxxxA
\else
??????\fi
}
\newcommand{\hatcurLCiblendA}[1]{\ifnum#1=43 %
\hatcurLCiblendAxxxxB
\else
??????\fi
}
\newcommand{\hatcurLCiblendB}[1]{\ifnum#1=43 %
\hatcurLCiblendBxxxxB
\else
??????\fi
}
\newcommand{\hatcurLCimp}[1]{\ifnum#1=42 %
\hatcurLCimpxxxxA
\else
\ifnum#1=43 %
\hatcurLCimpxxxxB
\else
??????\fi
\fi
}
\newcommand{\hatcurLCingdur}[1]{\ifnum#1=42 %
\hatcurLCingdurxxxxA
\else
\ifnum#1=43 %
\hatcurLCingdurxxxxB
\else
??????\fi
\fi
}
\newcommand{\hatcurLCP}[1]{\ifnum#1=42 %
\hatcurLCPxxxxA
\else
\ifnum#1=43 %
\hatcurLCPxxxxB
\else
??????\fi
\fi
}
\newcommand{\hatcurLCPprec}[1]{\ifnum#1=42 %
\hatcurLCPprecxxxxA
\else
\ifnum#1=43 %
\hatcurLCPprecxxxxB
\else
??????\fi
\fi
}
\newcommand{\hatcurLCPshort}[1]{\ifnum#1=42 %
\hatcurLCPshortxxxxA
\else
\ifnum#1=43 %
\hatcurLCPshortxxxxB
\else
??????\fi
\fi
}
\newcommand{\hatcurLCq}[1]{\ifnum#1=42 %
\hatcurLCqxxxxA
\else
\ifnum#1=43 %
\hatcurLCqxxxxB
\else
??????\fi
\fi
}
\newcommand{\hatcurLCqshort}[1]{\ifnum#1=42 %
\hatcurLCqshortxxxxA
\else
\ifnum#1=43 %
\hatcurLCqshortxxxxB
\else
??????\fi
\fi
}
\newcommand{\hatcurLCrprstar}[1]{\ifnum#1=42 %
\hatcurLCrprstarxxxxA
\else
\ifnum#1=43 %
\hatcurLCrprstarxxxxB
\else
??????\fi
\fi
}
\newcommand{\hatcurLCT}[1]{\ifnum#1=42 %
\hatcurLCTxxxxA
\else
\ifnum#1=43 %
\hatcurLCTxxxxB
\else
??????\fi
\fi
}
\newcommand{\hatcurLCTA}[1]{\ifnum#1=42 %
\hatcurLCTAxxxxA
\else
\ifnum#1=43 %
\hatcurLCTAxxxxB
\else
??????\fi
\fi
}
\newcommand{\hatcurLCTB}[1]{\ifnum#1=42 %
\hatcurLCTBxxxxA
\else
\ifnum#1=43 %
\hatcurLCTBxxxxB
\else
??????\fi
\fi
}
\newcommand{\hatcurLCzeta}[1]{\ifnum#1=42 %
\hatcurLCzetaxxxxA
\else
\ifnum#1=43 %
\hatcurLCzetaxxxxB
\else
??????\fi
\fi
}
\newcommand{\hatcurPPaequiv}[1]{\ifnum#1=42 %
\hatcurPPaequivxxxxA
\else
\ifnum#1=43 %
\hatcurPPaequivxxxxB
\else
??????\fi
\fi
}
\newcommand{\hatcurPPar}[1]{\ifnum#1=42 %
\hatcurPParxxxxA
\else
\ifnum#1=43 %
\hatcurPParxxxxB
\else
??????\fi
\fi
}
\newcommand{\hatcurPParel}[1]{\ifnum#1=42 %
\hatcurPParelxxxxA
\else
\ifnum#1=43 %
\hatcurPParelxxxxB
\else
??????\fi
\fi
}
\newcommand{\hatcurPPfluxap}[1]{\ifnum#1=42 %
\hatcurPPfluxapxxxxA
\else
\ifnum#1=43 %
\hatcurPPfluxapxxxxB
\else
??????\fi
\fi
}
\newcommand{\hatcurPPfluxapdim}[1]{\ifnum#1=42 %
\hatcurPPfluxapdimxxxxA
\else
\ifnum#1=43 %
\hatcurPPfluxapdimxxxxB
\else
??????\fi
\fi
}
\newcommand{\hatcurPPfluxavg}[1]{\ifnum#1=42 %
\hatcurPPfluxavgxxxxA
\else
\ifnum#1=43 %
\hatcurPPfluxavgxxxxB
\else
??????\fi
\fi
}
\newcommand{\hatcurPPfluxavgdim}[1]{\ifnum#1=42 %
\hatcurPPfluxavgdimxxxxA
\else
\ifnum#1=43 %
\hatcurPPfluxavgdimxxxxB
\else
??????\fi
\fi
}
\newcommand{\hatcurPPfluxperi}[1]{\ifnum#1=42 %
\hatcurPPfluxperixxxxA
\else
\ifnum#1=43 %
\hatcurPPfluxperixxxxB
\else
??????\fi
\fi
}
\newcommand{\hatcurPPfluxperidim}[1]{\ifnum#1=42 %
\hatcurPPfluxperidimxxxxA
\else
\ifnum#1=43 %
\hatcurPPfluxperidimxxxxB
\else
??????\fi
\fi
}
\newcommand{\hatcurPPg}[1]{\ifnum#1=42 %
\hatcurPPgxxxxA
\else
\ifnum#1=43 %
\hatcurPPgxxxxB
\else
??????\fi
\fi
}
\newcommand{\hatcurPPi}[1]{\ifnum#1=42 %
\hatcurPPixxxxA
\else
\ifnum#1=43 %
\hatcurPPixxxxB
\else
??????\fi
\fi
}
\newcommand{\hatcurPPlogg}[1]{\ifnum#1=42 %
\hatcurPPloggxxxxA
\else
\ifnum#1=43 %
\hatcurPPloggxxxxB
\else
??????\fi
\fi
}
\newcommand{\hatcurPPm}[1]{\ifnum#1=42 %
\hatcurPPmxxxxA
\else
\ifnum#1=43 %
\hatcurPPmxxxxB
\else
??????\fi
\fi
}
\newcommand{\hatcurPPme}[1]{\ifnum#1=42 %
\hatcurPPmexxxxA
\else
\ifnum#1=43 %
\hatcurPPmexxxxB
\else
??????\fi
\fi
}
\newcommand{\hatcurPPmelong}[1]{\ifnum#1=42 %
\hatcurPPmelongxxxxA
\else
\ifnum#1=43 %
\hatcurPPmelongxxxxB
\else
??????\fi
\fi
}
\newcommand{\hatcurPPmeshort}[1]{\ifnum#1=42 %
\hatcurPPmeshortxxxxA
\else
\ifnum#1=43 %
\hatcurPPmeshortxxxxB
\else
??????\fi
\fi
}
\newcommand{\hatcurPPmlong}[1]{\ifnum#1=42 %
\hatcurPPmlongxxxxA
\else
\ifnum#1=43 %
\hatcurPPmlongxxxxB
\else
??????\fi
\fi
}
\newcommand{\hatcurPPmrcorr}[1]{\ifnum#1=42 %
\hatcurPPmrcorrxxxxA
\else
\ifnum#1=43 %
\hatcurPPmrcorrxxxxB
\else
??????\fi
\fi
}
\newcommand{\hatcurPPmshort}[1]{\ifnum#1=42 %
\hatcurPPmshortxxxxA
\else
\ifnum#1=43 %
\hatcurPPmshortxxxxB
\else
??????\fi
\fi
}
\newcommand{\hatcurPPperi}[1]{\ifnum#1=42 %
\hatcurPPperixxxxA
\else
\ifnum#1=43 %
\hatcurPPperixxxxB
\else
??????\fi
\fi
}
\newcommand{\hatcurPPphiconj}[1]{\ifnum#1=42 %
\hatcurPPphiconjxxxxA
\else
\ifnum#1=43 %
\hatcurPPphiconjxxxxB
\else
??????\fi
\fi
}
\newcommand{\hatcurPPr}[1]{\ifnum#1=42 %
\hatcurPPrxxxxA
\else
\ifnum#1=43 %
\hatcurPPrxxxxB
\else
??????\fi
\fi
}
\newcommand{\hatcurPPre}[1]{\ifnum#1=42 %
\hatcurPPrexxxxA
\else
\ifnum#1=43 %
\hatcurPPrexxxxB
\else
??????\fi
\fi
}
\newcommand{\hatcurPPrelong}[1]{\ifnum#1=42 %
\hatcurPPrelongxxxxA
\else
\ifnum#1=43 %
\hatcurPPrelongxxxxB
\else
??????\fi
\fi
}
\newcommand{\hatcurPPreshort}[1]{\ifnum#1=42 %
\hatcurPPreshortxxxxA
\else
\ifnum#1=43 %
\hatcurPPreshortxxxxB
\else
??????\fi
\fi
}
\newcommand{\hatcurPPrho}[1]{\ifnum#1=42 %
\hatcurPPrhoxxxxA
\else
\ifnum#1=43 %
\hatcurPPrhoxxxxB
\else
??????\fi
\fi
}
\newcommand{\hatcurPPrlong}[1]{\ifnum#1=42 %
\hatcurPPrlongxxxxA
\else
\ifnum#1=43 %
\hatcurPPrlongxxxxB
\else
??????\fi
\fi
}
\newcommand{\hatcurPPrshort}[1]{\ifnum#1=42 %
\hatcurPPrshortxxxxA
\else
\ifnum#1=43 %
\hatcurPPrshortxxxxB
\else
??????\fi
\fi
}
\newcommand{\hatcurPPtcirc}[1]{\ifnum#1=42 %
\hatcurPPtcircxxxxA
\else
\ifnum#1=43 %
\hatcurPPtcircxxxxB
\else
??????\fi
\fi
}
\newcommand{\hatcurPPteff}[1]{\ifnum#1=42 %
\hatcurPPteffxxxxA
\else
\ifnum#1=43 %
\hatcurPPteffxxxxB
\else
??????\fi
\fi
}
\newcommand{\hatcurPPtheta}[1]{\ifnum#1=42 %
\hatcurPPthetaxxxxA
\else
\ifnum#1=43 %
\hatcurPPthetaxxxxB
\else
??????\fi
\fi
}
\newcommand{\hatcurPPtinfall}[1]{\ifnum#1=42 %
\hatcurPPtinfallxxxxA
\else
\ifnum#1=43 %
\hatcurPPtinfallxxxxB
\else
??????\fi
\fi
}
\newcommand{\hatcurRVeccen}[1]{\ifnum#1=42 %
\hatcurRVeccenxxxxA
\else
\ifnum#1=43 %
\hatcurRVeccenxxxxB
\else
??????\fi
\fi
}
\newcommand{\hatcurRVfitrms}[1]{\ifnum#1=43 %
\hatcurRVfitrmsxxxxB
\else
??????\fi
}
\newcommand{\hatcurRVfitrmsA}[1]{\ifnum#1=42 %
\hatcurRVfitrmsAxxxxA
\else
??????\fi
}
\newcommand{\hatcurRVfitrmsB}[1]{\ifnum#1=42 %
\hatcurRVfitrmsBxxxxA
\else
??????\fi
}
\newcommand{\hatcurRVgamma}[1]{\ifnum#1=43 %
\hatcurRVgammaxxxxB
\else
??????\fi
}
\newcommand{\hatcurRVgammaA}[1]{\ifnum#1=42 %
\hatcurRVgammaAxxxxA
\else
??????\fi
}
\newcommand{\hatcurRVgammaB}[1]{\ifnum#1=42 %
\hatcurRVgammaBxxxxA
\else
??????\fi
}
\newcommand{\hatcurRVh}[1]{\ifnum#1=42 %
\hatcurRVhxxxxA
\else
\ifnum#1=43 %
\hatcurRVhxxxxB
\else
??????\fi
\fi
}
\newcommand{\hatcurRVjitter}[1]{\ifnum#1=43 %
\hatcurRVjitterxxxxB
\else
??????\fi
}
\newcommand{\hatcurRVjitterA}[1]{\ifnum#1=42 %
\hatcurRVjitterAxxxxA
\else
??????\fi
}
\newcommand{\hatcurRVjitterB}[1]{\ifnum#1=42 %
\hatcurRVjitterBxxxxA
\else
??????\fi
}
\newcommand{\hatcurRVk}[1]{\ifnum#1=42 %
\hatcurRVkxxxxA
\else
\ifnum#1=43 %
\hatcurRVkxxxxB
\else
??????\fi
\fi
}
\newcommand{\hatcurRVK}[1]{\ifnum#1=42 %
\hatcurRVKxxxxA
\else
\ifnum#1=43 %
\hatcurRVKxxxxB
\else
??????\fi
\fi
}
\newcommand{\hatcurRVomega}[1]{\ifnum#1=42 %
\hatcurRVomegaxxxxA
\else
\ifnum#1=43 %
\hatcurRVomegaxxxxB
\else
??????\fi
\fi
}
\newcommand{\hatcurRVrh}[1]{\ifnum#1=42 %
\hatcurRVrhxxxxA
\else
\ifnum#1=43 %
\hatcurRVrhxxxxB
\else
??????\fi
\fi
}
\newcommand{\hatcurRVrk}[1]{\ifnum#1=42 %
\hatcurRVrkxxxxA
\else
\ifnum#1=43 %
\hatcurRVrkxxxxB
\else
??????\fi
\fi
}
\newcommand{\hatcurRVtrone}[1]{\ifnum#1=42 %
\hatcurRVtronexxxxA
\else
\ifnum#1=43 %
\hatcurRVtronexxxxB
\else
??????\fi
\fi
}
\newcommand{\hatcurRVtrtwo}[1]{\ifnum#1=42 %
\hatcurRVtrtwoxxxxA
\else
\ifnum#1=43 %
\hatcurRVtrtwoxxxxB
\else
??????\fi
\fi
}
\newcommand{\hatcurSMEiilogg}[1]{\ifnum#1=42 %
\hatcurSMEiiloggxxxxA
\else
\ifnum#1=43 %
\hatcurSMEiiloggxxxxB
\else
??????\fi
\fi
}
\newcommand{\hatcurSMEiiteff}[1]{\ifnum#1=42 %
\hatcurSMEiiteffxxxxA
\else
\ifnum#1=43 %
\hatcurSMEiiteffxxxxB
\else
??????\fi
\fi
}
\newcommand{\hatcurSMEiivmac}[1]{\ifnum#1=42 %
\hatcurSMEiivmacxxxxA
\else
\ifnum#1=43 %
\hatcurSMEiivmacxxxxB
\else
??????\fi
\fi
}
\newcommand{\hatcurSMEiivmic}[1]{\ifnum#1=42 %
\hatcurSMEiivmicxxxxA
\else
\ifnum#1=43 %
\hatcurSMEiivmicxxxxB
\else
??????\fi
\fi
}
\newcommand{\hatcurSMEiivsin}[1]{\ifnum#1=42 %
\hatcurSMEiivsinxxxxA
\else
\ifnum#1=43 %
\hatcurSMEiivsinxxxxB
\else
??????\fi
\fi
}
\newcommand{\hatcurSMEiizfeh}[1]{\ifnum#1=42 %
\hatcurSMEiizfehxxxxA
\else
\ifnum#1=43 %
\hatcurSMEiizfehxxxxB
\else
??????\fi
\fi
}
\newcommand{\hatcurSMEiizfehshort}[1]{\ifnum#1=42 %
\hatcurSMEiizfehshortxxxxA
\else
\ifnum#1=43 %
\hatcurSMEiizfehshortxxxxB
\else
??????\fi
\fi
}
\newcommand{\hatcurSMEilogg}[1]{\ifnum#1=42 %
\hatcurSMEiloggxxxxA
\else
\ifnum#1=43 %
\hatcurSMEiloggxxxxB
\else
??????\fi
\fi
}
\newcommand{\hatcurSMEiteff}[1]{\ifnum#1=42 %
\hatcurSMEiteffxxxxA
\else
\ifnum#1=43 %
\hatcurSMEiteffxxxxB
\else
??????\fi
\fi
}
\newcommand{\hatcurSMEivmac}[1]{\ifnum#1=42 %
\hatcurSMEivmacxxxxA
\else
\ifnum#1=43 %
\hatcurSMEivmacxxxxB
\else
??????\fi
\fi
}
\newcommand{\hatcurSMEivmic}[1]{\ifnum#1=42 %
\hatcurSMEivmicxxxxA
\else
\ifnum#1=43 %
\hatcurSMEivmicxxxxB
\else
??????\fi
\fi
}
\newcommand{\hatcurSMEivsin}[1]{\ifnum#1=42 %
\hatcurSMEivsinxxxxA
\else
\ifnum#1=43 %
\hatcurSMEivsinxxxxB
\else
??????\fi
\fi
}
\newcommand{\hatcurSMEizfeh}[1]{\ifnum#1=42 %
\hatcurSMEizfehxxxxA
\else
\ifnum#1=43 %
\hatcurSMEizfehxxxxB
\else
??????\fi
\fi
}
\newcommand{\hatcurSMEizfehshort}[1]{\ifnum#1=42 %
\hatcurSMEizfehshortxxxxA
\else
\ifnum#1=43 %
\hatcurSMEizfehshortxxxxB
\else
??????\fi
\fi
}
\newcommand{\hatcurTRESgamma}[1]{\ifnum#1=42 %
\hatcurTRESgammaxxxxA
\else
\ifnum#1=43 %
\hatcurTRESgammaxxxxB
\else
??????\fi
\fi
}
\newcommand{\hatcurTRESlogg}[1]{\ifnum#1=42 %
\hatcurTRESloggxxxxA
\else
\ifnum#1=43 %
\hatcurTRESloggxxxxB
\else
??????\fi
\fi
}
\newcommand{\hatcurTRESnumspec}[1]{\ifnum#1=42 %
\hatcurTRESnumspecxxxxA
\else
\ifnum#1=43 %
\hatcurTRESnumspecxxxxB
\else
??????\fi
\fi
}
\newcommand{\hatcurTRESrvrms}[1]{\ifnum#1=42 %
\hatcurTRESrvrmsxxxxA
\else
\ifnum#1=43 %
\hatcurTRESrvrmsxxxxB
\else
??????\fi
\fi
}
\newcommand{\hatcurTRESspan}[1]{\ifnum#1=42 %
\hatcurTRESspanxxxxA
\else
\ifnum#1=43 %
\hatcurTRESspanxxxxB
\else
??????\fi
\fi
}
\newcommand{\hatcurTRESteff}[1]{\ifnum#1=42 %
\hatcurTRESteffxxxxA
\else
\ifnum#1=43 %
\hatcurTRESteffxxxxB
\else
??????\fi
\fi
}
\newcommand{\hatcurTRESvsini}[1]{\ifnum#1=42 %
\hatcurTRESvsinixxxxA
\else
\ifnum#1=43 %
\hatcurTRESvsinixxxxB
\else
??????\fi
\fi
}
\newcommand{\hatcurTRESzfeh}[1]{\ifnum#1=42 %
\hatcurTRESzfehxxxxA
\else
\ifnum#1=43 %
\hatcurTRESzfehxxxxB
\else
??????\fi
\fi
}
\newcommand{\hatcurXAv}[1]{\ifnum#1=42 %
\hatcurXAvxxxxA
\else
\ifnum#1=43 %
\hatcurXAvxxxxB
\else
??????\fi
\fi
}
\newcommand{\hatcurXdist}[1]{\ifnum#1=42 %
\hatcurXdistxxxxA
\else
\ifnum#1=43 %
\hatcurXdistxxxxB
\else
??????\fi
\fi
}
\newcommand{\hatcurXdistred}[1]{\ifnum#1=42 %
\hatcurXdistredxxxxA
\else
\ifnum#1=43 %
\hatcurXdistredxxxxB
\else
??????\fi
\fi
}
\newcommand{\hatcurXEBV}[1]{\ifnum#1=42 %
\hatcurXEBVxxxxA
\else
\ifnum#1=43 %
\hatcurXEBVxxxxB
\else
??????\fi
\fi
}
\newcommand{\hatcurXjhisored}[1]{\ifnum#1=42 %
\hatcurXjhisoredxxxxA
\else
\ifnum#1=43 %
\hatcurXjhisoredxxxxB
\else
??????\fi
\fi
}
\newcommand{\hatcurXjkisored}[1]{\ifnum#1=42 %
\hatcurXjkisoredxxxxA
\else
\ifnum#1=43 %
\hatcurXjkisoredxxxxB
\else
??????\fi
\fi
}
\newcommand{\hatcurXmhisored}[1]{\ifnum#1=42 %
\hatcurXmhisoredxxxxA
\else
\ifnum#1=43 %
\hatcurXmhisoredxxxxB
\else
??????\fi
\fi
}
\newcommand{\hatcurXmiisored}[1]{\ifnum#1=42 %
\hatcurXmiisoredxxxxA
\else
\ifnum#1=43 %
\hatcurXmiisoredxxxxB
\else
??????\fi
\fi
}
\newcommand{\hatcurXmjisored}[1]{\ifnum#1=42 %
\hatcurXmjisoredxxxxA
\else
\ifnum#1=43 %
\hatcurXmjisoredxxxxB
\else
??????\fi
\fi
}
\newcommand{\hatcurXmkisored}[1]{\ifnum#1=42 %
\hatcurXmkisoredxxxxA
\else
\ifnum#1=43 %
\hatcurXmkisoredxxxxB
\else
??????\fi
\fi
}
\newcommand{\hatcurXmvisored}[1]{\ifnum#1=42 %
\hatcurXmvisoredxxxxA
\else
\ifnum#1=43 %
\hatcurXmvisoredxxxxB
\else
??????\fi
\fi
}
\newcommand{\hatcurXsecdur}[1]{\ifnum#1=42 %
\hatcurXsecdurxxxxA
\else
\ifnum#1=43 %
\hatcurXsecdurxxxxB
\else
??????\fi
\fi
}
\newcommand{\hatcurXsecingdur}[1]{\ifnum#1=42 %
\hatcurXsecingdurxxxxA
\else
\ifnum#1=43 %
\hatcurXsecingdurxxxxB
\else
??????\fi
\fi
}
\newcommand{\hatcurXsecondary}[1]{\ifnum#1=42 %
\hatcurXsecondaryxxxxA
\else
\ifnum#1=43 %
\hatcurXsecondaryxxxxB
\else
??????\fi
\fi
}
\newcommand{\hatcurXsecphase}[1]{\ifnum#1=42 %
\hatcurXsecphasexxxxA
\else
\ifnum#1=43 %
\hatcurXsecphasexxxxB
\else
??????\fi
\fi
}
\newcommand{\hatcurXviisored}[1]{\ifnum#1=42 %
\hatcurXviisoredxxxxA
\else
\ifnum#1=43 %
\hatcurXviisoredxxxxB
\else
??????\fi
\fi
}
\newcommand{\hatcurXvkisored}[1]{\ifnum#1=42 %
\hatcurXvkisoredxxxxA
\else
\ifnum#1=43 %
\hatcurXvkisoredxxxxB
\else
??????\fi
\fi
}

\newcommand{\hatcurhtrxxxxeccenA}{HTR366-001}                          % Original HTR name of target
\newcommand{\hatcurfieldxxxxeccenA}{366}                               % Original HTR field
\newcommand{\hatcurCCraxxxxeccenA}{\ensuremath{09^{\mathrm h}01^{\mathrm m}22.66{\mathrm s}}}                        % Right Ascension
\newcommand{\hatcurCCdecxxxxeccenA}{\ensuremath{+06{\arcdeg}05{\arcmin}50.0{\arcsec}}}                       % Declination
\newcommand{\hatcurCCmagxxxxeccenA}{\ensuremath{12.168\pm0.030}}                            % apparent V-band magnitude
\newcommand{\hatcurCCtwomassxxxxeccenA}{2MASS~09012265+0605500}        % 2MASS identifier
\newcommand{\hatcurCCgscxxxxeccenA}{GSC~0232-01451}                    % GSC(1.2) identifier
\newcommand{\hatcurCCtassmvxxxxeccenA}{\ensuremath{12.168\pm0.030}}                         % TASS V-band magnitude
\newcommand{\hatcurCCtassmbxxxxeccenA}{\ensuremath{12.827\pm0.020}}                         % TASS V-band magnitude
\newcommand{\hatcurCCtwomassJmagxxxxeccenA}{\ensuremath{10.960\pm0.023}} % 2MASS ORIG MAG
\newcommand{\hatcurCCtwomassHmagxxxxeccenA}{\ensuremath{10.677\pm0.024}} % 2MASS ORIG MAG
\newcommand{\hatcurCCtwomassKmagxxxxeccenA}{\ensuremath{10.626\pm0.026}} % 2MASS ORIG MAG
\newcommand{\hatcurCCcitJmagxxxxeccenA}{\ensuremath{10.979\pm0.023}}   % 2MASS CIT MAG
\newcommand{\hatcurCCcitHmagxxxxeccenA}{\ensuremath{10.672\pm0.024}}   % 2MASS CIT MAG
\newcommand{\hatcurCCcitKmagxxxxeccenA}{\ensuremath{10.650\pm0.026}}   % 2MASS CIT MAG
\newcommand{\hatcurCCbbJmagxxxxeccenA}{\ensuremath{11.025\pm0.025}}    % 2MASS BB MAG
\newcommand{\hatcurCCbbHmagxxxxeccenA}{\ensuremath{10.693\pm0.025}}    % 2MASS BB MAG
\newcommand{\hatcurCCbbKmagxxxxeccenA}{\ensuremath{10.670\pm0.026}}    % 2MASS BB MAG
\newcommand{\hatcurCCesoJmagxxxxeccenA}{\ensuremath{11.027\pm0.026}}   % 2MASS ESO MAG
\newcommand{\hatcurCCesoHmagxxxxeccenA}{\ensuremath{10.688\pm0.028}}   % 2MASS ESO MAG
\newcommand{\hatcurCCesoKmagxxxxeccenA}{\ensuremath{10.669\pm0.027}}   % 2MASS ESO MAG
\newcommand{\hatcurCCesoJHmagxxxxeccenA}{\ensuremath{0.339\pm0.036}}   % 2MASS ESO JH COLOR
\newcommand{\hatcurCCesoJKmagxxxxeccenA}{\ensuremath{0.359\pm0.037}}   % 2MASS ESO JK COLOR
\newcommand{\hatcurCCesoHKmagxxxxeccenA}{\ensuremath{0.019\pm0.039}}   % 2MASS ESO HK COLOR
\newcommand{\hatcurLCdipxxxxeccenA}{\ensuremath{7.6}}                  % BLS detected dip (mmag)
\newcommand{\hatcurLCrprstarxxxxeccenA}{\ensuremath{0.0859\pm0.0032}}  % Rp/R*
\newcommand{\hatcurLCbsqxxxxeccenA}{\ensuremath{0.328_{-0.134}^{+0.086}}} % impact parameter square
\newcommand{\hatcurLCimpxxxxeccenA}{\ensuremath{0.573_{-0.154}^{+0.069}}} % impact parameter
\newcommand{\hatcurLCzetaxxxxeccenA}{\ensuremath{13.41\pm0.14}}        % zeta/R*
\newcommand{\hatcurLCdurxxxxeccenA}{\ensuremath{0.1679\pm0.0039}}      % transit duration (days)
\newcommand{\hatcurLCdurshortxxxxeccenA}{\ensuremath{0.1679}}          % transit duration (days)
\newcommand{\hatcurLCdurhrxxxxeccenA}{\ensuremath{4.030\pm0.093}}      % transit duration (hours)
\newcommand{\hatcurLCdurhrshortxxxxeccenA}{\ensuremath{4.030}}         % transit duration (hours)
\newcommand{\hatcurLCqxxxxeccenA}{\ensuremath{0.0362\pm0.0008}}        % fractional transit duration (days)
\newcommand{\hatcurLCqshortxxxxeccenA}{\ensuremath{0.036}}             % fractional transit duration (days)
\newcommand{\hatcurLCingdurxxxxeccenA}{\ensuremath{0.0191\pm0.0036}}   % ingress/egress duration (days)
\newcommand{\hatcurLCPxxxxeccenA}{\ensuremath{4.641878\pm0.000030}}    % period (days)
\newcommand{\hatcurLCPprecxxxxeccenA}{\ensuremath{4.6418776}}          % period (days)
\newcommand{\hatcurLCPshortxxxxeccenA}{\ensuremath{4.6419}}            % period (days)
\newcommand{\hatcurLCTxxxxeccenA}{\ensuremath{2455952.52601\pm0.00075}} % epoch (BJD)
\newcommand{\hatcurLCTAxxxxeccenA}{\ensuremath{2455506.90575\pm0.00282}} % TA (BJD)
\newcommand{\hatcurLCTBxxxxeccenA}{\ensuremath{2455989.66103\pm0.00084}} % TB (BJD)
\newcommand{\hatcurLChatnetmxxxxeccenA}{\ensuremath{11.9999\pm0.0000}} % HATNet OOT level
\newcommand{\hatcurLCiblendxxxxeccenA}{\ensuremath{0.56\pm0.04}}       % HATNet iblend factor
\newcommand{\hatcurSMEiteffxxxxeccenA}{\ensuremath{5838\pm50}}         % Ini SME, stellar effective temperature
\newcommand{\hatcurSMEizfehxxxxeccenA}{\ensuremath{0.33\pm0.08}}       % Ini SME, stellar metallicity
\newcommand{\hatcurSMEizfehshortxxxxeccenA}{\ensuremath{0.33}}         % Ini SME, stellar metallicity
\newcommand{\hatcurSMEiloggxxxxeccenA}{\ensuremath{4.33\pm0.10}}       % Ini SME, stellar surface gravity
\newcommand{\hatcurSMEivsinxxxxeccenA}{\ensuremath{3.4\pm0.5}}         % Ini SME, stellar rotational velocity
\newcommand{\hatcurSMEivmacxxxxeccenA}{\ensuremath{NULL}}              % Ini SME, stellar macroturbulence
\newcommand{\hatcurSMEivmicxxxxeccenA}{\ensuremath{NULL}}              % Ini SME, stellar microturbulence
\newcommand{\hatcurSMEiiteffxxxxeccenA}{\ensuremath{5753\pm50}}        % Final SME, stellar effective temperature
\newcommand{\hatcurSMEiizfehxxxxeccenA}{\ensuremath{0.28\pm0.08}}      % Final SME, stellar metallicity
\newcommand{\hatcurSMEiizfehshortxxxxeccenA}{\ensuremath{0.28}}        % Final SME, stellar metallicity
\newcommand{\hatcurSMEiiloggxxxxeccenA}{\ensuremath{4.17\pm0.09}}      % Final SME, stellar surface gravity
\newcommand{\hatcurSMEiivsinxxxxeccenA}{\ensuremath{3.5\pm0.5}}        % Final SME, stellar rotational velocity
\newcommand{\hatcurSMEiivmacxxxxeccenA}{\ensuremath{NULL}}             % Final SME, stellar macroturbulence
\newcommand{\hatcurSMEiivmicxxxxeccenA}{\ensuremath{NULL}}             % Final SME, stellar microturbulence
\newcommand{\hatcurDSteffxxxxeccenA}{\ensuremath{NULL\pmNULL}}         % DS stellar effective temperature
\newcommand{\hatcurDSzfehxxxxeccenA}{\ensuremath{NULL\pmNULL}}         % DS stellar metallicity
\newcommand{\hatcurDSloggxxxxeccenA}{\ensuremath{NULL\pmNULL}}         % DS stellar surface gravity
\newcommand{\hatcurDSvsinixxxxeccenA}{\ensuremath{NULL\pmNULL}}        % DS stellar rotational velocity
\newcommand{\hatcurDSgammaxxxxeccenA}{\ensuremath{NULL\pmNULL}}        % DS absolute gamma velocity
\newcommand{\hatcurDSnumspecxxxxeccenA}{\ensuremath{0}}                % DS number of spectra
\newcommand{\hatcurDSspanxxxxeccenA}{\ensuremath{0}}                   % DS stellar surface gravity
\newcommand{\hatcurDSrvrmsxxxxeccenA}{\ensuremath{0.00}}               % DS rms of RV values [km/s]
\newcommand{\hatcurTRESteffxxxxeccenA}{\ensuremath{5838\pm50}}         % TRES stellar effective temperature
\newcommand{\hatcurTRESzfehxxxxeccenA}{\ensuremath{0.33\pm0.08}}       % TRES stellar metallicity
\newcommand{\hatcurTRESloggxxxxeccenA}{\ensuremath{4.33\pm0.10}}       % TRES stellar surface gravity
\newcommand{\hatcurTRESvsinixxxxeccenA}{\ensuremath{3.4\pm0.5}}        % TRES stellar rotational velocity
\newcommand{\hatcurTRESgammaxxxxeccenA}{\ensuremath{21.258\pm0.286}}   % TRES absolute gamma velocity
\newcommand{\hatcurTRESnumspecxxxxeccenA}{\ensuremath{13}}             % TRES number of spectra
\newcommand{\hatcurTRESspanxxxxeccenA}{\ensuremath{0}}                 % TRES stellar surface gravity
\newcommand{\hatcurTRESrvrmsxxxxeccenA}{\ensuremath{0.00}}             % TRES rms of RV values [km/s]
\newcommand{\hatcurFIESteffxxxxeccenA}{\ensuremath{NULL\pmNULL}}       % FIES stellar effective temperature
\newcommand{\hatcurFIESzfehxxxxeccenA}{\ensuremath{NULL\pmNULL}}       % FIES stellar metallicity
\newcommand{\hatcurFIESloggxxxxeccenA}{\ensuremath{NULL\pmNULL}}       % FIES stellar surface gravity
\newcommand{\hatcurFIESvsinixxxxeccenA}{\ensuremath{NULL\pmNULL}}      % FIES stellar rotational velocity
\newcommand{\hatcurFIESgammaxxxxeccenA}{\ensuremath{NULL\pmNULL}}      % FIES absolute gamma velocity
\newcommand{\hatcurFIESnumspecxxxxeccenA}{\ensuremath{0}}              % FIES number of spectra
\newcommand{\hatcurFIESspanxxxxeccenA}{\ensuremath{0}}                 % FIES stellar surface gravity
\newcommand{\hatcurFIESrvrmsxxxxeccenA}{\ensuremath{0.00}}             % FIES rms of RV values [km/s]
\newcommand{\hatcurLBizxxxxeccenA}{\ensuremath{0.2181}}                % Limb darkening parameters, Gamma1, z-band
\newcommand{\hatcurLBiizxxxxeccenA}{\ensuremath{0.3302}}               % Limb darkening parameters, Gamma2, z-band
\newcommand{\hatcurLBiixxxxeccenA}{\ensuremath{0.2852}}                % Limb darkening parameters, Gamma1, i-band
\newcommand{\hatcurLBiiixxxxeccenA}{\ensuremath{0.3283}}               % Limb darkening parameters, Gamma2, i-band
\newcommand{\hatcurLBiIxxxxeccenA}{\ensuremath{0.2626}}                % Limb darkening parameters, Gamma1, I-band
\newcommand{\hatcurLBiiIxxxxeccenA}{\ensuremath{0.3298}}               % Limb darkening parameters, Gamma2, I-band
\newcommand{\hatcurLBigxxxxeccenA}{\ensuremath{0.5871}}                % Limb darkening parameters, Gamma1, g-band
\newcommand{\hatcurLBiigxxxxeccenA}{\ensuremath{0.2130}}               % Limb darkening parameters, Gamma2, g-band
\newcommand{\hatcurLBirxxxxeccenA}{\ensuremath{0.3811}}                % Limb darkening parameters, Gamma1, r-band
\newcommand{\hatcurLBiirxxxxeccenA}{\ensuremath{0.3159}}               % Limb darkening parameters, Gamma2, r-band
\newcommand{\hatcurLBiRxxxxeccenA}{\ensuremath{0.3545}}                % Limb darkening parameters, Gamma1, R-band
\newcommand{\hatcurLBiiRxxxxeccenA}{\ensuremath{0.3204}}               % Limb darkening parameters, Gamma2, R-band
\newcommand{\hatcurLBikepxxxxeccenA}{\ensuremath{}}            % darkening parameters, Gamma1, Kep-band
\newcommand{\hatcurLBiikepxxxxeccenA}{\ensuremath{}}           % darkening parameters, Gamma2, Kep-band
\newcommand{\hatcurISOmxxxxeccenA}{\ensuremath{1.18\pm0.07}}           % stellar mass
\newcommand{\hatcurISOmshortxxxxeccenA}{\ensuremath{1.18}}             % stellar mass
\newcommand{\hatcurISOmlongxxxxeccenA}{\ensuremath{1.185\pm0.070}}     % stellar mass
\newcommand{\hatcurISOrxxxxeccenA}{\ensuremath{1.50\pm0.19}}           % stellar radius
\newcommand{\hatcurISOrshortxxxxeccenA}{\ensuremath{1.50}}             % stellar radius
\newcommand{\hatcurISOrlongxxxxeccenA}{\ensuremath{1.497\pm0.190}}     % stellar radius
\newcommand{\hatcurISOrhoxxxxeccenA}{\ensuremath{0.50_{-0.13}^{+0.25}}} % stellar density (cgs)
\newcommand{\hatcurISOloggxxxxeccenA}{\ensuremath{4.16\pm0.09}}        % stellar surface gravity from isochrones
\newcommand{\hatcurISOlumxxxxeccenA}{\ensuremath{2.20_{-0.48}^{+0.68}}} % stellar luminosity
\newcommand{\hatcurISOlumshortxxxxeccenA}{\ensuremath{2.20}}           % stellar luminosity
\newcommand{\hatcurISOmvxxxxeccenA}{\ensuremath{3.97\pm0.28}}          % stellar absolute magnitude
\newcommand{\hatcurISOvixxxxeccenA}{\ensuremath{0.710\pm0.016}}        % stellar V-I index
\newcommand{\hatcurISOagexxxxeccenA}{\ensuremath{4.9_{-0.7}^{+1.6}}}   % stellar age
\newcommand{\hatcurISOsigmaxxxxeccenA}{\ensuremath{0.00060\pm0.00015}} % system mass-correction sigma parameter
\newcommand{\hatcurISOMJxxxxeccenA}{\ensuremath{2.81\pm0.28}}          % stellar absolute J magnitude
\newcommand{\hatcurISOMHxxxxeccenA}{\ensuremath{2.48\pm0.28}}          % stellar absolute H magnitude
\newcommand{\hatcurISOMKxxxxeccenA}{\ensuremath{2.42\pm0.28}}          % stellar absolute K magnitude
\newcommand{\hatcurISOJKxxxxeccenA}{\ensuremath{0.39\pm0.01}}          % J-K color index from isochrones.
\newcommand{\hatcurISOspecxxxxeccenA}{G}                               % stellar spectral type
\newcommand{\hatcurRVKxxxxeccenA}{\ensuremath{104.4\pm13.8}}           % RV semi-amplitude [m/s]
\newcommand{\hatcurRVrkxxxxeccenA}{\ensuremath{-0.060\pm0.158}}        % sqrt(e)*cos(omega)
\newcommand{\hatcurRVrhxxxxeccenA}{\ensuremath{-0.055\pm0.226}}        % sqrt(e)*sin(omega)
\newcommand{\hatcurRVkxxxxeccenA}{\ensuremath{-0.012\pm0.052}}         % e*cos(omega)
\newcommand{\hatcurRVhxxxxeccenA}{\ensuremath{-0.008\pm0.084}}         % e*sin(omega)
\newcommand{\hatcurRVtronexxxxeccenA}{\ensuremath{0.0000\pm0.0000}}    % RV linear trend tr1 factor
\newcommand{\hatcurRVtrtwoxxxxeccenA}{\ensuremath{0.0000\pm0.0000}}    % RV linear trend tr2 factor
\newcommand{\hatcurRVgammaAxxxxeccenA}{\ensuremath{20559.4\pm14.2}}    % RV gamma velocity, relative scale
\newcommand{\hatcurRVjitterAxxxxeccenA}{\ensuremath{0.0}}              % RV jitter (m/s)
\newcommand{\hatcurRVfitrmsAxxxxeccenA}{\ensuremath{.1fym}}            % 
\newcommand{\hatcurRVgammaBxxxxeccenA}{\ensuremath{14.4\pm15.2}}       % RV gamma velocity, relative scale
\newcommand{\hatcurRVjitterBxxxxeccenA}{\ensuremath{35.0}}             % RV jitter (m/s)
\newcommand{\hatcurRVfitrmsBxxxxeccenA}{\ensuremath{.1fym}}            % 
\newcommand{\hatcurRVeccenxxxxeccenA}{\ensuremath{0.067\pm0.062}}      % eccentricity
\newcommand{\hatcurRVomegaxxxxeccenA}{\ensuremath{208\pm91}}           % argument of pericenter
\newcommand{\hatcurPPixxxxeccenA}{\ensuremath{86.2\pm1.2}}             % orbital inclination
\newcommand{\hatcurPPgxxxxeccenA}{\ensuremath{15.2_{-3.4}^{+5.7}}}     % planetary surface gravity (m/s^2)
\newcommand{\hatcurPPloggxxxxeccenA}{\ensuremath{3.18\pm0.12}}         % planetary surface gravity (log cgs)
\newcommand{\hatcurPParxxxxeccenA}{\ensuremath{8.28_{-0.83}^{+1.09}}}  % relative orbital radius (a/R*)
\newcommand{\hatcurPParelxxxxeccenA}{\ensuremath{0.0576\pm0.0011}}     % semimajor axis (AU)
\newcommand{\hatcurPPrhoxxxxeccenA}{\ensuremath{0.61_{-0.18}^{+0.39}}} % planetary density (cgs)
\newcommand{\hatcurPPmxxxxeccenA}{\ensuremath{0.96\pm0.14}}            % planetary mass (M_jup)
\newcommand{\hatcurPPmshortxxxxeccenA}{\ensuremath{0.96}}              % planetary mass (M_jup)
\newcommand{\hatcurPPmlongxxxxeccenA}{\ensuremath{0.957\pm0.137}}      % planetary mass (M_jup)
\newcommand{\hatcurPPmexxxxeccenA}{\ensuremath{304.2\pm43.6}}          % planetary mass (M_earth)
\newcommand{\hatcurPPmeshortxxxxeccenA}{\ensuremath{304.2}}            % planetary mass (M_earth)
\newcommand{\hatcurPPmelongxxxxeccenA}{\ensuremath{304.17\pm43.57}}    % planetary mass (M_earth)
\newcommand{\hatcurPPrxxxxeccenA}{\ensuremath{1.25\pm0.19}}            % planetary radius (R_jup)
\newcommand{\hatcurPPrshortxxxxeccenA}{\ensuremath{1.25}}              % planetary radius (R_jup)
\newcommand{\hatcurPPrlongxxxxeccenA}{\ensuremath{1.250\pm0.185}}      % planetary radius (R_jup)
\newcommand{\hatcurPPrexxxxeccenA}{\ensuremath{14.0\pm2.1}}            % planetary radius (R_earth)
\newcommand{\hatcurPPreshortxxxxeccenA}{\ensuremath{14.0}}             % planetary radius (R_earth)
\newcommand{\hatcurPPrelongxxxxeccenA}{\ensuremath{14.01\pm2.08}}      % planetary radius (R_earth)
\newcommand{\hatcurPPmrcorrxxxxeccenA}{\ensuremath{0.33}}              % mass/radius correlation
\newcommand{\hatcurPPteffxxxxeccenA}{\ensuremath{1415\pm81}}           % planetary temperature (K)
\newcommand{\hatcurPPthetaxxxxeccenA}{\ensuremath{0.074_{-0.012}^{+0.016}}} % Safranov number
\newcommand{\hatcurPPfluxperixxxxeccenA}{\ensuremath{10.37_{-1.74}^{+4.17}}} % flux @ periastron (CGS)
\newcommand{\hatcurPPfluxperidimxxxxeccenA}{\ensuremath{8}}            % flux @ periastron (CGS) units.
\newcommand{\hatcurPPfluxapxxxxeccenA}{\ensuremath{8.01\pm2.00}}       % flux @ apastron (CGS)
\newcommand{\hatcurPPfluxapdimxxxxeccenA}{\ensuremath{8}}              % flux @ apastron (CGS) units.
\newcommand{\hatcurPPfluxavgxxxxeccenA}{\ensuremath{9.05_{-1.78}^{+2.48}}} % flux on average (CGS)
\newcommand{\hatcurPPfluxavgdimxxxxeccenA}{\ensuremath{8}}             % flux average (CGS) units.
\newcommand{\hatcurXsecphasexxxxeccenA}{\ensuremath{0.4925\pm0.0334}}  % Phase of secondary eclipse
\newcommand{\hatcurXsecondaryxxxxeccenA}{\ensuremath{2455954.812\pm0.155}} % Secondary eclipse epoch
\newcommand{\hatcurXsecdurxxxxeccenA}{\ensuremath{0.1660\pm0.0176}}    % sec eclipse duration (days)
\newcommand{\hatcurXsecingdurxxxxeccenA}{\ensuremath{0.0179\pm0.0090}} % sec I/E duration (days)
\newcommand{\hatcurPPphiconjxxxxeccenA}{\ensuremath{-0.1175_{-0.2454}^{+0.3434}}} % phase diff between conjunction and periastron
\newcommand{\hatcurPPperixxxxeccenA}{\ensuremath{2455953.07\pm1.39}}   % time of periastron passage.
\newcommand{\hatcurPPaequivxxxxeccenA}{\ensuremath{0.0389_{-0.0039}^{+0.0052}}} % equivalent semi-major axis
\newcommand{\hatcurPPtcircxxxxeccenA}{\ensuremath{788.0_{-337.6}^{+1008.8}}} % circularization timescale
\newcommand{\hatcurPPtinfallxxxxeccenA}{\ensuremath{2137.0_{-828.7}^{+2492.2}}} % infall timescale
\newcommand{\hatcurXdistxxxxeccenA}{\ensuremath{445\pm57}}             % distance (pc), no reddenning correction
\newcommand{\hatcurXAvxxxxeccenA}{\ensuremath{0.000\pm0.023}}          % Av (mag)
\newcommand{\hatcurXdistredxxxxeccenA}{\ensuremath{439\pm56}}          % distance with Av correction (pc)
\newcommand{\hatcurXEBVxxxxeccenA}{\ensuremath{0.000\pm0.007}}         % E(B-V) (mag)
\newcommand{\hatcurXmvisoredxxxxeccenA}{\ensuremath{12.191\pm0.025}}   % Expected m_v with reddening (mag)
\newcommand{\hatcurXmiisoredxxxxeccenA}{\ensuremath{11.477\pm0.016}}   % Expected m_i with reddening (mag)
\newcommand{\hatcurXmjisoredxxxxeccenA}{\ensuremath{11.030\pm0.014}}   % Expected m_j with reddening (mag)
\newcommand{\hatcurXmhisoredxxxxeccenA}{\ensuremath{10.695\pm0.016}}   % Expected m_h with reddening (mag)
\newcommand{\hatcurXmkisoredxxxxeccenA}{\ensuremath{10.640\pm0.017}}   % Expected m_k with reddening (mag)
\newcommand{\hatcurXviisoredxxxxeccenA}{\ensuremath{0.714\pm0.015}}    % Expected V-I with reddening (mag)
\newcommand{\hatcurXvkisoredxxxxeccenA}{\ensuremath{1.551\pm0.031}}    % Expected V-K with reddening (mag)
\newcommand{\hatcurXjhisoredxxxxeccenA}{\ensuremath{0.334\pm0.009}}    % Expected J-H with reddening (mag)
\newcommand{\hatcurXjkisoredxxxxeccenA}{\ensuremath{0.390\pm0.010}}    % Expected J-K with reddening (mag)
\newcommand{\hatcurCCpmraxxxxeccenA}{\ensuremath{-6.2\pm1.9}}          % proper motion, in RA
\newcommand{\hatcurCCpmdecxxxxeccenA}{\ensuremath{-29.3\pm2.0}}        % proper motion, in DEC
\newcommand{\hatcurCCpmxxxxeccenA}{\ensuremath{29.9488\pm2.75862}}     % proper motion

\newcommand{\hatcurhtrxxxxeccenB}{HTR365-006}                          % Original HTR name of target
\newcommand{\hatcurfieldxxxxeccenB}{317}                               % Original HTR field
\newcommand{\hatcurCCraxxxxeccenB}{\ensuremath{08^{\mathrm h}35^{\mathrm m}42.18{\mathrm s}}}                        % Right Ascension
\newcommand{\hatcurCCdecxxxxeccenB}{\ensuremath{+10{\arcdeg}12{\arcmin}24.0{\arcsec}}}                       % Declination
\newcommand{\hatcurCCmagxxxxeccenB}{13.356}                            % apparent V-band magnitude
\newcommand{\hatcurCCtwomassxxxxeccenB}{2MASS~08354217+1012239}        % 2MASS identifier
\newcommand{\hatcurCCgscxxxxeccenB}{GSC~0801-00608}                    % GSC(1.2) identifier
\newcommand{\hatcurCCtassmvxxxxeccenB}{\ensuremath{13.356\pm0.030}}                         % note, APASS V-band magnitude
\newcommand{\hatcurCCtassmbxxxxeccenB}{\ensuremath{14.120\pm0.060}}                         % note, APASS B-band magnitude
\newcommand{\hatcurCCtwomassJmagxxxxeccenB}{\ensuremath{12.146\pm0.021}} % 2MASS ORIG MAG
\newcommand{\hatcurCCtwomassHmagxxxxeccenB}{\ensuremath{11.809\pm0.029}} % 2MASS ORIG MAG
\newcommand{\hatcurCCtwomassKmagxxxxeccenB}{\ensuremath{11.764\pm0.023}} % 2MASS ORIG MAG
\newcommand{\hatcurCCcitJmagxxxxeccenB}{\ensuremath{12.162\pm0.021}}   % 2MASS CIT MAG
\newcommand{\hatcurCCcitHmagxxxxeccenB}{\ensuremath{11.804\pm0.029}}   % 2MASS CIT MAG
\newcommand{\hatcurCCcitKmagxxxxeccenB}{\ensuremath{11.788\pm0.023}}   % 2MASS CIT MAG
\newcommand{\hatcurCCbbJmagxxxxeccenB}{\ensuremath{12.213\pm0.023}}    % 2MASS BB MAG
\newcommand{\hatcurCCbbHmagxxxxeccenB}{\ensuremath{11.825\pm0.030}}    % 2MASS BB MAG
\newcommand{\hatcurCCbbKmagxxxxeccenB}{\ensuremath{11.808\pm0.023}}    % 2MASS BB MAG
\newcommand{\hatcurCCesoJmagxxxxeccenB}{\ensuremath{12.215\pm0.025}}   % 2MASS ESO MAG
\newcommand{\hatcurCCesoHmagxxxxeccenB}{\ensuremath{11.819\pm0.033}}   % 2MASS ESO MAG
\newcommand{\hatcurCCesoKmagxxxxeccenB}{\ensuremath{11.807\pm0.024}}   % 2MASS ESO MAG
\newcommand{\hatcurCCesoJHmagxxxxeccenB}{\ensuremath{0.395\pm0.039}}   % 2MASS ESO JH COLOR
\newcommand{\hatcurCCesoJKmagxxxxeccenB}{\ensuremath{0.409\pm0.034}}   % 2MASS ESO JK COLOR
\newcommand{\hatcurCCesoHKmagxxxxeccenB}{\ensuremath{0.013\pm0.041}}   % 2MASS ESO HK COLOR
\newcommand{\hatcurLCdipxxxxeccenB}{\ensuremath{16.5}}                 % BLS detected dip (mmag)
\newcommand{\hatcurLCrprstarxxxxeccenB}{\ensuremath{0.1192\pm0.0019}}  % Rp/R*
\newcommand{\hatcurLCbsqxxxxeccenB}{\ensuremath{0.038_{-0.025}^{+0.058}}} % impact parameter square
\newcommand{\hatcurLCimpxxxxeccenB}{\ensuremath{0.195_{-0.096}^{+0.105}}} % impact parameter
\newcommand{\hatcurLCzetaxxxxeccenB}{\ensuremath{16.62\pm0.09}}        % zeta/R*
\newcommand{\hatcurLCdurxxxxeccenB}{\ensuremath{0.1354\pm0.0012}}      % transit duration (days)
\newcommand{\hatcurLCdurshortxxxxeccenB}{\ensuremath{0.1354}}          % transit duration (days)
\newcommand{\hatcurLCdurhrxxxxeccenB}{\ensuremath{3.249\pm0.028}}      % transit duration (hours)
\newcommand{\hatcurLCdurhrshortxxxxeccenB}{\ensuremath{3.249}}         % transit duration (hours)
\newcommand{\hatcurLCqxxxxeccenB}{\ensuremath{0.0406\pm0.0004}}        % fractional transit duration (days)
\newcommand{\hatcurLCqshortxxxxeccenB}{\ensuremath{0.041}}             % fractional transit duration (days)
\newcommand{\hatcurLCingdurxxxxeccenB}{\ensuremath{0.0150\pm0.0009}}   % ingress/egress duration (days)
\newcommand{\hatcurLCPxxxxeccenB}{\ensuremath{3.332687\pm0.000015}}    % period (days)
\newcommand{\hatcurLCPprecxxxxeccenB}{\ensuremath{3.3326874}}          % period (days)
\newcommand{\hatcurLCPshortxxxxeccenB}{\ensuremath{3.3327}}            % period (days)
\newcommand{\hatcurLCTxxxxeccenB}{\ensuremath{2455994.03838\pm0.00034}} % epoch (BJD)
\newcommand{\hatcurLCTAxxxxeccenB}{\ensuremath{2455504.13333\pm0.00215}} % TA (BJD)
\newcommand{\hatcurLCTBxxxxeccenB}{\ensuremath{2456010.70182\pm0.00038}} % TB (BJD)
\newcommand{\hatcurLChatnetmAxxxxeccenB}{\ensuremath{13.2562\pm0.0002}} % HATNet OOT level
\newcommand{\hatcurLCiblendAxxxxeccenB}{\ensuremath{0.77\pm0.07}}      % HATNet iblend factor
\newcommand{\hatcurLChatnetmBxxxxeccenB}{\ensuremath{13.2559\pm0.0002}} % HATNet OOT level
\newcommand{\hatcurLCiblendBxxxxeccenB}{\ensuremath{0.86\pm0.07}}      % HATNet iblend factor
\newcommand{\hatcurSMEiteffxxxxeccenB}{\ensuremath{5738\pm74}}         % Ini SME, stellar effective temperature
\newcommand{\hatcurSMEizfehxxxxeccenB}{\ensuremath{0.28\pm0.08}}       % Ini SME, stellar metallicity
\newcommand{\hatcurSMEizfehshortxxxxeccenB}{\ensuremath{0.28}}         % Ini SME, stellar metallicity
\newcommand{\hatcurSMEiloggxxxxeccenB}{\ensuremath{4.47\pm0.13}}       % Ini SME, stellar surface gravity
\newcommand{\hatcurSMEivsinxxxxeccenB}{\ensuremath{2.4\pm0.5}}         % Ini SME, stellar rotational velocity
\newcommand{\hatcurSMEivmacxxxxeccenB}{\ensuremath{NULL}}              % Ini SME, stellar macroturbulence
\newcommand{\hatcurSMEivmicxxxxeccenB}{\ensuremath{NULL}}              % Ini SME, stellar microturbulence
\newcommand{\hatcurSMEiiteffxxxxeccenB}{\ensuremath{5622\pm74}}        % Final SME, stellar effective temperature
\newcommand{\hatcurSMEiizfehxxxxeccenB}{\ensuremath{0.22\pm0.08}}      % Final SME, stellar metallicity
\newcommand{\hatcurSMEiizfehshortxxxxeccenB}{\ensuremath{0.22}}        % Final SME, stellar metallicity
\newcommand{\hatcurSMEiiloggxxxxeccenB}{\ensuremath{4.33\pm0.08}}      % Final SME, stellar surface gravity
\newcommand{\hatcurSMEiivsinxxxxeccenB}{\ensuremath{2.4\pm0.5}}        % Final SME, stellar rotational velocity
\newcommand{\hatcurSMEiivmacxxxxeccenB}{\ensuremath{NULL}}             % Final SME, stellar macroturbulence
\newcommand{\hatcurSMEiivmicxxxxeccenB}{\ensuremath{NULL}}             % Final SME, stellar microturbulence
\newcommand{\hatcurDSteffxxxxeccenB}{\ensuremath{NULL\pmNULL}}         % DS stellar effective temperature
\newcommand{\hatcurDSzfehxxxxeccenB}{\ensuremath{NULL\pmNULL}}         % DS stellar metallicity
\newcommand{\hatcurDSloggxxxxeccenB}{\ensuremath{NULL\pmNULL}}         % DS stellar surface gravity
\newcommand{\hatcurDSvsinixxxxeccenB}{\ensuremath{NULL\pmNULL}}        % DS stellar rotational velocity
\newcommand{\hatcurDSgammaxxxxeccenB}{\ensuremath{NULL\pmNULL}}        % DS absolute gamma velocity
\newcommand{\hatcurDSnumspecxxxxeccenB}{\ensuremath{0}}                % DS number of spectra
\newcommand{\hatcurDSspanxxxxeccenB}{\ensuremath{0}}                   % DS stellar surface gravity
\newcommand{\hatcurDSrvrmsxxxxeccenB}{\ensuremath{0.00}}               % DS rms of RV values [km/s]
\newcommand{\hatcurTRESteffxxxxeccenB}{\ensuremath{5738\pm74}}         % TRES stellar effective temperature
\newcommand{\hatcurTRESzfehxxxxeccenB}{\ensuremath{0.28\pm0.1}}        % TRES stellar metallicity
\newcommand{\hatcurTRESloggxxxxeccenB}{\ensuremath{4.47\pm0.13}}       % TRES stellar surface gravity
\newcommand{\hatcurTRESvsinixxxxeccenB}{\ensuremath{2.4\pm0.5}}        % TRES stellar rotational velocity
\newcommand{\hatcurTRESgammaxxxxeccenB}{\ensuremath{-4.42\pm0.16}}     % TRES absolute gamma velocity
\newcommand{\hatcurTRESnumspecxxxxeccenB}{\ensuremath{2}}              % TRES number of spectra
\newcommand{\hatcurTRESspanxxxxeccenB}{\ensuremath{0}}                 % TRES stellar surface gravity
\newcommand{\hatcurTRESrvrmsxxxxeccenB}{\ensuremath{0.00}}             % TRES rms of RV values [km/s]
\newcommand{\hatcurFIESteffxxxxeccenB}{\ensuremath{5738\pm74}}         % FIES stellar effective temperature
\newcommand{\hatcurFIESzfehxxxxeccenB}{\ensuremath{0.28\pm0.08}}       % FIES stellar metallicity
\newcommand{\hatcurFIESloggxxxxeccenB}{\ensuremath{4.47\pm0.13}}       % FIES stellar surface gravity
\newcommand{\hatcurFIESvsinixxxxeccenB}{\ensuremath{2.4\pm0.5}}        % FIES stellar rotational velocity
\newcommand{\hatcurFIESgammaxxxxeccenB}{\ensuremath{-4.42\pm0.16}}     % FIES absolute gamma velocity
\newcommand{\hatcurFIESnumspecxxxxeccenB}{\ensuremath{2}}              % FIES number of spectra
\newcommand{\hatcurFIESspanxxxxeccenB}{\ensuremath{0}}                 % FIES stellar surface gravity
\newcommand{\hatcurFIESrvrmsxxxxeccenB}{\ensuremath{0.00}}             % FIES rms of RV values [km/s]
\newcommand{\hatcurLBizxxxxeccenB}{\ensuremath{0.2338}}                % Limb darkening parameters, Gamma1, z-band
\newcommand{\hatcurLBiizxxxxeccenB}{\ensuremath{0.3204}}               % Limb darkening parameters, Gamma2, z-band
\newcommand{\hatcurLBiixxxxeccenB}{\ensuremath{0.3036}}                % Limb darkening parameters, Gamma1, i-band
\newcommand{\hatcurLBiiixxxxeccenB}{\ensuremath{0.3158}}               % Limb darkening parameters, Gamma2, i-band
\newcommand{\hatcurLBiIxxxxeccenB}{\ensuremath{0.2803}}                % Limb darkening parameters, Gamma1, I-band
\newcommand{\hatcurLBiiIxxxxeccenB}{\ensuremath{0.3179}}               % Limb darkening parameters, Gamma2, I-band
\newcommand{\hatcurLBigxxxxeccenB}{\ensuremath{0.6151}}                % Limb darkening parameters, Gamma1, g-band
\newcommand{\hatcurLBiigxxxxeccenB}{\ensuremath{0.1903}}               % Limb darkening parameters, Gamma2, g-band
\newcommand{\hatcurLBirxxxxeccenB}{\ensuremath{0.4035}}                % Limb darkening parameters, Gamma1, r-band
\newcommand{\hatcurLBiirxxxxeccenB}{\ensuremath{0.3004}}               % Limb darkening parameters, Gamma2, r-band
\newcommand{\hatcurLBiRxxxxeccenB}{\ensuremath{0.3757}}                % Limb darkening parameters, Gamma1, R-band
\newcommand{\hatcurLBiiRxxxxeccenB}{\ensuremath{0.3056}}               % Limb darkening parameters, Gamma2, R-band
\newcommand{\hatcurLBikepxxxxeccenB}{\ensuremath{}}            % darkening parameters, Gamma1, Kep-band
\newcommand{\hatcurLBiikepxxxxeccenB}{\ensuremath{}}           % darkening parameters, Gamma2, Kep-band
\newcommand{\hatcurISOmxxxxeccenB}{\ensuremath{1.04\pm0.05}}           % stellar mass
\newcommand{\hatcurISOmshortxxxxeccenB}{\ensuremath{1.04}}             % stellar mass
\newcommand{\hatcurISOmlongxxxxeccenB}{\ensuremath{1.044\pm0.050}}     % stellar mass
\newcommand{\hatcurISOrxxxxeccenB}{\ensuremath{1.15_{-0.08}^{+0.20}}}  % stellar radius
\newcommand{\hatcurISOrshortxxxxeccenB}{\ensuremath{1.15}}             % stellar radius
\newcommand{\hatcurISOrlongxxxxeccenB}{\ensuremath{1.151_{-0.081}^{+0.200}}} % stellar radius
\newcommand{\hatcurISOrhoxxxxeccenB}{\ensuremath{0.96\pm0.27}}         % stellar density (cgs)
\newcommand{\hatcurISOloggxxxxeccenB}{\ensuremath{4.33\pm0.09}}        % stellar surface gravity from isochrones
\newcommand{\hatcurISOlumxxxxeccenB}{\ensuremath{1.19_{-0.18}^{+0.50}}} % stellar luminosity
\newcommand{\hatcurISOlumshortxxxxeccenB}{\ensuremath{1.19}}           % stellar luminosity
\newcommand{\hatcurISOmvxxxxeccenB}{\ensuremath{4.66\pm0.27}}          % stellar absolute magnitude
\newcommand{\hatcurISOvixxxxeccenB}{\ensuremath{0.751\pm0.023}}        % stellar V-I index
\newcommand{\hatcurISOagexxxxeccenB}{\ensuremath{6.5_{-1.8}^{+2.5}}}   % stellar age
\newcommand{\hatcurISOsigmaxxxxeccenB}{\ensuremath{0.00070\pm0.00015}} % system mass-correction sigma parameter
\newcommand{\hatcurISOMJxxxxeccenB}{\ensuremath{3.44\pm0.26}}          % stellar absolute J magnitude
\newcommand{\hatcurISOMHxxxxeccenB}{\ensuremath{3.08\pm0.26}}          % stellar absolute H magnitude
\newcommand{\hatcurISOMKxxxxeccenB}{\ensuremath{3.02\pm0.26}}          % stellar absolute K magnitude
\newcommand{\hatcurISOJKxxxxeccenB}{\ensuremath{0.42\pm0.02}}          % J-K color index from isochrones.
\newcommand{\hatcurISOspecxxxxeccenB}{G}                               % stellar spectral type
\newcommand{\hatcurRVKxxxxeccenB}{\ensuremath{89.7\pm13.0}}            % RV semi-amplitude [m/s]
\newcommand{\hatcurRVrkxxxxeccenB}{\ensuremath{-0.123\pm0.181}}        % sqrt(e)*cos(omega)
\newcommand{\hatcurRVrhxxxxeccenB}{\ensuremath{0.146\pm0.228}}         % sqrt(e)*sin(omega)
\newcommand{\hatcurRVkxxxxeccenB}{\ensuremath{-0.031_{-0.090}^{+0.055}}} % e*cos(omega)
\newcommand{\hatcurRVhxxxxeccenB}{\ensuremath{0.035_{-0.070}^{+0.130}}} % e*sin(omega)
\newcommand{\hatcurRVtronexxxxeccenB}{\ensuremath{0.0000\pm0.0000}}    % RV linear trend tr1 factor
\newcommand{\hatcurRVtrtwoxxxxeccenB}{\ensuremath{0.0000\pm0.0000}}    % RV linear trend tr2 factor
\newcommand{\hatcurRVgammaxxxxeccenB}{\ensuremath{-5082.5\pm9.1}}      % RV gamma velocity, relative scale
\newcommand{\hatcurRVjitterxxxxeccenB}{\ensuremath{0.0}}               % RV jitter (m/s)
\newcommand{\hatcurRVfitrmsxxxxeccenB}{\ensuremath{10.0}}              % RVfitrms
\newcommand{\hatcurRVeccenxxxxeccenB}{\ensuremath{0.091\pm0.094}}      % eccentricity
\newcommand{\hatcurRVomegaxxxxeccenB}{\ensuremath{135\pm80}}           % argument of pericenter
\newcommand{\hatcurPPixxxxeccenB}{\ensuremath{88.5_{-1.1}^{+0.7}}}     % orbital inclination
\newcommand{\hatcurPPgxxxxeccenB}{\ensuremath{9.1\pm1.8}}              % planetary surface gravity (m/s^2)
\newcommand{\hatcurPPloggxxxxeccenB}{\ensuremath{2.96_{-0.10}^{+0.08}}} % planetary surface gravity (log cgs)
\newcommand{\hatcurPParxxxxeccenB}{\ensuremath{8.25_{-1.01}^{+0.64}}}  % relative orbital radius (a/R*)
\newcommand{\hatcurPParelxxxxeccenB}{\ensuremath{0.0443\pm0.0007}}     % semimajor axis (AU)
\newcommand{\hatcurPPrhoxxxxeccenB}{\ensuremath{0.34\pm0.10}}          % planetary density (cgs)
\newcommand{\hatcurPPmxxxxeccenB}{\ensuremath{0.67\pm0.10}}            % planetary mass (M_jup)
\newcommand{\hatcurPPmshortxxxxeccenB}{\ensuremath{0.67}}              % planetary mass (M_jup)
\newcommand{\hatcurPPmlongxxxxeccenB}{\ensuremath{0.674\pm0.097}}      % planetary mass (M_jup)
\newcommand{\hatcurPPmexxxxeccenB}{\ensuremath{214.1\pm30.9}}          % planetary mass (M_earth)
\newcommand{\hatcurPPmeshortxxxxeccenB}{\ensuremath{214.1}}            % planetary mass (M_earth)
\newcommand{\hatcurPPmelongxxxxeccenB}{\ensuremath{214.07\pm30.92}}    % planetary mass (M_earth)
\newcommand{\hatcurPPrxxxxeccenB}{\ensuremath{1.34_{-0.10}^{+0.24}}}   % planetary radius (R_jup)
\newcommand{\hatcurPPrshortxxxxeccenB}{\ensuremath{1.34}}              % planetary radius (R_jup)
\newcommand{\hatcurPPrlongxxxxeccenB}{\ensuremath{1.340_{-0.099}^{+0.236}}} % planetary radius (R_jup)
\newcommand{\hatcurPPrexxxxeccenB}{\ensuremath{15.0_{-1.1}^{+2.6}}}    % planetary radius (R_earth)
\newcommand{\hatcurPPreshortxxxxeccenB}{\ensuremath{15.0}}             % planetary radius (R_earth)
\newcommand{\hatcurPPrelongxxxxeccenB}{\ensuremath{15.02_{-1.11}^{+2.64}}} % planetary radius (R_earth)
\newcommand{\hatcurPPmrcorrxxxxeccenB}{\ensuremath{0.50}}              % mass/radius correlation
\newcommand{\hatcurPPteffxxxxeccenB}{\ensuremath{1386_{-52}^{+114}}}   % planetary temperature (K)
\newcommand{\hatcurPPthetaxxxxeccenB}{\ensuremath{0.042\pm0.006}}      % Safranov number
\newcommand{\hatcurPPfluxperixxxxeccenB}{\ensuremath{9.73_{-1.61}^{+13.16}}} % flux @ periastron (CGS)
\newcommand{\hatcurPPfluxperidimxxxxeccenB}{\ensuremath{8}}            % flux @ periastron (CGS) units.
\newcommand{\hatcurPPfluxapxxxxeccenB}{\ensuremath{7.22_{-1.16}^{+0.82}}} % flux @ apastron (CGS)
\newcommand{\hatcurPPfluxapdimxxxxeccenB}{\ensuremath{8}}              % flux @ apastron (CGS) units.
\newcommand{\hatcurPPfluxavgxxxxeccenB}{\ensuremath{8.33_{-1.15}^{+3.58}}} % flux on average (CGS)
\newcommand{\hatcurPPfluxavgdimxxxxeccenB}{\ensuremath{8}}             % flux average (CGS) units.
\newcommand{\hatcurXsecphasexxxxeccenB}{\ensuremath{0.4802\pm0.0482}}  % Phase of secondary eclipse
\newcommand{\hatcurXsecondaryxxxxeccenB}{\ensuremath{2455995.639\pm0.161}} % Secondary eclipse epoch
\newcommand{\hatcurXsecdurxxxxeccenB}{\ensuremath{0.1451\pm0.0341}}    % sec eclipse duration (days)
\newcommand{\hatcurXsecingdurxxxxeccenB}{\ensuremath{0.0164\pm0.0056}} % sec I/E duration (days)
\newcommand{\hatcurPPphiconjxxxxeccenB}{\ensuremath{-0.0634\pm0.2115}} % phase diff between conjunction and periastron
\newcommand{\hatcurPPperixxxxeccenB}{\ensuremath{2455994.25\pm0.70}}   % time of periastron passage.
\newcommand{\hatcurPPaequivxxxxeccenB}{\ensuremath{0.0405_{-0.0050}^{+0.0034}}} % equivalent semi-major axis
\newcommand{\hatcurPPtcircxxxxeccenB}{\ensuremath{83.2\pm41.0}}        % circularization timescale
\newcommand{\hatcurPPtinfallxxxxeccenB}{\ensuremath{1877.9_{-814.2}^{+1152.3}}} % infall timescale
\newcommand{\hatcurXdistxxxxeccenB}{\ensuremath{571_{-41}^{+99}}}      % distance (pc), no reddenning correction
\newcommand{\hatcurXAvxxxxeccenB}{\ensuremath{0.000\pm0.014}}          % Av (mag)
\newcommand{\hatcurXdistredxxxxeccenB}{\ensuremath{563_{-42}^{+98}}}   % distance with Av correction (pc)
\newcommand{\hatcurXEBVxxxxeccenB}{\ensuremath{0.000\pm0.005}}         % E(B-V) (mag)
\newcommand{\hatcurXmvisoredxxxxeccenB}{\ensuremath{13.418\pm0.041}}   % Expected m_v with reddening (mag)
\newcommand{\hatcurXmiisoredxxxxeccenB}{\ensuremath{12.668\pm0.022}}   % Expected m_i with reddening (mag)
\newcommand{\hatcurXmjisoredxxxxeccenB}{\ensuremath{12.197\pm0.014}}   % Expected m_j with reddening (mag)
\newcommand{\hatcurXmhisoredxxxxeccenB}{\ensuremath{11.834\pm0.020}}   % Expected m_h with reddening (mag)
\newcommand{\hatcurXmkisoredxxxxeccenB}{\ensuremath{11.775\pm0.022}}   % Expected m_k with reddening (mag)
\newcommand{\hatcurXviisoredxxxxeccenB}{\ensuremath{0.752\pm0.021}}    % Expected V-I with reddening (mag)
\newcommand{\hatcurXvkisoredxxxxeccenB}{\ensuremath{1.643\pm0.056}}    % Expected V-K with reddening (mag)
\newcommand{\hatcurXjhisoredxxxxeccenB}{\ensuremath{0.362\pm0.015}}    % Expected J-H with reddening (mag)
\newcommand{\hatcurXjkisoredxxxxeccenB}{\ensuremath{0.422\pm0.018}}    % Expected J-K with reddening (mag)
\newcommand{\hatcurCCpmraxxxxeccenB}{\ensuremath{-10.3\pm2.6}}         % proper motion, in RA
\newcommand{\hatcurCCpmdecxxxxeccenB}{\ensuremath{-16.0\pm3.2}}        % proper motion, in DEC
\newcommand{\hatcurCCpmxxxxeccenB}{\ensuremath{19.0287\pm4.12311}}     % proper motion

\newcommand{\hatcurCCbbHmageccen}[1]{\ifnum#1=42 %
\hatcurCCbbHmagxxxxeccenA
\else
\ifnum#1=43 %
\hatcurCCbbHmagxxxxeccenB
\else
??????\fi
\fi
}
\newcommand{\hatcurCCbbJmageccen}[1]{\ifnum#1=42 %
\hatcurCCbbJmagxxxxeccenA
\else
\ifnum#1=43 %
\hatcurCCbbJmagxxxxeccenB
\else
??????\fi
\fi
}
\newcommand{\hatcurCCbbKmageccen}[1]{\ifnum#1=42 %
\hatcurCCbbKmagxxxxeccenA
\else
\ifnum#1=43 %
\hatcurCCbbKmagxxxxeccenB
\else
??????\fi
\fi
}
\newcommand{\hatcurCCcitHmageccen}[1]{\ifnum#1=42 %
\hatcurCCcitHmagxxxxeccenA
\else
\ifnum#1=43 %
\hatcurCCcitHmagxxxxeccenB
\else
??????\fi
\fi
}
\newcommand{\hatcurCCcitJmageccen}[1]{\ifnum#1=42 %
\hatcurCCcitJmagxxxxeccenA
\else
\ifnum#1=43 %
\hatcurCCcitJmagxxxxeccenB
\else
??????\fi
\fi
}
\newcommand{\hatcurCCcitKmageccen}[1]{\ifnum#1=42 %
\hatcurCCcitKmagxxxxeccenA
\else
\ifnum#1=43 %
\hatcurCCcitKmagxxxxeccenB
\else
??????\fi
\fi
}
\newcommand{\hatcurCCdececcen}[1]{\ifnum#1=42 %
\hatcurCCdecxxxxeccenA
\else
\ifnum#1=43 %
\hatcurCCdecxxxxeccenB
\else
??????\fi
\fi
}
\newcommand{\hatcurCCesoHKmageccen}[1]{\ifnum#1=42 %
\hatcurCCesoHKmagxxxxeccenA
\else
\ifnum#1=43 %
\hatcurCCesoHKmagxxxxeccenB
\else
??????\fi
\fi
}
\newcommand{\hatcurCCesoHmageccen}[1]{\ifnum#1=42 %
\hatcurCCesoHmagxxxxeccenA
\else
\ifnum#1=43 %
\hatcurCCesoHmagxxxxeccenB
\else
??????\fi
\fi
}
\newcommand{\hatcurCCesoJHmageccen}[1]{\ifnum#1=42 %
\hatcurCCesoJHmagxxxxeccenA
\else
\ifnum#1=43 %
\hatcurCCesoJHmagxxxxeccenB
\else
??????\fi
\fi
}
\newcommand{\hatcurCCesoJKmageccen}[1]{\ifnum#1=42 %
\hatcurCCesoJKmagxxxxeccenA
\else
\ifnum#1=43 %
\hatcurCCesoJKmagxxxxeccenB
\else
??????\fi
\fi
}
\newcommand{\hatcurCCesoJmageccen}[1]{\ifnum#1=42 %
\hatcurCCesoJmagxxxxeccenA
\else
\ifnum#1=43 %
\hatcurCCesoJmagxxxxeccenB
\else
??????\fi
\fi
}
\newcommand{\hatcurCCesoKmageccen}[1]{\ifnum#1=42 %
\hatcurCCesoKmagxxxxeccenA
\else
\ifnum#1=43 %
\hatcurCCesoKmagxxxxeccenB
\else
??????\fi
\fi
}
\newcommand{\hatcurCCgsceccen}[1]{\ifnum#1=42 %
\hatcurCCgscxxxxeccenA
\else
\ifnum#1=43 %
\hatcurCCgscxxxxeccenB
\else
??????\fi
\fi
}
\newcommand{\hatcurCCmageccen}[1]{\ifnum#1=42 %
\hatcurCCmagxxxxeccenA
\else
\ifnum#1=43 %
\hatcurCCmagxxxxeccenB
\else
??????\fi
\fi
}
\newcommand{\hatcurCCpmeccen}[1]{\ifnum#1=42 %
\hatcurCCpmxxxxeccenA
\else
\ifnum#1=43 %
\hatcurCCpmxxxxeccenB
\else
??????\fi
\fi
}
\newcommand{\hatcurCCpmdececcen}[1]{\ifnum#1=42 %
\hatcurCCpmdecxxxxeccenA
\else
\ifnum#1=43 %
\hatcurCCpmdecxxxxeccenB
\else
??????\fi
\fi
}
\newcommand{\hatcurCCpmraeccen}[1]{\ifnum#1=42 %
\hatcurCCpmraxxxxeccenA
\else
\ifnum#1=43 %
\hatcurCCpmraxxxxeccenB
\else
??????\fi
\fi
}
\newcommand{\hatcurCCraeccen}[1]{\ifnum#1=42 %
\hatcurCCraxxxxeccenA
\else
\ifnum#1=43 %
\hatcurCCraxxxxeccenB
\else
??????\fi
\fi
}
\newcommand{\hatcurCCtassmveccen}[1]{\ifnum#1=42 %
\hatcurCCtassmvxxxxeccenA
\else
\ifnum#1=43 %
\hatcurCCtassmvxxxxeccenB
\else
??????\fi
\fi
}
\newcommand{\hatcurCCtassmbeccen}[1]{\ifnum#1=42 %
\hatcurCCtassmbxxxxeccenA
\else
\ifnum#1=43 %
\hatcurCCtassmbxxxxeccenB
\else
??????\fi
\fi
}
\newcommand{\hatcurCCtwomasseccen}[1]{\ifnum#1=42 %
\hatcurCCtwomassxxxxeccenA
\else
\ifnum#1=43 %
\hatcurCCtwomassxxxxeccenB
\else
??????\fi
\fi
}
\newcommand{\hatcurCCtwomassHmageccen}[1]{\ifnum#1=42 %
\hatcurCCtwomassHmagxxxxeccenA
\else
\ifnum#1=43 %
\hatcurCCtwomassHmagxxxxeccenB
\else
??????\fi
\fi
}
\newcommand{\hatcurCCtwomassJmageccen}[1]{\ifnum#1=42 %
\hatcurCCtwomassJmagxxxxeccenA
\else
\ifnum#1=43 %
\hatcurCCtwomassJmagxxxxeccenB
\else
??????\fi
\fi
}
\newcommand{\hatcurCCtwomassKmageccen}[1]{\ifnum#1=42 %
\hatcurCCtwomassKmagxxxxeccenA
\else
\ifnum#1=43 %
\hatcurCCtwomassKmagxxxxeccenB
\else
??????\fi
\fi
}
\newcommand{\hatcurDSgammaeccen}[1]{\ifnum#1=42 %
\hatcurDSgammaxxxxeccenA
\else
\ifnum#1=43 %
\hatcurDSgammaxxxxeccenB
\else
??????\fi
\fi
}
\newcommand{\hatcurDSloggeccen}[1]{\ifnum#1=42 %
\hatcurDSloggxxxxeccenA
\else
\ifnum#1=43 %
\hatcurDSloggxxxxeccenB
\else
??????\fi
\fi
}
\newcommand{\hatcurDSnumspececcen}[1]{\ifnum#1=42 %
\hatcurDSnumspecxxxxeccenA
\else
\ifnum#1=43 %
\hatcurDSnumspecxxxxeccenB
\else
??????\fi
\fi
}
\newcommand{\hatcurDSrvrmseccen}[1]{\ifnum#1=42 %
\hatcurDSrvrmsxxxxeccenA
\else
\ifnum#1=43 %
\hatcurDSrvrmsxxxxeccenB
\else
??????\fi
\fi
}
\newcommand{\hatcurDSspaneccen}[1]{\ifnum#1=42 %
\hatcurDSspanxxxxeccenA
\else
\ifnum#1=43 %
\hatcurDSspanxxxxeccenB
\else
??????\fi
\fi
}
\newcommand{\hatcurDSteffeccen}[1]{\ifnum#1=42 %
\hatcurDSteffxxxxeccenA
\else
\ifnum#1=43 %
\hatcurDSteffxxxxeccenB
\else
??????\fi
\fi
}
\newcommand{\hatcurDSvsinieccen}[1]{\ifnum#1=42 %
\hatcurDSvsinixxxxeccenA
\else
\ifnum#1=43 %
\hatcurDSvsinixxxxeccenB
\else
??????\fi
\fi
}
\newcommand{\hatcurDSzfeheccen}[1]{\ifnum#1=42 %
\hatcurDSzfehxxxxeccenA
\else
\ifnum#1=43 %
\hatcurDSzfehxxxxeccenB
\else
??????\fi
\fi
}
\newcommand{\hatcurfieldeccen}[1]{\ifnum#1=42 %
\hatcurfieldxxxxeccenA
\else
\ifnum#1=43 %
\hatcurfieldxxxxeccenB
\else
??????\fi
\fi
}
\newcommand{\hatcurFIESgammaeccen}[1]{\ifnum#1=42 %
\hatcurFIESgammaxxxxeccenA
\else
\ifnum#1=43 %
\hatcurFIESgammaxxxxeccenB
\else
??????\fi
\fi
}
\newcommand{\hatcurFIESloggeccen}[1]{\ifnum#1=42 %
\hatcurFIESloggxxxxeccenA
\else
\ifnum#1=43 %
\hatcurFIESloggxxxxeccenB
\else
??????\fi
\fi
}
\newcommand{\hatcurFIESnumspececcen}[1]{\ifnum#1=42 %
\hatcurFIESnumspecxxxxeccenA
\else
\ifnum#1=43 %
\hatcurFIESnumspecxxxxeccenB
\else
??????\fi
\fi
}
\newcommand{\hatcurFIESrvrmseccen}[1]{\ifnum#1=42 %
\hatcurFIESrvrmsxxxxeccenA
\else
\ifnum#1=43 %
\hatcurFIESrvrmsxxxxeccenB
\else
??????\fi
\fi
}
\newcommand{\hatcurFIESspaneccen}[1]{\ifnum#1=42 %
\hatcurFIESspanxxxxeccenA
\else
\ifnum#1=43 %
\hatcurFIESspanxxxxeccenB
\else
??????\fi
\fi
}
\newcommand{\hatcurFIESteffeccen}[1]{\ifnum#1=42 %
\hatcurFIESteffxxxxeccenA
\else
\ifnum#1=43 %
\hatcurFIESteffxxxxeccenB
\else
??????\fi
\fi
}
\newcommand{\hatcurFIESvsinieccen}[1]{\ifnum#1=42 %
\hatcurFIESvsinixxxxeccenA
\else
\ifnum#1=43 %
\hatcurFIESvsinixxxxeccenB
\else
??????\fi
\fi
}
\newcommand{\hatcurFIESzfeheccen}[1]{\ifnum#1=42 %
\hatcurFIESzfehxxxxeccenA
\else
\ifnum#1=43 %
\hatcurFIESzfehxxxxeccenB
\else
??????\fi
\fi
}
\newcommand{\hatcurhtreccen}[1]{\ifnum#1=42 %
\hatcurhtrxxxxeccenA
\else
\ifnum#1=43 %
\hatcurhtrxxxxeccenB
\else
??????\fi
\fi
}
\newcommand{\hatcurISOageeccen}[1]{\ifnum#1=42 %
\hatcurISOagexxxxeccenA
\else
\ifnum#1=43 %
\hatcurISOagexxxxeccenB
\else
??????\fi
\fi
}
\newcommand{\hatcurISOJKeccen}[1]{\ifnum#1=42 %
\hatcurISOJKxxxxeccenA
\else
\ifnum#1=43 %
\hatcurISOJKxxxxeccenB
\else
??????\fi
\fi
}
\newcommand{\hatcurISOloggeccen}[1]{\ifnum#1=42 %
\hatcurISOloggxxxxeccenA
\else
\ifnum#1=43 %
\hatcurISOloggxxxxeccenB
\else
??????\fi
\fi
}
\newcommand{\hatcurISOlumeccen}[1]{\ifnum#1=42 %
\hatcurISOlumxxxxeccenA
\else
\ifnum#1=43 %
\hatcurISOlumxxxxeccenB
\else
??????\fi
\fi
}
\newcommand{\hatcurISOlumshorteccen}[1]{\ifnum#1=42 %
\hatcurISOlumshortxxxxeccenA
\else
\ifnum#1=43 %
\hatcurISOlumshortxxxxeccenB
\else
??????\fi
\fi
}
\newcommand{\hatcurISOmeccen}[1]{\ifnum#1=42 %
\hatcurISOmxxxxeccenA
\else
\ifnum#1=43 %
\hatcurISOmxxxxeccenB
\else
??????\fi
\fi
}
\newcommand{\hatcurISOMHeccen}[1]{\ifnum#1=42 %
\hatcurISOMHxxxxeccenA
\else
\ifnum#1=43 %
\hatcurISOMHxxxxeccenB
\else
??????\fi
\fi
}
\newcommand{\hatcurISOMJeccen}[1]{\ifnum#1=42 %
\hatcurISOMJxxxxeccenA
\else
\ifnum#1=43 %
\hatcurISOMJxxxxeccenB
\else
??????\fi
\fi
}
\newcommand{\hatcurISOMKeccen}[1]{\ifnum#1=42 %
\hatcurISOMKxxxxeccenA
\else
\ifnum#1=43 %
\hatcurISOMKxxxxeccenB
\else
??????\fi
\fi
}
\newcommand{\hatcurISOmlongeccen}[1]{\ifnum#1=42 %
\hatcurISOmlongxxxxeccenA
\else
\ifnum#1=43 %
\hatcurISOmlongxxxxeccenB
\else
??????\fi
\fi
}
\newcommand{\hatcurISOmshorteccen}[1]{\ifnum#1=42 %
\hatcurISOmshortxxxxeccenA
\else
\ifnum#1=43 %
\hatcurISOmshortxxxxeccenB
\else
??????\fi
\fi
}
\newcommand{\hatcurISOmveccen}[1]{\ifnum#1=42 %
\hatcurISOmvxxxxeccenA
\else
\ifnum#1=43 %
\hatcurISOmvxxxxeccenB
\else
??????\fi
\fi
}
\newcommand{\hatcurISOreccen}[1]{\ifnum#1=42 %
\hatcurISOrxxxxeccenA
\else
\ifnum#1=43 %
\hatcurISOrxxxxeccenB
\else
??????\fi
\fi
}
\newcommand{\hatcurISOrhoeccen}[1]{\ifnum#1=42 %
\hatcurISOrhoxxxxeccenA
\else
\ifnum#1=43 %
\hatcurISOrhoxxxxeccenB
\else
??????\fi
\fi
}
\newcommand{\hatcurISOrlongeccen}[1]{\ifnum#1=42 %
\hatcurISOrlongxxxxeccenA
\else
\ifnum#1=43 %
\hatcurISOrlongxxxxeccenB
\else
??????\fi
\fi
}
\newcommand{\hatcurISOrshorteccen}[1]{\ifnum#1=42 %
\hatcurISOrshortxxxxeccenA
\else
\ifnum#1=43 %
\hatcurISOrshortxxxxeccenB
\else
??????\fi
\fi
}
\newcommand{\hatcurISOsigmaeccen}[1]{\ifnum#1=42 %
\hatcurISOsigmaxxxxeccenA
\else
\ifnum#1=43 %
\hatcurISOsigmaxxxxeccenB
\else
??????\fi
\fi
}
\newcommand{\hatcurISOspececcen}[1]{\ifnum#1=42 %
\hatcurISOspecxxxxeccenA
\else
\ifnum#1=43 %
\hatcurISOspecxxxxeccenB
\else
??????\fi
\fi
}
\newcommand{\hatcurISOvieccen}[1]{\ifnum#1=42 %
\hatcurISOvixxxxeccenA
\else
\ifnum#1=43 %
\hatcurISOvixxxxeccenB
\else
??????\fi
\fi
}
\newcommand{\hatcurLBigeccen}[1]{\ifnum#1=42 %
\hatcurLBigxxxxeccenA
\else
\ifnum#1=43 %
\hatcurLBigxxxxeccenB
\else
??????\fi
\fi
}
\newcommand{\hatcurLBiieccen}[1]{\ifnum#1=42 %
\hatcurLBiixxxxeccenA
\else
\ifnum#1=43 %
\hatcurLBiixxxxeccenB
\else
??????\fi
\fi
}
\newcommand{\hatcurLBiIeccen}[1]{\ifnum#1=42 %
\hatcurLBiIxxxxeccenA
\else
\ifnum#1=43 %
\hatcurLBiIxxxxeccenB
\else
??????\fi
\fi
}
\newcommand{\hatcurLBiigeccen}[1]{\ifnum#1=42 %
\hatcurLBiigxxxxeccenA
\else
\ifnum#1=43 %
\hatcurLBiigxxxxeccenB
\else
??????\fi
\fi
}
\newcommand{\hatcurLBiiieccen}[1]{\ifnum#1=42 %
\hatcurLBiiixxxxeccenA
\else
\ifnum#1=43 %
\hatcurLBiiixxxxeccenB
\else
??????\fi
\fi
}
\newcommand{\hatcurLBiiIeccen}[1]{\ifnum#1=42 %
\hatcurLBiiIxxxxeccenA
\else
\ifnum#1=43 %
\hatcurLBiiIxxxxeccenB
\else
??????\fi
\fi
}
\newcommand{\hatcurLBiikepeccen}[1]{\ifnum#1=42 %
\hatcurLBiikepxxxxeccenA
\else
\ifnum#1=43 %
\hatcurLBiikepxxxxeccenB
\else
??????\fi
\fi
}
\newcommand{\hatcurLBiireccen}[1]{\ifnum#1=42 %
\hatcurLBiirxxxxeccenA
\else
\ifnum#1=43 %
\hatcurLBiirxxxxeccenB
\else
??????\fi
\fi
}
\newcommand{\hatcurLBiiReccen}[1]{\ifnum#1=42 %
\hatcurLBiiRxxxxeccenA
\else
\ifnum#1=43 %
\hatcurLBiiRxxxxeccenB
\else
??????\fi
\fi
}
\newcommand{\hatcurLBiizeccen}[1]{\ifnum#1=42 %
\hatcurLBiizxxxxeccenA
\else
\ifnum#1=43 %
\hatcurLBiizxxxxeccenB
\else
??????\fi
\fi
}
\newcommand{\hatcurLBikepeccen}[1]{\ifnum#1=42 %
\hatcurLBikepxxxxeccenA
\else
\ifnum#1=43 %
\hatcurLBikepxxxxeccenB
\else
??????\fi
\fi
}
\newcommand{\hatcurLBireccen}[1]{\ifnum#1=42 %
\hatcurLBirxxxxeccenA
\else
\ifnum#1=43 %
\hatcurLBirxxxxeccenB
\else
??????\fi
\fi
}
\newcommand{\hatcurLBiReccen}[1]{\ifnum#1=42 %
\hatcurLBiRxxxxeccenA
\else
\ifnum#1=43 %
\hatcurLBiRxxxxeccenB
\else
??????\fi
\fi
}
\newcommand{\hatcurLBizeccen}[1]{\ifnum#1=42 %
\hatcurLBizxxxxeccenA
\else
\ifnum#1=43 %
\hatcurLBizxxxxeccenB
\else
??????\fi
\fi
}
\newcommand{\hatcurLCbsqeccen}[1]{\ifnum#1=42 %
\hatcurLCbsqxxxxeccenA
\else
\ifnum#1=43 %
\hatcurLCbsqxxxxeccenB
\else
??????\fi
\fi
}
\newcommand{\hatcurLCdipeccen}[1]{\ifnum#1=42 %
\hatcurLCdipxxxxeccenA
\else
\ifnum#1=43 %
\hatcurLCdipxxxxeccenB
\else
??????\fi
\fi
}
\newcommand{\hatcurLCdureccen}[1]{\ifnum#1=42 %
\hatcurLCdurxxxxeccenA
\else
\ifnum#1=43 %
\hatcurLCdurxxxxeccenB
\else
??????\fi
\fi
}
\newcommand{\hatcurLCdurhreccen}[1]{\ifnum#1=42 %
\hatcurLCdurhrxxxxeccenA
\else
\ifnum#1=43 %
\hatcurLCdurhrxxxxeccenB
\else
??????\fi
\fi
}
\newcommand{\hatcurLCdurhrshorteccen}[1]{\ifnum#1=42 %
\hatcurLCdurhrshortxxxxeccenA
\else
\ifnum#1=43 %
\hatcurLCdurhrshortxxxxeccenB
\else
??????\fi
\fi
}
\newcommand{\hatcurLCdurshorteccen}[1]{\ifnum#1=42 %
\hatcurLCdurshortxxxxeccenA
\else
\ifnum#1=43 %
\hatcurLCdurshortxxxxeccenB
\else
??????\fi
\fi
}
\newcommand{\hatcurLChatnetmeccen}[1]{\ifnum#1=42 %
\hatcurLChatnetmxxxxeccenA
\else
??????\fi
}
\newcommand{\hatcurLChatnetmAeccen}[1]{\ifnum#1=43 %
\hatcurLChatnetmAxxxxeccenB
\else
??????\fi
}
\newcommand{\hatcurLChatnetmBeccen}[1]{\ifnum#1=43 %
\hatcurLChatnetmBxxxxeccenB
\else
??????\fi
}
\newcommand{\hatcurLCiblendeccen}[1]{\ifnum#1=42 %
\hatcurLCiblendxxxxeccenA
\else
??????\fi
}
\newcommand{\hatcurLCiblendAeccen}[1]{\ifnum#1=43 %
\hatcurLCiblendAxxxxeccenB
\else
??????\fi
}
\newcommand{\hatcurLCiblendBeccen}[1]{\ifnum#1=43 %
\hatcurLCiblendBxxxxeccenB
\else
??????\fi
}
\newcommand{\hatcurLCimpeccen}[1]{\ifnum#1=42 %
\hatcurLCimpxxxxeccenA
\else
\ifnum#1=43 %
\hatcurLCimpxxxxeccenB
\else
??????\fi
\fi
}
\newcommand{\hatcurLCingdureccen}[1]{\ifnum#1=42 %
\hatcurLCingdurxxxxeccenA
\else
\ifnum#1=43 %
\hatcurLCingdurxxxxeccenB
\else
??????\fi
\fi
}
\newcommand{\hatcurLCPeccen}[1]{\ifnum#1=42 %
\hatcurLCPxxxxeccenA
\else
\ifnum#1=43 %
\hatcurLCPxxxxeccenB
\else
??????\fi
\fi
}
\newcommand{\hatcurLCPprececcen}[1]{\ifnum#1=42 %
\hatcurLCPprecxxxxeccenA
\else
\ifnum#1=43 %
\hatcurLCPprecxxxxeccenB
\else
??????\fi
\fi
}
\newcommand{\hatcurLCPshorteccen}[1]{\ifnum#1=42 %
\hatcurLCPshortxxxxeccenA
\else
\ifnum#1=43 %
\hatcurLCPshortxxxxeccenB
\else
??????\fi
\fi
}
\newcommand{\hatcurLCqeccen}[1]{\ifnum#1=42 %
\hatcurLCqxxxxeccenA
\else
\ifnum#1=43 %
\hatcurLCqxxxxeccenB
\else
??????\fi
\fi
}
\newcommand{\hatcurLCqshorteccen}[1]{\ifnum#1=42 %
\hatcurLCqshortxxxxeccenA
\else
\ifnum#1=43 %
\hatcurLCqshortxxxxeccenB
\else
??????\fi
\fi
}
\newcommand{\hatcurLCrprstareccen}[1]{\ifnum#1=42 %
\hatcurLCrprstarxxxxeccenA
\else
\ifnum#1=43 %
\hatcurLCrprstarxxxxeccenB
\else
??????\fi
\fi
}
\newcommand{\hatcurLCTeccen}[1]{\ifnum#1=42 %
\hatcurLCTxxxxeccenA
\else
\ifnum#1=43 %
\hatcurLCTxxxxeccenB
\else
??????\fi
\fi
}
\newcommand{\hatcurLCTAeccen}[1]{\ifnum#1=42 %
\hatcurLCTAxxxxeccenA
\else
\ifnum#1=43 %
\hatcurLCTAxxxxeccenB
\else
??????\fi
\fi
}
\newcommand{\hatcurLCTBeccen}[1]{\ifnum#1=42 %
\hatcurLCTBxxxxeccenA
\else
\ifnum#1=43 %
\hatcurLCTBxxxxeccenB
\else
??????\fi
\fi
}
\newcommand{\hatcurLCzetaeccen}[1]{\ifnum#1=42 %
\hatcurLCzetaxxxxeccenA
\else
\ifnum#1=43 %
\hatcurLCzetaxxxxeccenB
\else
??????\fi
\fi
}
\newcommand{\hatcurPPaequiveccen}[1]{\ifnum#1=42 %
\hatcurPPaequivxxxxeccenA
\else
\ifnum#1=43 %
\hatcurPPaequivxxxxeccenB
\else
??????\fi
\fi
}
\newcommand{\hatcurPPareccen}[1]{\ifnum#1=42 %
\hatcurPParxxxxeccenA
\else
\ifnum#1=43 %
\hatcurPParxxxxeccenB
\else
??????\fi
\fi
}
\newcommand{\hatcurPPareleccen}[1]{\ifnum#1=42 %
\hatcurPParelxxxxeccenA
\else
\ifnum#1=43 %
\hatcurPParelxxxxeccenB
\else
??????\fi
\fi
}
\newcommand{\hatcurPPfluxapeccen}[1]{\ifnum#1=42 %
\hatcurPPfluxapxxxxeccenA
\else
\ifnum#1=43 %
\hatcurPPfluxapxxxxeccenB
\else
??????\fi
\fi
}
\newcommand{\hatcurPPfluxapdimeccen}[1]{\ifnum#1=42 %
\hatcurPPfluxapdimxxxxeccenA
\else
\ifnum#1=43 %
\hatcurPPfluxapdimxxxxeccenB
\else
??????\fi
\fi
}
\newcommand{\hatcurPPfluxavgeccen}[1]{\ifnum#1=42 %
\hatcurPPfluxavgxxxxeccenA
\else
\ifnum#1=43 %
\hatcurPPfluxavgxxxxeccenB
\else
??????\fi
\fi
}
\newcommand{\hatcurPPfluxavgdimeccen}[1]{\ifnum#1=42 %
\hatcurPPfluxavgdimxxxxeccenA
\else
\ifnum#1=43 %
\hatcurPPfluxavgdimxxxxeccenB
\else
??????\fi
\fi
}
\newcommand{\hatcurPPfluxperieccen}[1]{\ifnum#1=42 %
\hatcurPPfluxperixxxxeccenA
\else
\ifnum#1=43 %
\hatcurPPfluxperixxxxeccenB
\else
??????\fi
\fi
}
\newcommand{\hatcurPPfluxperidimeccen}[1]{\ifnum#1=42 %
\hatcurPPfluxperidimxxxxeccenA
\else
\ifnum#1=43 %
\hatcurPPfluxperidimxxxxeccenB
\else
??????\fi
\fi
}
\newcommand{\hatcurPPgeccen}[1]{\ifnum#1=42 %
\hatcurPPgxxxxeccenA
\else
\ifnum#1=43 %
\hatcurPPgxxxxeccenB
\else
??????\fi
\fi
}
\newcommand{\hatcurPPieccen}[1]{\ifnum#1=42 %
\hatcurPPixxxxeccenA
\else
\ifnum#1=43 %
\hatcurPPixxxxeccenB
\else
??????\fi
\fi
}
\newcommand{\hatcurPPloggeccen}[1]{\ifnum#1=42 %
\hatcurPPloggxxxxeccenA
\else
\ifnum#1=43 %
\hatcurPPloggxxxxeccenB
\else
??????\fi
\fi
}
\newcommand{\hatcurPPmeccen}[1]{\ifnum#1=42 %
\hatcurPPmxxxxeccenA
\else
\ifnum#1=43 %
\hatcurPPmxxxxeccenB
\else
??????\fi
\fi
}
\newcommand{\hatcurPPmeeccen}[1]{\ifnum#1=42 %
\hatcurPPmexxxxeccenA
\else
\ifnum#1=43 %
\hatcurPPmexxxxeccenB
\else
??????\fi
\fi
}
\newcommand{\hatcurPPmelongeccen}[1]{\ifnum#1=42 %
\hatcurPPmelongxxxxeccenA
\else
\ifnum#1=43 %
\hatcurPPmelongxxxxeccenB
\else
??????\fi
\fi
}
\newcommand{\hatcurPPmeshorteccen}[1]{\ifnum#1=42 %
\hatcurPPmeshortxxxxeccenA
\else
\ifnum#1=43 %
\hatcurPPmeshortxxxxeccenB
\else
??????\fi
\fi
}
\newcommand{\hatcurPPmlongeccen}[1]{\ifnum#1=42 %
\hatcurPPmlongxxxxeccenA
\else
\ifnum#1=43 %
\hatcurPPmlongxxxxeccenB
\else
??????\fi
\fi
}
\newcommand{\hatcurPPmrcorreccen}[1]{\ifnum#1=42 %
\hatcurPPmrcorrxxxxeccenA
\else
\ifnum#1=43 %
\hatcurPPmrcorrxxxxeccenB
\else
??????\fi
\fi
}
\newcommand{\hatcurPPmshorteccen}[1]{\ifnum#1=42 %
\hatcurPPmshortxxxxeccenA
\else
\ifnum#1=43 %
\hatcurPPmshortxxxxeccenB
\else
??????\fi
\fi
}
\newcommand{\hatcurPPperieccen}[1]{\ifnum#1=42 %
\hatcurPPperixxxxeccenA
\else
\ifnum#1=43 %
\hatcurPPperixxxxeccenB
\else
??????\fi
\fi
}
\newcommand{\hatcurPPphiconjeccen}[1]{\ifnum#1=42 %
\hatcurPPphiconjxxxxeccenA
\else
\ifnum#1=43 %
\hatcurPPphiconjxxxxeccenB
\else
??????\fi
\fi
}
\newcommand{\hatcurPPreccen}[1]{\ifnum#1=42 %
\hatcurPPrxxxxeccenA
\else
\ifnum#1=43 %
\hatcurPPrxxxxeccenB
\else
??????\fi
\fi
}
\newcommand{\hatcurPPreeccen}[1]{\ifnum#1=42 %
\hatcurPPrexxxxeccenA
\else
\ifnum#1=43 %
\hatcurPPrexxxxeccenB
\else
??????\fi
\fi
}
\newcommand{\hatcurPPrelongeccen}[1]{\ifnum#1=42 %
\hatcurPPrelongxxxxeccenA
\else
\ifnum#1=43 %
\hatcurPPrelongxxxxeccenB
\else
??????\fi
\fi
}
\newcommand{\hatcurPPreshorteccen}[1]{\ifnum#1=42 %
\hatcurPPreshortxxxxeccenA
\else
\ifnum#1=43 %
\hatcurPPreshortxxxxeccenB
\else
??????\fi
\fi
}
\newcommand{\hatcurPPrhoeccen}[1]{\ifnum#1=42 %
\hatcurPPrhoxxxxeccenA
\else
\ifnum#1=43 %
\hatcurPPrhoxxxxeccenB
\else
??????\fi
\fi
}
\newcommand{\hatcurPPrlongeccen}[1]{\ifnum#1=42 %
\hatcurPPrlongxxxxeccenA
\else
\ifnum#1=43 %
\hatcurPPrlongxxxxeccenB
\else
??????\fi
\fi
}
\newcommand{\hatcurPPrshorteccen}[1]{\ifnum#1=42 %
\hatcurPPrshortxxxxeccenA
\else
\ifnum#1=43 %
\hatcurPPrshortxxxxeccenB
\else
??????\fi
\fi
}
\newcommand{\hatcurPPtcirceccen}[1]{\ifnum#1=42 %
\hatcurPPtcircxxxxeccenA
\else
\ifnum#1=43 %
\hatcurPPtcircxxxxeccenB
\else
??????\fi
\fi
}
\newcommand{\hatcurPPteffeccen}[1]{\ifnum#1=42 %
\hatcurPPteffxxxxeccenA
\else
\ifnum#1=43 %
\hatcurPPteffxxxxeccenB
\else
??????\fi
\fi
}
\newcommand{\hatcurPPthetaeccen}[1]{\ifnum#1=42 %
\hatcurPPthetaxxxxeccenA
\else
\ifnum#1=43 %
\hatcurPPthetaxxxxeccenB
\else
??????\fi
\fi
}
\newcommand{\hatcurPPtinfalleccen}[1]{\ifnum#1=42 %
\hatcurPPtinfallxxxxeccenA
\else
\ifnum#1=43 %
\hatcurPPtinfallxxxxeccenB
\else
??????\fi
\fi
}
\newcommand{\hatcurRVecceneccen}[1]{\ifnum#1=42 %
\hatcurRVeccenxxxxeccenA
\else
\ifnum#1=43 %
\hatcurRVeccenxxxxeccenB
\else
??????\fi
\fi
}
\newcommand{\hatcurRVfitrmseccen}[1]{\ifnum#1=43 %
\hatcurRVfitrmsxxxxeccenB
\else
??????\fi
}
\newcommand{\hatcurRVfitrmsAeccen}[1]{\ifnum#1=42 %
\hatcurRVfitrmsAxxxxeccenA
\else
??????\fi
}
\newcommand{\hatcurRVfitrmsBeccen}[1]{\ifnum#1=42 %
\hatcurRVfitrmsBxxxxeccenA
\else
??????\fi
}
\newcommand{\hatcurRVgammaeccen}[1]{\ifnum#1=43 %
\hatcurRVgammaxxxxeccenB
\else
??????\fi
}
\newcommand{\hatcurRVgammaAeccen}[1]{\ifnum#1=42 %
\hatcurRVgammaAxxxxeccenA
\else
??????\fi
}
\newcommand{\hatcurRVgammaBeccen}[1]{\ifnum#1=42 %
\hatcurRVgammaBxxxxeccenA
\else
??????\fi
}
\newcommand{\hatcurRVheccen}[1]{\ifnum#1=42 %
\hatcurRVhxxxxeccenA
\else
\ifnum#1=43 %
\hatcurRVhxxxxeccenB
\else
??????\fi
\fi
}
\newcommand{\hatcurRVjittereccen}[1]{\ifnum#1=43 %
\hatcurRVjitterxxxxeccenB
\else
??????\fi
}
\newcommand{\hatcurRVjitterAeccen}[1]{\ifnum#1=42 %
\hatcurRVjitterAxxxxeccenA
\else
??????\fi
}
\newcommand{\hatcurRVjitterBeccen}[1]{\ifnum#1=42 %
\hatcurRVjitterBxxxxeccenA
\else
??????\fi
}
\newcommand{\hatcurRVkeccen}[1]{\ifnum#1=42 %
\hatcurRVkxxxxeccenA
\else
\ifnum#1=43 %
\hatcurRVkxxxxeccenB
\else
??????\fi
\fi
}
\newcommand{\hatcurRVKeccen}[1]{\ifnum#1=42 %
\hatcurRVKxxxxeccenA
\else
\ifnum#1=43 %
\hatcurRVKxxxxeccenB
\else
??????\fi
\fi
}
\newcommand{\hatcurRVomegaeccen}[1]{\ifnum#1=42 %
\hatcurRVomegaxxxxeccenA
\else
\ifnum#1=43 %
\hatcurRVomegaxxxxeccenB
\else
??????\fi
\fi
}
\newcommand{\hatcurRVrheccen}[1]{\ifnum#1=42 %
\hatcurRVrhxxxxeccenA
\else
\ifnum#1=43 %
\hatcurRVrhxxxxeccenB
\else
??????\fi
\fi
}
\newcommand{\hatcurRVrkeccen}[1]{\ifnum#1=42 %
\hatcurRVrkxxxxeccenA
\else
\ifnum#1=43 %
\hatcurRVrkxxxxeccenB
\else
??????\fi
\fi
}
\newcommand{\hatcurRVtroneeccen}[1]{\ifnum#1=42 %
\hatcurRVtronexxxxeccenA
\else
\ifnum#1=43 %
\hatcurRVtronexxxxeccenB
\else
??????\fi
\fi
}
\newcommand{\hatcurRVtrtwoeccen}[1]{\ifnum#1=42 %
\hatcurRVtrtwoxxxxeccenA
\else
\ifnum#1=43 %
\hatcurRVtrtwoxxxxeccenB
\else
??????\fi
\fi
}
\newcommand{\hatcurSMEiiloggeccen}[1]{\ifnum#1=42 %
\hatcurSMEiiloggxxxxeccenA
\else
\ifnum#1=43 %
\hatcurSMEiiloggxxxxeccenB
\else
??????\fi
\fi
}
\newcommand{\hatcurSMEiiteffeccen}[1]{\ifnum#1=42 %
\hatcurSMEiiteffxxxxeccenA
\else
\ifnum#1=43 %
\hatcurSMEiiteffxxxxeccenB
\else
??????\fi
\fi
}
\newcommand{\hatcurSMEiivmaceccen}[1]{\ifnum#1=42 %
\hatcurSMEiivmacxxxxeccenA
\else
\ifnum#1=43 %
\hatcurSMEiivmacxxxxeccenB
\else
??????\fi
\fi
}
\newcommand{\hatcurSMEiivmiceccen}[1]{\ifnum#1=42 %
\hatcurSMEiivmicxxxxeccenA
\else
\ifnum#1=43 %
\hatcurSMEiivmicxxxxeccenB
\else
??????\fi
\fi
}
\newcommand{\hatcurSMEiivsineccen}[1]{\ifnum#1=42 %
\hatcurSMEiivsinxxxxeccenA
\else
\ifnum#1=43 %
\hatcurSMEiivsinxxxxeccenB
\else
??????\fi
\fi
}
\newcommand{\hatcurSMEiizfeheccen}[1]{\ifnum#1=42 %
\hatcurSMEiizfehxxxxeccenA
\else
\ifnum#1=43 %
\hatcurSMEiizfehxxxxeccenB
\else
??????\fi
\fi
}
\newcommand{\hatcurSMEiizfehshorteccen}[1]{\ifnum#1=42 %
\hatcurSMEiizfehshortxxxxeccenA
\else
\ifnum#1=43 %
\hatcurSMEiizfehshortxxxxeccenB
\else
??????\fi
\fi
}
\newcommand{\hatcurSMEiloggeccen}[1]{\ifnum#1=42 %
\hatcurSMEiloggxxxxeccenA
\else
\ifnum#1=43 %
\hatcurSMEiloggxxxxeccenB
\else
??????\fi
\fi
}
\newcommand{\hatcurSMEiteffeccen}[1]{\ifnum#1=42 %
\hatcurSMEiteffxxxxeccenA
\else
\ifnum#1=43 %
\hatcurSMEiteffxxxxeccenB
\else
??????\fi
\fi
}
\newcommand{\hatcurSMEivmaceccen}[1]{\ifnum#1=42 %
\hatcurSMEivmacxxxxeccenA
\else
\ifnum#1=43 %
\hatcurSMEivmacxxxxeccenB
\else
??????\fi
\fi
}
\newcommand{\hatcurSMEivmiceccen}[1]{\ifnum#1=42 %
\hatcurSMEivmicxxxxeccenA
\else
\ifnum#1=43 %
\hatcurSMEivmicxxxxeccenB
\else
??????\fi
\fi
}
\newcommand{\hatcurSMEivsineccen}[1]{\ifnum#1=42 %
\hatcurSMEivsinxxxxeccenA
\else
\ifnum#1=43 %
\hatcurSMEivsinxxxxeccenB
\else
??????\fi
\fi
}
\newcommand{\hatcurSMEizfeheccen}[1]{\ifnum#1=42 %
\hatcurSMEizfehxxxxeccenA
\else
\ifnum#1=43 %
\hatcurSMEizfehxxxxeccenB
\else
??????\fi
\fi
}
\newcommand{\hatcurSMEizfehshorteccen}[1]{\ifnum#1=42 %
\hatcurSMEizfehshortxxxxeccenA
\else
\ifnum#1=43 %
\hatcurSMEizfehshortxxxxeccenB
\else
??????\fi
\fi
}
\newcommand{\hatcurTRESgammaeccen}[1]{\ifnum#1=42 %
\hatcurTRESgammaxxxxeccenA
\else
\ifnum#1=43 %
\hatcurTRESgammaxxxxeccenB
\else
??????\fi
\fi
}
\newcommand{\hatcurTRESloggeccen}[1]{\ifnum#1=42 %
\hatcurTRESloggxxxxeccenA
\else
\ifnum#1=43 %
\hatcurTRESloggxxxxeccenB
\else
??????\fi
\fi
}
\newcommand{\hatcurTRESnumspececcen}[1]{\ifnum#1=42 %
\hatcurTRESnumspecxxxxeccenA
\else
\ifnum#1=43 %
\hatcurTRESnumspecxxxxeccenB
\else
??????\fi
\fi
}
\newcommand{\hatcurTRESrvrmseccen}[1]{\ifnum#1=42 %
\hatcurTRESrvrmsxxxxeccenA
\else
\ifnum#1=43 %
\hatcurTRESrvrmsxxxxeccenB
\else
??????\fi
\fi
}
\newcommand{\hatcurTRESspaneccen}[1]{\ifnum#1=42 %
\hatcurTRESspanxxxxeccenA
\else
\ifnum#1=43 %
\hatcurTRESspanxxxxeccenB
\else
??????\fi
\fi
}
\newcommand{\hatcurTRESteffeccen}[1]{\ifnum#1=42 %
\hatcurTRESteffxxxxeccenA
\else
\ifnum#1=43 %
\hatcurTRESteffxxxxeccenB
\else
??????\fi
\fi
}
\newcommand{\hatcurTRESvsinieccen}[1]{\ifnum#1=42 %
\hatcurTRESvsinixxxxeccenA
\else
\ifnum#1=43 %
\hatcurTRESvsinixxxxeccenB
\else
??????\fi
\fi
}
\newcommand{\hatcurTRESzfeheccen}[1]{\ifnum#1=42 %
\hatcurTRESzfehxxxxeccenA
\else
\ifnum#1=43 %
\hatcurTRESzfehxxxxeccenB
\else
??????\fi
\fi
}
\newcommand{\hatcurXAveccen}[1]{\ifnum#1=42 %
\hatcurXAvxxxxeccenA
\else
\ifnum#1=43 %
\hatcurXAvxxxxeccenB
\else
??????\fi
\fi
}
\newcommand{\hatcurXdisteccen}[1]{\ifnum#1=42 %
\hatcurXdistxxxxeccenA
\else
\ifnum#1=43 %
\hatcurXdistxxxxeccenB
\else
??????\fi
\fi
}
\newcommand{\hatcurXdistredeccen}[1]{\ifnum#1=42 %
\hatcurXdistredxxxxeccenA
\else
\ifnum#1=43 %
\hatcurXdistredxxxxeccenB
\else
??????\fi
\fi
}
\newcommand{\hatcurXEBVeccen}[1]{\ifnum#1=42 %
\hatcurXEBVxxxxeccenA
\else
\ifnum#1=43 %
\hatcurXEBVxxxxeccenB
\else
??????\fi
\fi
}
\newcommand{\hatcurXjhisoredeccen}[1]{\ifnum#1=42 %
\hatcurXjhisoredxxxxeccenA
\else
\ifnum#1=43 %
\hatcurXjhisoredxxxxeccenB
\else
??????\fi
\fi
}
\newcommand{\hatcurXjkisoredeccen}[1]{\ifnum#1=42 %
\hatcurXjkisoredxxxxeccenA
\else
\ifnum#1=43 %
\hatcurXjkisoredxxxxeccenB
\else
??????\fi
\fi
}
\newcommand{\hatcurXmhisoredeccen}[1]{\ifnum#1=42 %
\hatcurXmhisoredxxxxeccenA
\else
\ifnum#1=43 %
\hatcurXmhisoredxxxxeccenB
\else
??????\fi
\fi
}
\newcommand{\hatcurXmiisoredeccen}[1]{\ifnum#1=42 %
\hatcurXmiisoredxxxxeccenA
\else
\ifnum#1=43 %
\hatcurXmiisoredxxxxeccenB
\else
??????\fi
\fi
}
\newcommand{\hatcurXmjisoredeccen}[1]{\ifnum#1=42 %
\hatcurXmjisoredxxxxeccenA
\else
\ifnum#1=43 %
\hatcurXmjisoredxxxxeccenB
\else
??????\fi
\fi
}
\newcommand{\hatcurXmkisoredeccen}[1]{\ifnum#1=42 %
\hatcurXmkisoredxxxxeccenA
\else
\ifnum#1=43 %
\hatcurXmkisoredxxxxeccenB
\else
??????\fi
\fi
}
\newcommand{\hatcurXmvisoredeccen}[1]{\ifnum#1=42 %
\hatcurXmvisoredxxxxeccenA
\else
\ifnum#1=43 %
\hatcurXmvisoredxxxxeccenB
\else
??????\fi
\fi
}
\newcommand{\hatcurXsecdureccen}[1]{\ifnum#1=42 %
\hatcurXsecdurxxxxeccenA
\else
\ifnum#1=43 %
\hatcurXsecdurxxxxeccenB
\else
??????\fi
\fi
}
\newcommand{\hatcurXsecingdureccen}[1]{\ifnum#1=42 %
\hatcurXsecingdurxxxxeccenA
\else
\ifnum#1=43 %
\hatcurXsecingdurxxxxeccenB
\else
??????\fi
\fi
}
\newcommand{\hatcurXsecondaryeccen}[1]{\ifnum#1=42 %
\hatcurXsecondaryxxxxeccenA
\else
\ifnum#1=43 %
\hatcurXsecondaryxxxxeccenB
\else
??????\fi
\fi
}
\newcommand{\hatcurXsecphaseeccen}[1]{\ifnum#1=42 %
\hatcurXsecphasexxxxeccenA
\else
\ifnum#1=43 %
\hatcurXsecphasexxxxeccenB
\else
??????\fi
\fi
}
\newcommand{\hatcurXviisoredeccen}[1]{\ifnum#1=42 %
\hatcurXviisoredxxxxeccenA
\else
\ifnum#1=43 %
\hatcurXviisoredxxxxeccenB
\else
??????\fi
\fi
}
\newcommand{\hatcurXvkisoredeccen}[1]{\ifnum#1=42 %
\hatcurXvkisoredxxxxeccenA
\else
\ifnum#1=43 %
\hatcurXvkisoredxxxxeccenB
\else
??????\fi
\fi
}

\newcommand{\hatcurxxxxA}{HAT-P-42}
\newcommand{\hatcurbxxxxA}{HAT-P-42b}
\newcommand{\hatcurcxxxxA}{HAT-P-42c}
\newcommand{\hatcurplanetnumxxxxA}{42}
\newcommand{\hatcurRVgammaabsxxxxA}{\hatcurDSgamma{\hatcurplanetnumxxxxA}}                           
\newcommand{\hatcurRVgammarelxxxxA}{\hatcurRVgamma{\hatcurplanetnumxxxxA}}      
\newcommand{\hatcurCCtassvixxxxA}{\ensuremath{NULL}}   
\newcommand{\hatcurSMEversionxxxxA}{ii}      
\newcommand{\hatcurisoshortxxxxA}{YY}
\newcommand{\hatcurisofullxxxxA}{Yonsei-Yale (YY)}
\newcommand{\hatcurisocitexxxxA}{yi:2001}
\newcommand{\hatcurlumindxxxxA}{\arstar}
\newcommand{\hatcurjhkfilsetxxxxA}{ESO}
\newcommand{\hatcurSMEteffxxxxA}{\ifthenelse{\equal{\hatcurSMEversionxxxxA}{i}}{\hatcurSMEiteff{\hatcurplanetnumxxxxA}}{\hatcurSMEiiteff{\hatcurplanetnumxxxxA}}}
\newcommand{\hatcurSMEzfehxxxxA}{\ifthenelse{\equal{\hatcurSMEversionxxxxA}{i}}{\hatcurSMEizfeh{\hatcurplanetnumxxxxA}}{\hatcurSMEiizfeh{\hatcurplanetnumxxxxA}}}
\newcommand{\hatcurSMEzfehshortxxxxA}{\ifthenelse{\equal{\hatcurSMEversionxxxxA}{i}}{\hatcurSMEizfehshort{\hatcurplanetnumxxxxA}}{\hatcurSMEiizfehshort{\hatcurplanetnumxxxxA}}}
\newcommand{\hatcurSMEloggxxxxA}{\ifthenelse{\equal{\hatcurSMEversionxxxxA}{i}}{\hatcurSMEilogg{\hatcurplanetnumxxxxA}}{\hatcurSMEiilogg{\hatcurplanetnumxxxxA}}}
\newcommand{\hatcurSMEvsinxxxxA}{\ifthenelse{\equal{\hatcurSMEversionxxxxA}{i}}{\hatcurSMEivsin{\hatcurplanetnumxxxxA}}{\hatcurSMEiivsin{\hatcurplanetnumxxxxA}}}
\newcommand{\hatcurSMEvmacxxxxA}{\ifthenelse{\equal{\hatcurSMEversionxxxxA}{i}}{\hatcurSMEivmac{\hatcurplanetnumxxxxA}}{\hatcurSMEiivmac{\hatcurplanetnumxxxxA}}}
\newcommand{\hatcurSMEvmicxxxxA}{\ifthenelse{\equal{\hatcurSMEversionxxxxA}{i}}{\hatcurSMEivmic{\hatcurplanetnumxxxxA}}{\hatcurSMEiivmic{\hatcurplanetnumxxxxA}}}

\newcommand{\hatcurxxxxB}{HAT-P-43}
\newcommand{\hatcurbxxxxB}{HAT-P-43b}
\newcommand{\hatcurcxxxxB}{HAT-P-43c}

\newcommand{\hatcurplanetnumxxxxB}{43}

\newcommand{\hatcurRVgammaabsxxxxB}{\hatcurDSgamma{\hatcurplanetnumxxxxB}}                           % Absolute Gamma velocity

\newcommand{\hatcurRVgammarelxxxxB}{\hatcurRVgamma{\hatcurplanetnumxxxxB}}                           % Relative Gamma velocity. Typically that of the Keck RVs.

\newcommand{\hatcurCCtassvixxxxB}{\ensuremath{NULL}}                  % TASS V-I

\newcommand{\hatcurSMEversionxxxxB}{ii}                                       % Final SME version:i or ii?

\newcommand{\hatcurisoshortxxxxB}{YY}
\newcommand{\hatcurisofullxxxxB}{Yonsei-Yale (YY)}
\newcommand{\hatcurisocitexxxxB}{yi:2001}
\newcommand{\hatcurlumindxxxxB}{\arstar}
\newcommand{\hatcurjhkfilsetxxxxB}{ESO}
\newcommand{\hatcurSMEteffxxxxB}{\ifthenelse{\equal{\hatcurSMEversionxxxxB}{i}}{\hatcurSMEiteff{\hatcurplanetnumxxxxB}}{\hatcurSMEiiteff{\hatcurplanetnumxxxxB}}}
\newcommand{\hatcurSMEzfehxxxxB}{\ifthenelse{\equal{\hatcurSMEversionxxxxB}{i}}{\hatcurSMEizfeh{\hatcurplanetnumxxxxB}}{\hatcurSMEiizfeh{\hatcurplanetnumxxxxB}}}
\newcommand{\hatcurSMEzfehshortxxxxB}{\ifthenelse{\equal{\hatcurSMEversionxxxxB}{i}}{\hatcurSMEizfehshort{\hatcurplanetnumxxxxB}}{\hatcurSMEiizfehshort{\hatcurplanetnumxxxxB}}}
\newcommand{\hatcurSMEloggxxxxB}{\ifthenelse{\equal{\hatcurSMEversionxxxxB}{i}}{\hatcurSMEilogg{\hatcurplanetnumxxxxB}}{\hatcurSMEiilogg{\hatcurplanetnumxxxxB}}}
\newcommand{\hatcurSMEvsinxxxxB}{\ifthenelse{\equal{\hatcurSMEversionxxxxB}{i}}{\hatcurSMEivsin{\hatcurplanetnumxxxxB}}{\hatcurSMEiivsin{\hatcurplanetnumxxxxB}}}
\newcommand{\hatcurSMEvmacxxxxB}{\ifthenelse{\equal{\hatcurSMEversionxxxxB}{i}}{\hatcurSMEivmac{\hatcurplanetnumxxxxB}}{\hatcurSMEiivmac{\hatcurplanetnumxxxxB}}}
\newcommand{\hatcurSMEvmicxxxxB}{\ifthenelse{\equal{\hatcurSMEversionxxxxB}{i}}{\hatcurSMEivmic{\hatcurplanetnumxxxxB}}{\hatcurSMEiivmic{\hatcurplanetnumxxxxB}}}
\newcommand{\hatcur}[1]{\ifnum#1=42 %
\hatcurxxxxA
\else
\ifnum#1=43 %
\hatcurxxxxB
\else
??????\fi
\fi
}
\newcommand{\hatcurb}[1]{\ifnum#1=42 %
\hatcurbxxxxA
\else
\ifnum#1=43 %
\hatcurbxxxxB
\else
??????\fi
\fi
}
\newcommand{\hatcurc}[1]{\ifnum#1=42 %
\hatcurcxxxxA
\else
\ifnum#1=43 %
\hatcurcxxxxB
\else
??????\fi
\fi
}
\newcommand{\hatcurCCtassvi}[1]{\ifnum#1=42 %
\hatcurCCtassvixxxxA
\else
\ifnum#1=43 %
\hatcurCCtassvixxxxB
\else
??????\fi
\fi
}
\newcommand{\hatcurisocite}[1]{\ifnum#1=42 %
\hatcurisocitexxxxA
\else
\ifnum#1=43 %
\hatcurisocitexxxxB
\else
??????\fi
\fi
}
\newcommand{\hatcurisofull}[1]{\ifnum#1=42 %
\hatcurisofullxxxxA
\else
\ifnum#1=43 %
\hatcurisofullxxxxB
\else
??????\fi
\fi
}
\newcommand{\hatcurisoshort}[1]{\ifnum#1=42 %
\hatcurisoshortxxxxA
\else
\ifnum#1=43 %
\hatcurisoshortxxxxB
\else
??????\fi
\fi
}
\newcommand{\hatcurjhkfilset}[1]{\ifnum#1=42 %
\hatcurjhkfilsetxxxxA
\else
\ifnum#1=43 %
\hatcurjhkfilsetxxxxB
\else
??????\fi
\fi
}
\newcommand{\hatcurlumind}[1]{\ifnum#1=42 %
\hatcurlumindxxxxA
\else
\ifnum#1=43 %
\hatcurlumindxxxxB
\else
??????\fi
\fi
}
\newcommand{\hatcurplanetnum}[1]{\ifnum#1=42 %
\hatcurplanetnumxxxxA
\else
\ifnum#1=43 %
\hatcurplanetnumxxxxB
\else
??????\fi
\fi
}
\newcommand{\hatcurRVgammaabs}[1]{\ifnum#1=42 %
\hatcurRVgammaabsxxxxA
\else
\ifnum#1=43 %
\hatcurRVgammaabsxxxxB
\else
??????\fi
\fi
}
\newcommand{\hatcurRVgammarel}[1]{\ifnum#1=42 %
\hatcurRVgammarelxxxxA
\else
\ifnum#1=43 %
\hatcurRVgammarelxxxxB
\else
??????\fi
\fi
}
\newcommand{\hatcurSMElogg}[1]{\ifnum#1=42 %
\hatcurSMEloggxxxxA
\else
\ifnum#1=43 %
\hatcurSMEloggxxxxB
\else
??????\fi
\fi
}
\newcommand{\hatcurSMEteff}[1]{\ifnum#1=42 %
\hatcurSMEteffxxxxA
\else
\ifnum#1=43 %
\hatcurSMEteffxxxxB
\else
??????\fi
\fi
}
\newcommand{\hatcurSMEversion}[1]{\ifnum#1=42 %
\hatcurSMEversionxxxxA
\else
\ifnum#1=43 %
\hatcurSMEversionxxxxB
\else
??????\fi
\fi
}
\newcommand{\hatcurSMEvmac}[1]{\ifnum#1=42 %
\hatcurSMEvmacxxxxA
\else
\ifnum#1=43 %
\hatcurSMEvmacxxxxB
\else
??????\fi
\fi
}
\newcommand{\hatcurSMEvmic}[1]{\ifnum#1=42 %
\hatcurSMEvmicxxxxA
\else
\ifnum#1=43 %
\hatcurSMEvmicxxxxB
\else
??????\fi
\fi
}
\newcommand{\hatcurSMEvsin}[1]{\ifnum#1=42 %
\hatcurSMEvsinxxxxA
\else
\ifnum#1=43 %
\hatcurSMEvsinxxxxB
\else
??????\fi
\fi
}
\newcommand{\hatcurSMEzfeh}[1]{\ifnum#1=42 %
\hatcurSMEzfehxxxxA
\else
\ifnum#1=43 %
\hatcurSMEzfehxxxxB
\else
??????\fi
\fi
}
\newcommand{\hatcurSMEzfehshort}[1]{\ifnum#1=42 %
\hatcurSMEzfehshortxxxxA
\else
\ifnum#1=43 %
\hatcurSMEzfehshortxxxxB
\else
??????\fi
\fi
}

\begin{document}
   \titlerunning{HAT-P-42b and HAT-P-43b}
   \authorrunning{Boisse et al.}

   \title{HAT-P-42b and HAT-P-43b}
   \subtitle{Two Inflated Transiting Hot Jupiters from the HATNet Survey}
   \author{
	I.~Boisse\inst{1}, 
	J.~D.~Hartman\inst{2}, 
	G.~\'{A}.~Bakos\inst{2,12,13},
	K.~Penev\inst{2},
	Z.~Csubry\inst{2},
	B.~B\'eky\inst{3},
	D.~W.~Latham\inst{3},
	A.~Bieryla\inst{3}, 
	G.~Torres\inst{3},
	G.~Kov\'acs\inst{4,11},
	L.~A.~Buchhave\inst{5}, 
	T.~Hansen\inst{5}, 
	M.~Everett\inst{3}, 
	G.~A.~Esquerdo\inst{3}, 
	T.~Szklen\'ar\inst{3}, 
	E.~Falco\inst{3},
	A.~Shporer\inst{6,7,8},
	B.~J.~Fulton\inst{9,7}, 
	R.~W.~Noyes\inst{3},
	R.~P.~Stefanik\inst{3},
	J.~L\'az\'ar\inst{10},
	I.~Papp\inst{10} and
	P.~S\'ari\inst{10}
\fnmsep\thanks{
The photometric/spectroscopic data presented in this paper are based in part on observations carried out by the Hungarian-made
Automated Telescope Network, using telescopes operated at the Fred
Lawrence Whipple Observatory (FLWO) of the Smithsonian Astrophysical
Observatory (SAO), and at the Submillimeter Array (SMA) of SAO, by the Tillinghast Reflector 1.5\,m
telescope and the 1.2\,m telescope, both operated by SAO at FLWO, by the SOPHIE
spectrograph mounted on the 1.93\,m telescope at Observatoire de Haute
Provence, France (runs DDT-Dec2011), by the Nordic Optical Telescope, operated on the island of La
Palma jointly by Denmark, Finland, Iceland, Norway, and Sweden, in the
Spanish Observatorio del Roque de los Muchachos of the Instituto de
Astrofisica de Canarias, and by the 
facilities of the Las Cumbres Observatory Global Telescope.
}}
\institute{
	Centro de Astrof\'isica, Universidade do Porto, 
	Rua das Estrelas, 4150-762 
	Porto, Portugal\\
	\email{Isabelle.Boisse@astro.up.pt}
\and
	Princeton University, Department of Astrophysical Sciences, 
	Princeton, NJ, USA
	%\email{gbakos@astro.princeton.edu}
\and
	Harvard-Smithsonian Center for Astrophysics, Cambridge, MA, USA
\and
	Konkoly Observatory, Budapest, Hungary
\and
	Niels Bohr Institute, University of Copenhagen, DK-2100, Copenhagen, Denmark and Centre for Star and Planet Formation, Natural History Museum of Denmark, University of Copenhagen, DK-1350 Copenhagen, Denmark
\and
	Division of Geological and Planetary Sciences, California Institute of
Technology, Pasadena, CA 91125, USA
\and
	LCOGT, 6740 Cortona Drive, Santa Barbara, CA, USA
\and	
	Department of Physics, Broida Hall, University of California, Santa
Barbara, CA 93106, USA
\and
	Institute for Astronomy, University of Hawaii, Honolulu, HI, USA
\and
	Hungarian Astronomical Association, Budapest, Hungary
\and
         Department of Physics and Astrophysics, University of North Dakota, Grand Forks, ND, USA
\and
	Alfred P.~Sloan Fellow
\and
	Packard Fellow
}
\date{Received ***; accepted ***}

% \abstract{}{}{}{}{} 
% 5 {} token are mandatory
 
\abstract
  {}
  {
	We announce the discovery of two new transiting planets, and provide
	their accurate initial characterization.
  }
   {First identified from the HATNet wide-field photometric survey, these
	candidate transiting planets were then followed-up with a variety
	of photometric observations.  Determining the planetary nature of
	the objects and characterizing the parameters of the systems were
	mainly done with the SOPHIE spectrograph at the 1.93\,m telescope at OHP 
	and the TRES spectrograph at the 1.5\,m telescope at FLWO.}
%% results heading (mandatory)
   {\hatcur{42}b and \hatcur{43}b are typical hot Jupiters on circular orbits around early-G/late-F main sequence host stars, with
   periods of $\hatcurLCP{42}$ and $\hatcurLCP{43}$ days, masses of
   $\hatcurPPmlong{42}$ and $\hatcurPPmlong{43}$ $\mjup$, and radii of
   $\hatcurPPrlong{42}$ and $\hatcurPPrlong{43}$ $\rjup$, respectively.  
   These discoveries increase the sample of planets with measured mean
   densities,
   which is needed to constrain theories of planetary interiors and
   atmospheres.  Moreover, their hosts are relatively bright ($V < 13.5$)
   facilitating further follow-up studies.}
   {}

   \maketitle
%
%________________________________________________________________

%_______ Figures ________________________________________________

%% ____ Figure: HATNet light curve  ____ %%
   \begin{figure*}
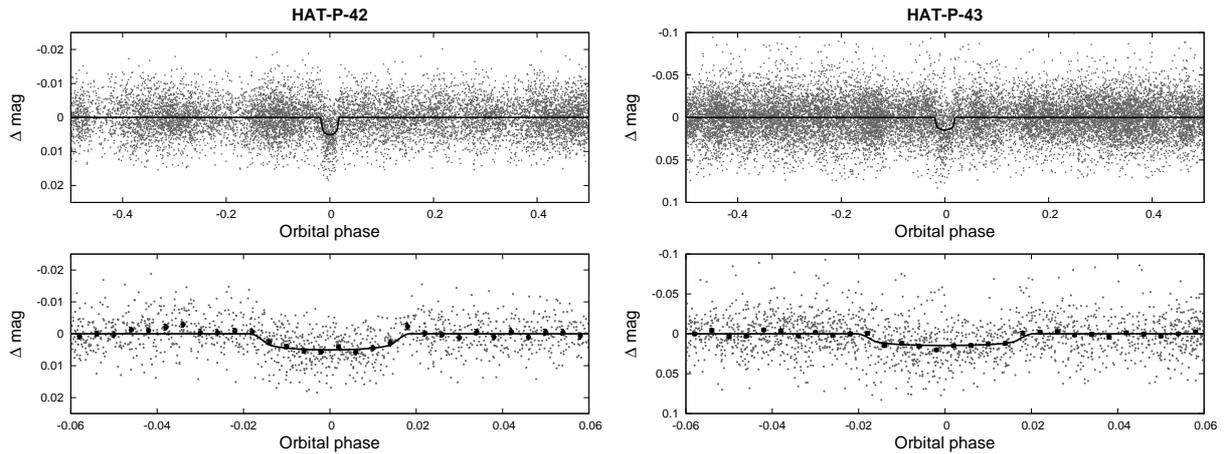

   \centering
   \includegraphics[width=80mm]{\hatcurhtr{42}-hatnet.eps}
   \includegraphics[width=80mm]{\hatcurhtr{43}-hatnet.eps}
   \caption{
     HATNet \lcs{} of \hatcur{42} (top) and \hatcur{43} (bottom) phase
     folded with the transit period.  In both cases we show two panels: the
     top shows the unbinned light curve, while the bottom shows the region
     zoomed-in on the transit, with dark filled circles for the light curve
     binned in phase with a binsize of 0.002.  The solid line shows the
     model fit to the light curve.
}
    \label{fig:hatnet}%
    \end{figure*}

%% ____ Figure: PHFU curves ____ %%
   \begin{figure*}
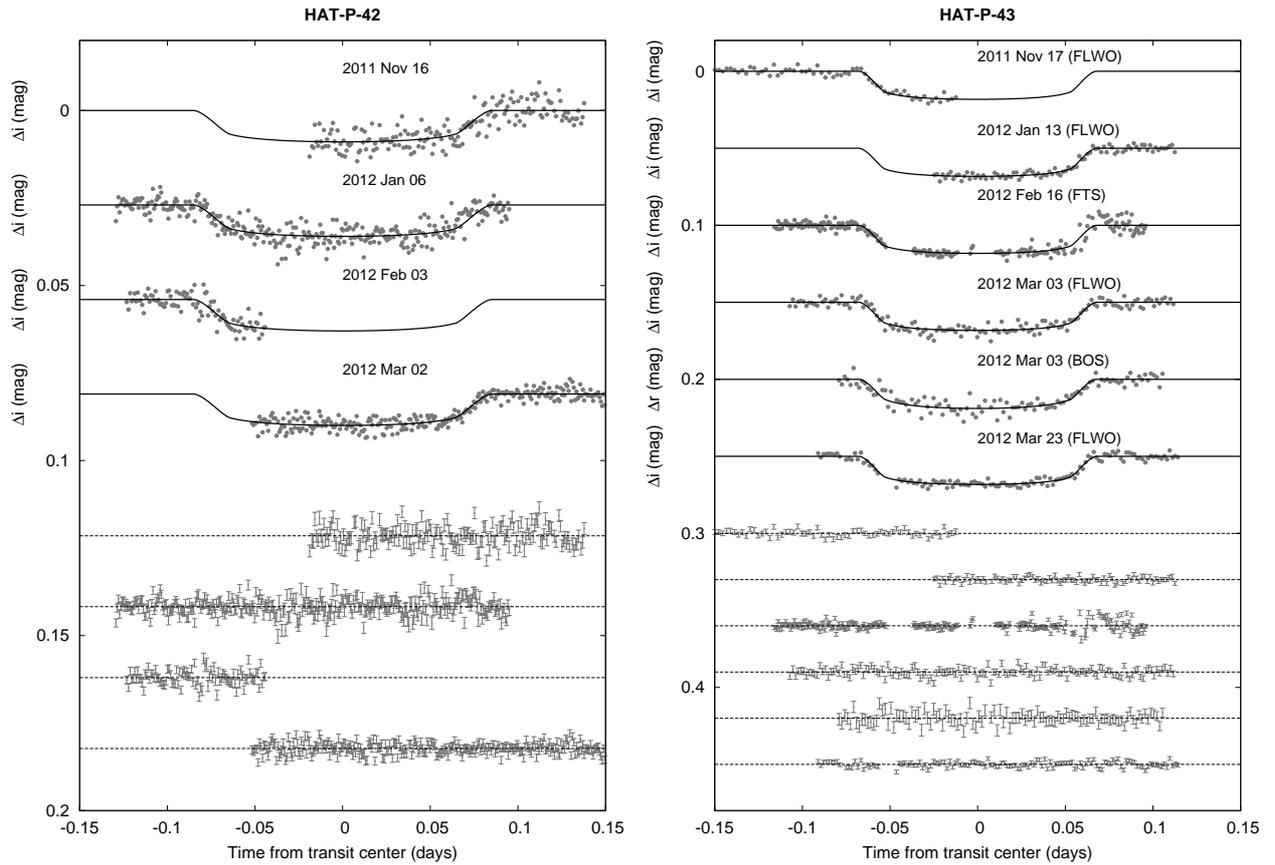

   \centering
   \mbox{
     \subfigure{\includegraphics[width=80mm]{\hatcurhtr{42}-lc.eps}}\quad
     \subfigure{\includegraphics[width=80mm]{\hatcurhtr{43}-lc.eps}}
   }
   \caption{
     Follow-up light curves for \hatcur{42} (left) and \hatcur{43}
     (right). For \hatcur{42} all observations were obtained with
     Keplercam on the FLWO~1.2\,m telescope. For \hatcur{43}
     observations were obtained with Keplercam on the FLWO~1.2\,m
     (indicated by FLWO in the figure), the Spectral CCD on the
     FTS~2.0\,m (FTS), and CCD imager on the BOS~0.8\,m (BOS). The
     Keplercam light curves have been corrected for trends during the
     modeling. The BOS and FTS light curves have also been corrected
     for trends, though for these light curves we only apply the
     External Parameter Decorrelation (EPD) procedure, and not the
     Trend Filtering Algorithm (TFA) (see for
       details Bakos et al. 2010). The dates of the events are indicated. 
     Curves after the first are displaced vertically for clarity.  Our best
     fit from the global modeling is shown by the solid lines.  Residuals
     from the fits are displayed at the bottom, in the same order as the top
     curves.  The error bars represent the photon and background shot noise,
     plus the readout noise.  They are only plotted on the residuals for
     reason of readability.
}
              \label{fig:lc}%
    \end{figure*}

%% ____ Figure: RV curves ____ %%
   \begin{figure*}
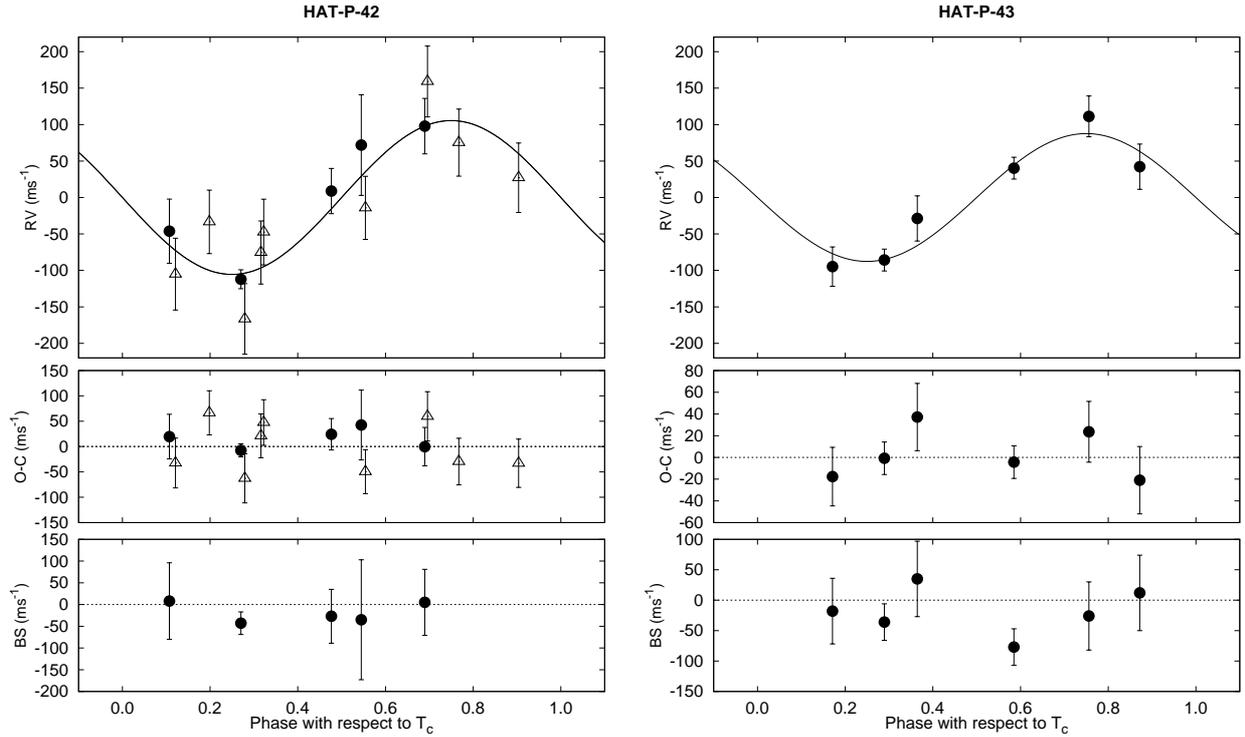

   \centering
   \mbox{
     \subfigure{\includegraphics[width=80mm]{\hatcurhtr{42}-rv.eps}}\quad
     \subfigure{\includegraphics[width=80mm]{\hatcurhtr{43}-rv.eps}}
   }
   \caption{
     High precision RV measurements and bisector spans (BS) for
     \hatcur{42} (left) and \hatcur{43} (right). Filled circles show
     OHP~1.93\,m/SOPHIE observations, while open triangles show
     FLWO~1.5\,m/TRES observations. The top panels show the RV
     measurements as a function of orbital phase, along with our
     best-fit circular orbit model. Zero phase corresponds to the time
     of mid-transit. The center-of-mass velocity has been subtracted
     for each system. The middle panels show velocity $O\!-\!C$
     residuals from the best fit circular orbit model. The bottom
     panels show the BS values, with the mean values subtracted. Note
     the different vertical scales of the panels.
}
              \label{fig:rv}%
    \end{figure*}

%% ____ Figure: RV-BIS diagram ____ %%
   \begin{figure*}
   \centering
   \mbox{
     \subfigure{\includegraphics[width=80mm]{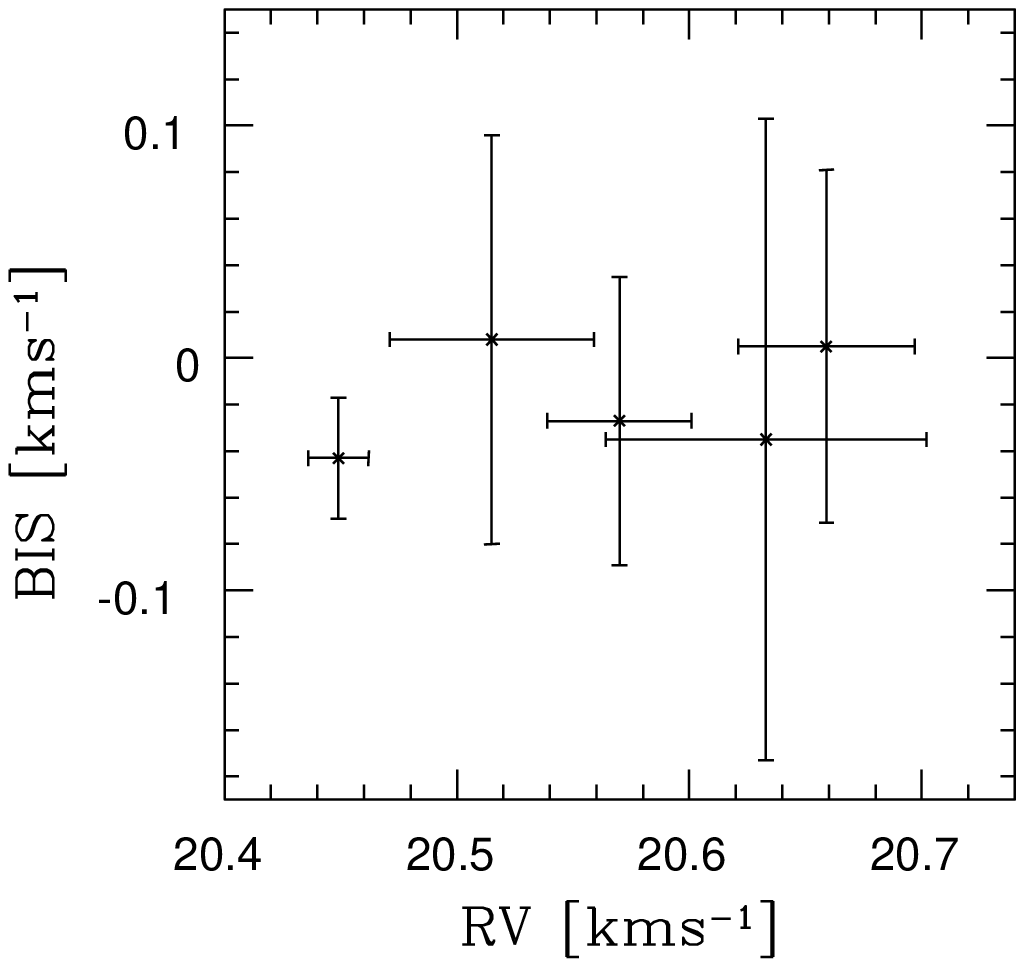}}\quad
     \subfigure{\includegraphics[width=80mm]{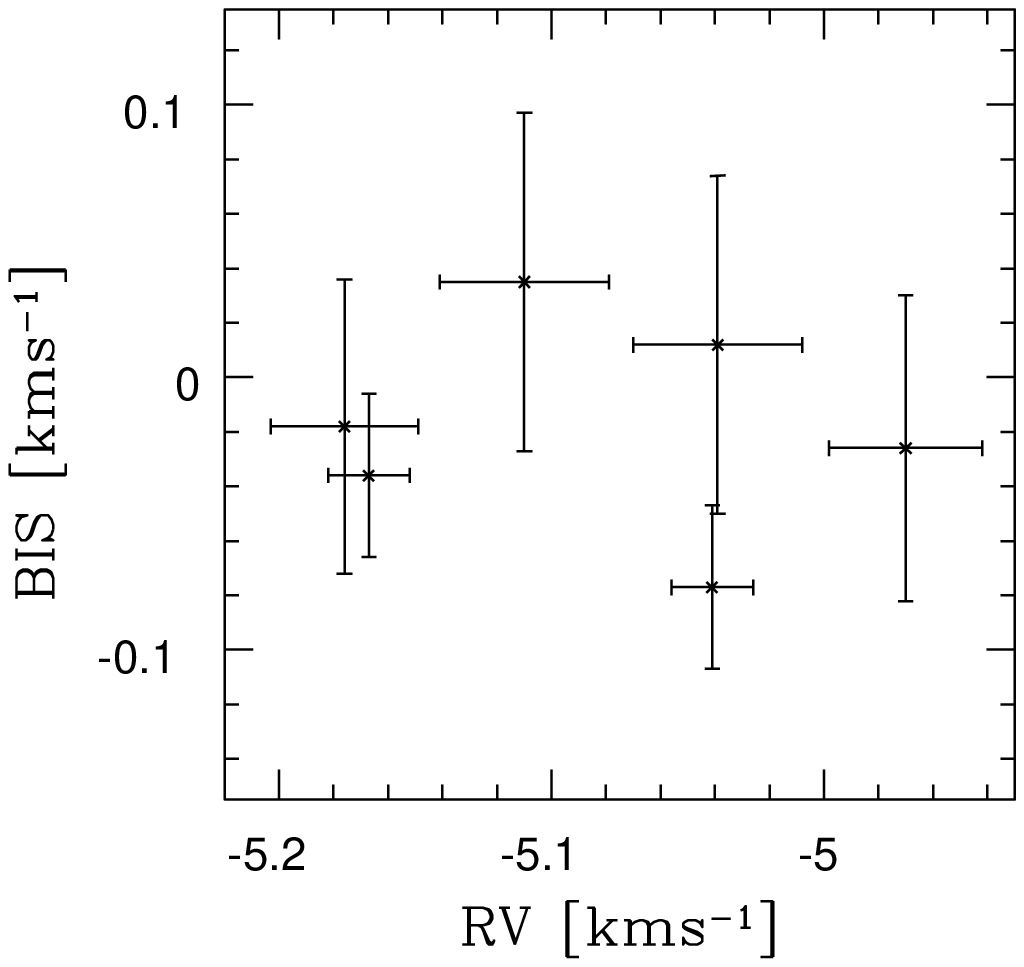}}
   }
   \caption{
     Bisector spans (BIS) as a function of the high precision RV
     measurements for \hatcur{42} (left) and \hatcur{43} (right) from the
     OHP~1.93\,m/SOPHIE observations. In both panels the scale is the same in the $x$ and $y$ axes. The 
     mean values of the BS have been subtracted for each system.
}
              \label{fig:rvbis}%
    \end{figure*}

%% ____ Figure: HR diagrams  ____ %%
   \begin{figure*}
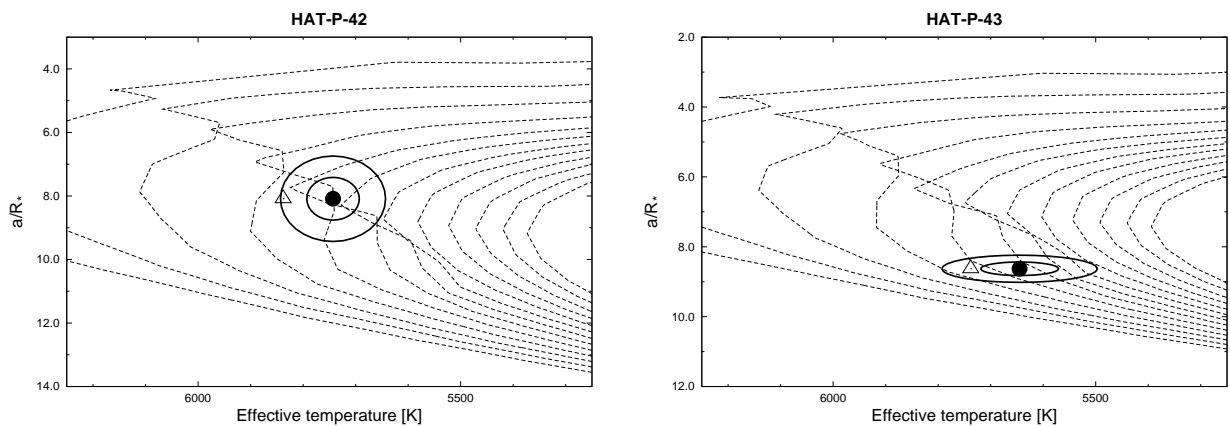

   \centering
   \mbox{
     \subfigure{\includegraphics[width=80mm]{\hatcurhtr{42}-iso-ar.eps}}\quad
     \subfigure{\includegraphics[width=80mm]{\hatcurhtr{43}-iso-ar.eps}}
   }
   \caption{
     Model isochrones from \cite{\hatcurisocite{42}} for the
     metallicities of \hatcur{42} (left) and \hatcur{43} (right) and
     ages of 1 to 13\,Gyr in 1\,Gyr increments (left to right). The
     adopted values of $\teffstar$ and \arstar\ are shown together
     with their 1$\sigma$ and 2$\sigma$ confidence ellipsoids. In each
     plot the initial values of \teffstar\ and \arstar\ from the first
     SPC and \lc\ analyses are represented with a triangle.
}
    \label{fig:iso}%
    \end{figure*}

%% ____ Figure: MassRadius  ____ %%
   \begin{figure*}
   \centering
   \includegraphics[width=80mm]{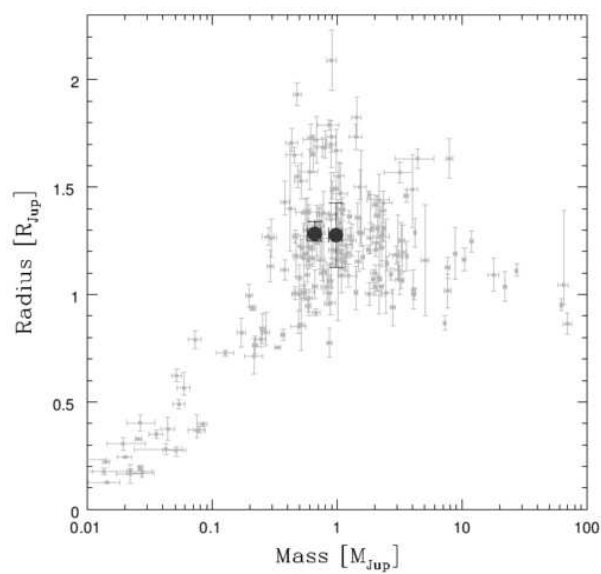}
   \caption{
     Mass radius plot of transiting extrasolar planets with determined mass.
     \hatcur{42} and \hatcur{43} are marked with the large filled circles.
     }
    \label{fig:MassRadius}%
    \end{figure*}

%% ____ Figure: TeqRadius  ____ %%
   \begin{figure*}
   \centering
   \includegraphics[width=80mm]{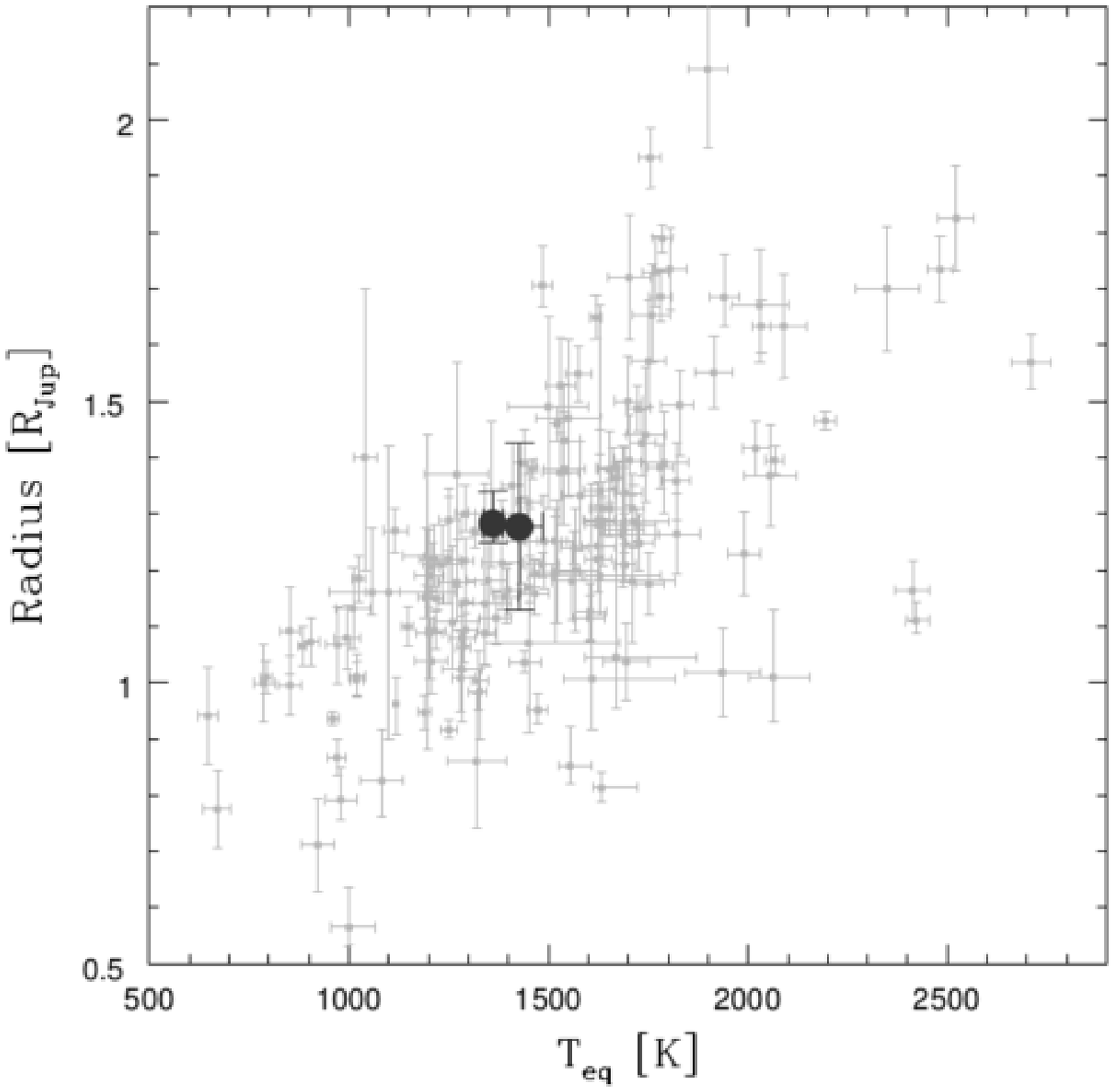}
   \caption{
     Radius of transiting extrasolar planets as a function of their
     equilibrium temperature ($T_{eq}$).  \hatcur{42} and \hatcur{43} are
     marked with big filled circles. Their respective 
     $T_{eq}$ values are calculated by assuming
     zero Bond albedos, and also that the flux is re-radiated from the full
     planet surface.  It is not the case for all planets and we advise the
     reader that the $T_{eq}$ sample is not homogeneous.
}
    \label{fig:TeqRadius}%
    \end{figure*}

%
%________________________________________________________________

\section{Introduction}
\label{sec:introduction}

Currently, more than 800 confirmed exoplanets are known, including about 200
well-characterized transiting planets (i.e.~with planetary masses and radii measured
to better than 10\% accuracy).  Transiting extrasolar planets are crucial
for exploring the physical properties of exoplanets, such as their 
mass-radius relationships,
atmospheric composition or orbital obliquity.  They provide an important first step
toward a valuable comparison of extrasolar planets.  These objects will be
the targets of next space missions designed for atmospheric
characterization.

RV surveys with well defined sample limits estimated an occurrence rate 
of hot Jupiters in the Solar
neighborhood to be around one percent (Udry \& Santos 2007; Wright et al. 2012). 
With a transit probability around $\sim10\%$, a huge number of stars have to
be monitored to detect such rare objects.  This is why majority of the transiting planets
were discovered by photometric surveys targeting tens of thousands of stars
per night.  Ground-based surveys are biased toward the detection of short
period and relatively large planets, i.e.~hot Jupiters 
(with some exceptions, e.g.~HAT-P-11, Bakos et al. 2010).  
Complementarily, space-based surveys (i.e.~CoRoT and Kepler) have excelled
in the detection of longer periods and smaller photometric transit
depths, and hence smaller planet sizes.  But, because the target stars are often relatively faint, radial-velocity follow-up to measure the planetary mass and confirm
the planetary nature of the object is often not feasible for the space-based
discoveries. 
  
We present here two new transiting hot Jupiters, first identified from
the Hungarian-made Automated Telescope Network wide-field photometric
ground-based survey (HATNet; Bakos et al. 2004).  
Since 2006, HATNet has announced and published 41 transiting
exoplanets, i.e.~$\sim20\%$ of the well-characterized sample.  These
two new typical hot Jupiters, \hatcur{42}b and \hatcur{43}b, were
partially confirmed thanks to spectroscopic observations made with
SOPHIE mounted at the 1.93\,m telescope at the Observatoire de
Haute-Provence (OHP) in France and TRES at the 1.5\,m telescope at FLWO.

The layout of the paper is as follows. 
Section 2 presents the diverse photometric and spectroscopic observations
that lead to the detection and characterization of these systems.  In the third
section, we derive the stellar parameters and the planetary orbits.  The
characteristics of these two hot Jupiters are discussed in the last
section.

%__________________________________________________________________

\section{Observations}
\label{sec:obs}

%_________________________________________________________________

\subsection{Photometric Detection}
\label{sec:detection}

%% --------------------------------------------------------------------
%% Table listing HATNet and photometric follow-up observations
%%
\begin{table*}
\caption{\label{tab:photobs}Summary of Photometric Observations.}
\centering
\begin{tabular}{llrrr}
\hline\hline
~~~~~~~~Instrument/Field\tablefootmark{a}~~~~~~~~& Date(s) & Number of Images &
	Approx.~Cadence (min) & Filter\\
\hline
\multicolumn{5}{l}{\bf HAT-P-42}\\
~~~~HAT-6/G366 & 2010~Nov--2011~Apr & 5325 & 3.7 & Sloan~r\\
~~~~HAT-9/G366 & 2010~Nov--2011~Apr & 4859 & 3.7 & Sloan~r\\
~~~~Keplercam & 2011~Nov~16 & 176 & 1.2 & Sloan~i\\
~~~~Keplercam & 2012~Jan~6 & 310 & 1.0 & Sloan~i\\
~~~~Keplercam & 2012~Feb~3 & 93 & 1.2 & Sloan~i\\
~~~~Keplercam & 2012~Mar~2 & 278 & 1.1 & Sloan~i\\
\multicolumn{5}{l}{\bf HAT-P-43}\\
~~~~HAT-5/G317 & 2010~Nov--2011~Apr & 4174 & 3.7 & Sloan~r\\
~~~~HAT-8/G317 & 2010~Nov--2011~Apr & 4300 & 3.7 & Sloan~r\\
~~~~HAT-7/G365 & 2010~Nov--2011~May & 8262 & 3.7 & Sloan~r\\
~~~~HAT-8/G365 & 2011~Apr--2011~May &  464 & 3.7 & Sloan~r\\
~~~~Keplercam & 2011~Nov~17 & 68 & 3.2 & Sloan~i\\
~~~~Keplercam & 2012~Jan~13 & 87 & 2.2 & Sloan~i\\
~~~~FTS & 2012-Feb-16 & 226 & 1.0 & Sloan~i\\
~~~~Keplercam & 2012~Mar~3 & 141 & 2.2 & Sloan~i\\
~~~~BOS & 2012~Mar~3 & 109 & 2.4 & Sloan~r\\
~~~~Keplercam & 2012~Mar~23 & 125 & 2.2 & Sloan~i\\
\hline
\end{tabular}
\tablefoot{
\tablefoottext{a}{HAT-5, -6, -7, and -10 are located at FLWO in AZ, USA.
HAT-8 and -9 are located at Mauna Kea Observatory in HI, USA.}
}
\end{table*}
%% --------------------------------------------------------------------

\hatcur{42} and \hatcur{43} were identified as candidate transiting
planets based on photometric observations conducted by the HATNet
survey (Bakos et al. 2004). These observations are summarized,
together with the follow-up photometric observations, in
\reftabl{photobs}. The data were processed and searched for transits
following the procedure of Bakos et al. (2010), see also
Kovacs et al. (2002, 2005). \reffigl{hatnet} shows the
folded HATNet light curves for both systems. We give
cross-identifications, and catalog photometry on an absolute scale for
each system in~\reftabl{stellar}.

%_________________________________________________________________
\subsection{Follow-up}

\begin{table}
\caption{
	\label{tab:phfu42}
	High-precision differential photometry of \hatcur{42}.
}
\centering
\begin{tabular}{lrrrr}
\hline\hline
BJD\tablefootmark{a} & Mag\tablefootmark{b} & \ensuremath{\sigma_{\rm Mag}} & 
	Mag(orig)\tablefootmark{c} & Filter \\
(2,400,000$+$) & & & \\
\hline
$ 55882.87909 $ & $   0.01318 $ & $   0.00170 $ & $  11.46340 $ & $ i$\\
$ 55882.88049 $ & $   0.01025 $ & $   0.00191 $ & $  11.46150 $ & $ i$\\
$ 55882.88128 $ & $   0.01064 $ & $   0.00192 $ & $  11.45990 $ & $ i$\\
$ 55882.88208 $ & $   0.00410 $ & $   0.00192 $ & $  11.45190 $ & $ i$\\
$ 55882.88303 $ & $   0.00898 $ & $   0.00173 $ & $  11.45770 $ & $ i$\\
$ 55882.88389 $ & $   0.01010 $ & $   0.00180 $ & $  11.46170 $ & $ i$\\
$ 55882.88473 $ & $   0.00376 $ & $   0.00187 $ & $  11.45210 $ & $ i$\\
$ 55882.88558 $ & $   0.00885 $ & $   0.00187 $ & $  11.45610 $ & $ i$\\
$ 55882.88644 $ & $   0.01057 $ & $   0.00188 $ & $  11.45900 $ & $ i$\\
$ 55882.88729 $ & $   0.00559 $ & $   0.00168 $ & $  11.45500 $ & $ i$\\
\hline
\end{tabular}
\tablefoot{
This table is available in a machine-readable form in the on-line journal. A
portion is shown here for guidance regarding its form and content.
\tablefoottext{a}{Barycentric Julian Date calculated directly from UTC, {\em
without} correction for leap seconds.}
\tablefoottext{b}{The out-of-transit level has been subtracted. These
magnitudes have been subjected to the EPD and TFA procedures, carried out
simultaneously with the transit fit.}
\tablefoottext{c}{Raw magnitude values without application of the EPD or TFA
procedures.}
}
\end{table}

\begin{table}
\caption{\label{tab:phfu43}High-precision differential photometry of
\hatcur{43}.}
\centering
\begin{tabular}{lrrrr}
\hline\hline
BJD\tablefootmark{a} & Mag\tablefootmark{b} & \ensuremath{\sigma_{\rm Mag}} & 
	Mag(orig)\tablefootmark{c} & Filter \\
(2,400,000$+$) & & & \\
\hline
$ 55883.89634 $ & $   0.00397 $ & $   0.00167 $ & $  12.40490 $ & $ i$\\
$ 55883.89856 $ & $   0.00002 $ & $   0.00157 $ & $  12.40130 $ & $ i$\\
$ 55883.90082 $ & $  -0.00162 $ & $   0.00153 $ & $  12.39990 $ & $ i$\\
$ 55883.90306 $ & $   0.00397 $ & $   0.00150 $ & $  12.40600 $ & $ i$\\
$ 55883.90530 $ & $  -0.00219 $ & $   0.00158 $ & $  12.39980 $ & $ i$\\
$ 55883.90753 $ & $   0.00141 $ & $   0.00150 $ & $  12.40250 $ & $ i$\\
$ 55883.90976 $ & $  -0.00085 $ & $   0.00155 $ & $  12.40180 $ & $ i$\\
$ 55883.91201 $ & $  -0.00087 $ & $   0.00153 $ & $  12.40340 $ & $ i$\\
$ 55883.91426 $ & $   0.00004 $ & $   0.00151 $ & $  12.40220 $ & $ i$\\
$ 55883.91650 $ & $  -0.00165 $ & $   0.00166 $ & $  12.39820 $ & $ i$\\
\hline
\end{tabular}
\tablefoot{
This table is available in a machine-readable form in the on-line journal. A
portion is shown here for guidance regarding its form and content.
\tablefoottext{a}{Barycentric Julian Date calculated directly from UTC, {\em
without} correction for leap seconds.}
\tablefoottext{b}{The out-of-transit level has been subtracted. These
magnitudes have been subjected to the EPD and, in some cases, TFA
procedures, carried out simultaneously with the transit fit.  The TFA
procedure has not been applied to observations with $2455973 < {\rm BJD} <
2455975$ or to observations in the $r$ filter with $2455990 < {\rm BJD} <
2455991$.}
\tablefoottext{c}{Raw magnitude values without application of the EPD or TFA
procedures.}
}
\end{table}

We conducted follow-up spectroscopic and photometric observations of both
systems to confirm their planetary natures and determine their physical
properties.  See Latham et al. (2009) for a more detailed description of our
follow-up procedure.  Below we provide specific details relevant to the
discoveries of \hatcurb{42} and \hatcurb{43}.

%_________________________________________________________________
\subsubsection{Reconnaissance Spectroscopy}

\begin{table*}
\caption{\label{tab:specobs}Summary of Spectroscopic Observations.}
\centering
\begin{tabular}{llrrrr}
\hline\hline
\multicolumn{1}{c}{Instrument} & \multicolumn{1}{c}{Date(s)} & \multicolumn{1}{c}{Number of} & \multicolumn{1}{c}{Resolution} & \multicolumn{1}{c}{Wavelength} & \multicolumn{1}{c}{Velocity\tablefootmark{a}} \\
& & \multicolumn{1}{c}{Observations} & $\lambda\,/\,\Delta\lambda$ & \multicolumn{1}{c}{Coverage} [$\AA$] & \multicolumn{1}{c}{Precision} \\
\hline
\multicolumn{6}{l}{\bf HAT-P-42}\\
~~~~FLWO~1.5\,m/TRES & 2011~Oct--2012~Jan & 13 & 44,000 & 3900--8900 & $\sim (30+35)$\tablefootmark{b}\,\ms \\
~~~~OHP~1.93\,m/SOPHIE & 2011~Dec~7--13 & 5 & 39,000 & 3900--6900 & $\sim (6+35)$\tablefootmark{c}\,\ms \\
\multicolumn{6}{l}{\bf HAT-P-43}\\
~~~~NOT~2.5\,m/FIES & 2011~Oct~24--27 & 2 & 46,000 & 3700--7300 & $\sim 100$\,\ms \\
~~~~FLWO~1.5\,m/TRES & 2011~Nov--2012~Jan & 2 & 44,000 & 3900--8900 & $\sim 100$\,\ms \\
~~~~OHP~1.93\,m/SOPHIE & 2011~Dec~6--13 & 6 & 39,000 & 3900--6900 & $\sim (10+25)$\tablefootmark{c}\,\ms \\
\hline
\end{tabular}
\tablefoot{
\tablefoottext{a}{
We give the median velocity precision of the observations, and not the
best precision that can be attained with the given instrument. The
FIES and TRES observations of \hatcur{43} were done in reconnaissance
mode (lower S/N) which is why the velocities are less precise than the
SOPHIE observations.
}
\tablefoottext{b}{
A precision of 30\ms, plus 35\ms\ added in quadrature to achieve a \chisq\ per
degree of freedom of unity for the Keplerian fit of \hatcur{42}.  }
\tablefoottext{c}{
Values in the parentheses include the sum of the theoretical noise
due to the instrument (6 and 10\ms, respectively), plus empirical errors 
taking into account the correction due to Moon light in the spectra.
}
}
\end{table*}

We obtained initial high-resolution, low S/N ``reconnaissance''
spectra of \hatcur{42} and \hatcur{43} using the Tillinghast Reflector
Echelle Spectrograph (TRES; Furesz 2008) on the 1.5\,m
Tillinghast Reflector at the Fred Lawrence Whipple Observatory (FLWO)
in AZ, USA, and using the FIbre-fed Echelle Spectrograph (FIES) on the
2.5\,m Nordic Optical Telescope (NOT) on the island of La Palma, Spain
(Djupvik \& Andersen 2010). These observations are summarized in
\reftabl{specobs}. The data were analyzed via cross-correlation
against synthetic templates as described by
Buchhave et al. (2010) and Quinn et al. (2012). Both stars were found to be slowly
rotating G~dwarfs, with no evidence for composite spectra, and with
low velocity variations indicating that neither object is an undiluted
eclipsing binary star system.

%_________________________________________________________________________

\subsubsection{Photometry}

Photometric follow-up observations were obtained with Keplercam on the
FLWO~1.2\,m, the Spectral CCD on the 2.0\,m Faulkes Telescope South
(FTS) at Siding Springs Observatory in Australia, and with the CCD
imager on the 0.8\,m Byrne Observatory at Sedgwick (BOS) telescope at
Sedgwick Reserve in the Santa Ynez Valley, CA, USA. These observations are
summarized in \reftabl{photobs}. Both FTS and BOS are operated by the
Las Cumbres Observatory Global Telescope
(LCOGT\footnote{http://lcogt.net}; Shporer et al. 2011; Brown et
al.~2013, in preparation). The Keplercam observations were reduced to
light curves following the procedure of Bakos et al. (2010), while
the FTS and BOS observations were reduced following the procedure of
Fulton et al. (2011).

\reffigl{lc} shows the photometric follow-up light curves for both systems
together with our best-fit models.  \hatcur{42} and \hatcur{43} have
photometric transit depths of $\hatcurLCdip{42}$ and
$\hatcurLCdip{43}$\,mmag, respectively.  The data are provided in
machine-readable format in \reftabl{phfu42} for \hatcur{42}, and in
\reftabl{phfu43} for \hatcur{43}.

%_________________________________________________________________________
\subsubsection{Confirmation Spectroscopy}

\begin{table}
\caption{\label{tab:rvs42}Relative radial velocities and bisector spans of
\hatcur{42}.}
\centering
\begin{tabular}{lrrrrr}
\hline\hline
   BJD\tablefootmark{a} & 
   RV\tablefootmark{b} & 
   \ensuremath{\sigma_{\rm RV}}\tablefootmark{c} & 
   BIS\tablefootmark{d} & 
   Phase \\
   (2,454,000$+$) &
   (\ms) &
   (\ms) &
   (\ms) &
   \\
\hline
$ 1889.03539 $ & $   -47.4 $ & $    28.4 $ & $ \cdots $      & $   0.322 $ \\
$ 1900.05380 $ & $   159.1 $ & $    33.8 $ & $ \cdots $      & $   0.696 $ \\
$ 1901.01864 $ & $    27.1 $ & $    32.4 $ & $ \cdots $      & $   0.904 $ \\
$ 1902.02682 $ & $  -105.1 $ & $    34.6 $ & $ \cdots $      & $   0.121 $ \\
$ 1902.71983 $ & $  -112.1 $ & $    13.0 $ & $   -43.0 $    & $   0.270 $ \\
$ 1902.93220 $ & $   -75.4 $ & $    25.4 $ & $ \cdots $      & $   0.316 $ \\
$ 1903.67805 $ & $     8.9 $ & $    31.0 $ & $   -27.0 $ &    $   0.477 $ \\
$ 1904.03850 $ & $   -14.2 $ & $    25.4 $ & $ \cdots $      & $   0.554 $ \\
$ 1904.66696 $ & $    97.9 $ & $    38.0 $ & $     5.0 $ &     $   0.690 $ \\
$ 1905.02718 $ & $    75.3 $ & $    29.9 $ & $ \cdots $      & $   0.767 $ \\
$ 1906.60475 $ & $   -46.1 $ & $    44.0 $ & $     8.0 $ &     $   0.107 $ \\
$ 1907.02711 $ & $   -33.4 $ & $    25.9 $ & $ \cdots $      & $   0.198 $ \\
$ 1908.63608 $ & $    71.9 $ & $    69.0 $ & $   -35.0 $     & $   0.545 $ \\
$ 1939.89586 $ & $  -166.5 $ & $    33.4 $ & $ \cdots $      & $   0.279 $ \\
\hline
\end{tabular}
\tablefoot{Observations without a BS value were obtained with the
TRES instrument, observations with a BS value were obtained with SOPHIE.
\tablefoottext{a}{Barycentric Julian Date calculated directly from UTC, {\em
without} correction for leap seconds.}
\tablefoottext{b}{A zero-level $\gamma$ velocity, fitted independently
  for each instrument, has been subtracted from these measurements.
  For reference, the SOPHIE $\gamma$ velocity from the fit was
  $20.561\pm0.014$\kms, which is an estimate of the true recession
  velocity of the center of mass of the system. For TRES the value is
  $0.014 \pm 0.015$\kms. As an artifact of the
  reduction procedure the TRES RVs were measured relative to an
  arbitrary zero-point, the fitted $\gamma$ velocity in this case has
  no physical meaning.}
\tablefoottext{c}{Internal errors excluding any component of astrophysical
jitter.
\tablefoottext{d}{We use the relation $\sigma_{\rm BS} = 2\sigma_{\rm RV}$
to estimate the BS uncertainties.}
}
}
\end{table}

\begin{table}
\caption{\label{tab:rvs43}Relative radial velocities and bisector spans of
\hatcur{43}.}
\centering
\begin{tabular}{lrrrrr}
\hline\hline
   BJD\tablefootmark{a} & 
   RV\tablefootmark{b} & 
   \ensuremath{\sigma_{\rm RV}}\tablefootmark{c} & 
   BIS\tablefootmark{d} & 
   Phase \\
   (2,454,000$+$) &
   (\ms) &
   (\ms) &
   (\ms) &
   \\
\hline
$ 1901.68768 $ & $   -85.8 $ & $    15.0 $ & $   -36.0 $ & $   0.289 $ \\
$ 1902.67245 $ & $    40.3 $ & $    15.0 $ & $   -77.0 $ & $   0.585 $ \\
$ 1903.62875 $ & $    42.3 $ & $    31.0 $ & $    12.0 $ & $   0.872 $ \\
$ 1904.62601 $ & $   -94.8 $ & $    27.0 $ & $   -18.0 $ & $   0.171 $ \\
$ 1906.57509 $ & $   111.3 $ & $    28.0 $ & $   -26.0 $ & $   0.756 $ \\
$ 1908.60322 $ & $   -28.8 $ & $    31.0 $ & $    35.0 $ & $   0.364 $ \\
\hline
\end{tabular}
\tablefoot{All observations listed were obtained with SOPHIE.
\tablefoottext{a}{Barycentric Julian Date calculated directly from UTC, {\em
without} correction for leap seconds.}
\tablefoottext{b}{For reference, the SOPHIE $\gamma$ velocity from the fit was
$-5.082\pm0.009$\kms.}
\tablefoottext{c}{Internal errors excluding any component of astrophysical
jitter.}
\tablefoottext{d}{We use the relation $\sigma_{\rm BS} = 2\sigma_{\rm RV}$
to estimate the BS uncertainties.}
}
\end{table}

We obtained high-resolution, high-S/N spectra of \hatcur{42} and \hatcur{43}
using the SOPHIE spectrograph on the 1.93\,m telescope at OHP
(Bouchy et al. 2009), and of \hatcur{42} using TRES on the FLWO~1.5\,m
telescope.  These observations are summarized in \reftabl{specobs}.  The TRES
observations of \hatcur{42} were simply a continuation of the first two
``reconnaissance'' TRES observations of this target which were already of
sufficient S/N to detect the $\sim 100$\,\ms\ orbital variation.  Of the 13
TRES measurements, 4 with too low S/N (i.e.~precision $>100$\ms) were
excluded from the orbital analysis.  While we have previously used SOPHIE to
confirm HATNet planets (e.g. Bakos et al. 2007, Shporer et al. 2009), there have
been significant changes to the instrument and reduction procedure.
These significantly increased the RV accuracy
(Perruchot et al. 2011, Bouchy et al. 2012), thus we provide a description of our
observing and reduction procedures below.

We observed \hatcur{42} and \hatcur{43} with the {\it High Efficiency}
fibers, yielding a resolving power of 
R=$\lambda/\Delta\lambda\,\approx\,39\,000$ (at 550~nm).  The spectra were
correlated with a G2 numerical mask in order to calculate the
cross-correlation function (CCF).  This CCF is fitted by a Gaussian and its
center gives the RV measurement (Baranne et al. 1996, Pepe et al. 2002).  The
bisector spans values were calculated as described in Queloz et al. (2001). 
The error bars were derived from the CCF, following the procedure in
Boisse et al. (2010), yielding a mean precision of $\sim6\ms$ for \hatcur{42}
and $\sim10\ms$ for \hatcur{43}.  Since most of our spectra were polluted by
Moon light, we corrected them following the method presented in
H\'ebrard et al. (2008), and increased the error bars as determined empirically,
leading to a mean precision of $\sim40\ms$ for \hatcur{42} and $\sim30\ms$
for \hatcur{43}.  A thorium-argon calibration was done each two hours to monitor the instrumental drift mainly due to small variation of temperature and pressure in the spectrograph tank. SOPHIE is environmentally stabilized and this intrinsic drift is less than $2\ms$ per hour. Each measurement was also corrected for this drift.

We computed the CCF using masks corresponding to different spectral types (F0, G2, K5, M5) and found no significant variation in the RV semi-amplitude. If a planet induces a RV shift, the amplitude of the variation should remain constant regardless of the spectral lines used to measure the RV. In contrast, for the case of a blend consisting of a bright target and a faint binary having different spectral types, the relative contributions to the CCF from the different components will change when a different mask is used (Santos et al. 2002, Collier Cameron et al. 2007). This result thus favors the planetary hypothesis.

\reffigl{rv} shows the high-precision RV curves for both systems, together
with our best-fit circular-orbit models.  \reffigl{rvbis} shows the bisector
spans as a function of the RV from the SOPHIE spectra.  The data are
provided in \reftabl{rvs42} for \hatcur{42}, and in \reftabl{rvs43} for
\hatcur{43}.

% ==========================================================================
% 
\section{Analysis}
\label{sec:analysis}

%_________________________________________________________________

To rule out the possibility that either object might actually be a
blended stellar eclipsing binary system, we conducted a blend analysis
similar to that done in Hartman et al. (2012).  We find that in both
cases we cannot rule out blend scenarios based on the photometric
observations alone, however all scenarios which fit the photometric
data predict significant RV and BS variations (greater than $300$\,\ms\
for \hatcur{42} and greater than $1$\,\kms\ for \hatcur{43}). Such
variations are
ruled out by our spectroscopic observations.  We conclude that both
objects are transiting planet systems.

Having confirmed the planetary nature of both systems, we proceeded
with their analysis following the methods described in
Bakos et al. (2010), with some modifications as described in
Hartman et al. (2012).  Briefly, this consists of: (1) inferring the
stellar atmospheric parameters from the available high-resolution
spectra (we used the TRES spectra of \hatcur{42}, and the TRES and FIES
spectra of \hatcur{43}, together with the Stellar Parameter
Classification (SPC) method of Buchhave et al. (2012); (2) conducting
a global Markov-Chain Monte Carlo (MCMC)-based modeling of the
available photometric light curves and RVs (we fix the limb darkening
coefficients using the tables by Claret 2004); (3) using the
spectroscopically inferred stellar effective temperatures and
metallicities, together with the stellar densities determined from the
light curve modeling, and the Yonsei-Yale theoretical stellar evolution
models Yi et al. (2001), to determine the stellar masses, radii and
ages, as well as the planetary parameters (e.g.~mass and radius) which
depend on these values (\reffigl{iso}); (4) re-analyzing the
high-resolution spectra fixing the stellar surface gravities to the
values found in (3), and then re-iterating steps (2) and (3).

In conducting the analysis we inflated the TRES RV errors for
\hatcur{42} by adding in quadrature a value of 35\,\ms\ to the internal
errors.  This ``jitter'' is needed to achieve a $\chi^{2}$ per degree
of freedom of unity for the TRES RVs in the best-fit model.  We do not
add jitter to the SOPHIE RVs of \hatcur{42} or \hatcur{43} because in
both cases $\chi^{2}$ per degree of freedom is less than one for these
RVs, indicating that the SOPHIE formal uncertainties may be
overestimated.

For both systems we conducted the analysis fixing the eccentricities to
zero, as well as allowing the eccentricities to vary.  For each system
we find that the eccentricity is consistent with zero (the 95\%
confidence upper limits on the eccentricity are $e < 0.2$ for
\hatcurb{42}, and $e < 0.29$ for \hatcurb{43}).  Following the
suggestion of Anderson et al. (2012) we adopt the parameter values
associated with the fixed circular orbits.  The adopted stellar
parameters are given in \reftabl{stellar}, while the adopted planetary
parameters are given in \reftabl{planetparam}.  For completeness we
also provide the parameters which result when the eccentricities are
allowed to be non-zero in Tables~\ref{tab:stellareccen}
and~\ref{tab:planetparameccen}.  Note that all the eccentric
parameters are within $1\sigma$ of the circular
orbit values.

We searched for sinusoidal signals and additional transits in the light curves. Neither present a significant peak in the L-S periodogram (false alarm probability of 50\% or more). After removing the detected in-transit data, neither light curves shows a transit signal that would had passed our automated or by-eye selections.

%% --------------------------------------------------------------------
%% Table of stellar parameters, assuming circular orbits 
%%
\begin{table*}
\caption{\label{tab:stellar}Adopted stellar parameters for \hatcur{42}
and \hatcur{43} assuming circular orbits.}
\centering
\begin{tabular}{lccr}
\hline\hline
    &{\bf HAT-P-42}&{\bf HAT-P-43}&\\
~~~~~~~~Parameter~~~~~~~~&Value&Value&Source\\
\hline
\multicolumn{4}{l}{\em Astrometric properties}\\
~~~~GSC~ID\dotfill                 & \hatcurCCgsc{42}      & \hatcurCCgsc{43}      &\\
~~~~2MASS-ID\dotfill               & \hatcurCCtwomass{42}  & \hatcurCCtwomass{43}  &\\
~~~~R.A. (J2000)\dotfill           & $\hatcurCCra{42}$     & $\hatcurCCra{43}$     & 2MASS\\
~~~~Dec. (J2000)\dotfill           & $\hatcurCCdec{42}$    & $\hatcurCCdec{43}$    & 2MASS\\
~~~~$\mu_{\rm RA}$ ($\masyr$)\dotfill    & $\hatcurCCpmra{42}$   & $\hatcurCCpmra{43}$  & UCAC4\\
~~~~$\mu_{\rm Dec}$ ($\masyr$)\dotfill    & $\hatcurCCpmdec{42}$   & $\hatcurCCpmdec{43}$  & UCAC4\\
\multicolumn{4}{l}{\em Spectroscopic properties}\\
~~~~$\teffstar$ (K)\dotfill         &  \hatcurSMEteff{42}   &  \hatcurSMEteff{43}   & SPC\tablefootmark{a}\\
~~~~$\feh$\dotfill                  &  \hatcurSMEzfeh{42}   &  \hatcurSMEzfeh{43}   & SPC                 \\
~~~~$\vsini$ (\kms)\dotfill         &  \hatcurSMEvsin{42}   &  \hatcurSMEvsin{43}   & SPC                 \\
~~~~$\gamma_{\rm RV}$ (\kms)\dotfill&  20.26 $\pm$ 0.15\tablefootmark{b}   & -4.86 $\pm$ 0.15 \tablefootmark{b}   & TRES                  \\
\multicolumn{4}{l}{\em Photometric properties}\\
~~~~$B$ (mag)\dotfill               &  \hatcurCCtassmb{42}  &  \hatcurCCtassmb{43}  & APASS                \\
~~~~$V$ (mag)\dotfill               &  \hatcurCCtassmv{42}  &  \hatcurCCtassmv{43}  & APASS                \\
~~~~$J$ (mag)\dotfill               &  \hatcurCCtwomassJmag{42} &  \hatcurCCtwomassJmag{43} & 2MASS           \\
~~~~$H$ (mag)\dotfill               &  \hatcurCCtwomassHmag{42} &  \hatcurCCtwomassHmag{43} & 2MASS           \\
~~~~$K_s$ (mag)\dotfill             &  \hatcurCCtwomassKmag{42} &  \hatcurCCtwomassKmag{43} & 2MASS           \\
\multicolumn{4}{l}{\em Derived properties}\\
~~~~$\mstar$ ($\msun$)\dotfill      &  \hatcurISOmlong{42}   &  \hatcurISOmlong{43}   & \hatcurisoshort{42}+\hatcurlumind{42}+SPC\tablefootmark{c}\\
~~~~$\rstar$ ($\rsun$)\dotfill      &  \hatcurISOrlong{42}   &  \hatcurISOrlong{43}   & \hatcurisoshort{42}+\hatcurlumind{42}+SPC         \\
~~~~$\loggstar$ (cgs)\dotfill       &  \hatcurISOlogg{42}    &  \hatcurISOlogg{43}    & \hatcurisoshort{42}+\hatcurlumind{42}+SPC         \\
~~~~$\lstar$ ($\lsun$)\dotfill      &  \hatcurISOlum{42}     &  \hatcurISOlum{43}     & \hatcurisoshort{42}+\hatcurlumind{42}+SPC         \\
~~~~$M_V$ (mag)\dotfill             &  \hatcurISOmv{42}      &  \hatcurISOmv{43}      & \hatcurisoshort{42}+\hatcurlumind{42}+SPC         \\
~~~~$M_K$ (mag,\hatcurjhkfilset{42})\dotfill &  \hatcurISOMK{42} &  \hatcurISOMK{43} & \hatcurisoshort{42}+\hatcurlumind{42}+SPC         \\
~~~~Age (Gyr)\dotfill               &  \hatcurISOage{42}     &  \hatcurISOage{43}     & \hatcurisoshort{42}+\hatcurlumind{42}+SPC         \\
~~~~$A_{V}$ (mag)\tablefootmark{d}\dotfill           &  \hatcurXAv{42}  &  \hatcurXAv{43}    & \hatcurisoshort{42}+\hatcurlumind{42}+SPC\\
~~~~Distance (pc)\dotfill           &  \hatcurXdistred{42}  &  \hatcurXdistred{43}  & \hatcurisoshort{42}+\hatcurlumind{42}+SPC\\
\hline
\end{tabular}
\tablefoot{
\tablefoottext{a}{
    SPC = ``Stellar Parameter Classification'' method, described by
   Buchhave et al. (2012), which derives stellar atmospheric parameters
    from high-resolution spectra.  These parameters rely primarily on
    SPC, but have a small dependence also on the iterative analysis
    incorporating the isochrone search and global modeling of the data,
    as described in the text.
}
\tablefoottext{b}{
  These velocities corresponds to an absolute scale references thanks to nightly observations of HD182488 and correction of gravitational redshift of the Sun.
}
\tablefoottext{b}{
    \hatcurisoshort{42}+\hatcurlumind{42}+SPC = Based on the
    \hatcurisoshort{42}\ isochrones (Yi et al. 2001),
    \hatcurlumind{42}\ as a luminosity indicator, and the SPC results.
}
\tablefoottext{c}{
  \band{V} extinction determined by comparing the measured 2MASS and
  APASS photometry for the star to the expected magnitudes from the
  \hatcurisoshort{42}+\hatcurlumind{42}+SPC model for the star.  We use
  the Cardelli et al. (1989) extinction law.
}
}
\end{table*}
%% --------------------------------------------------------------------

%% --------------------------------------------------------------------
%% Table of stellar parameters, assuming eccentric orbits 
%%
\begin{table*}
\caption{\label{tab:stellareccen}Derived stellar parameters for
\hatcur{42} and \hatcur{43} allowing nonzero eccentricty.}
\centering
\begin{tabular}{lcc}
\hline\hline
    &{\bf HAT-P-42}&{\bf HAT-P-43}\\
~~~~~~~~Parameter~~~~~~~~&Value&Value\\
\hline
~~~~$\mstar$ ($\msun$)\dotfill      &  \hatcurISOmlongeccen{42}   &  \hatcurISOmlongeccen{43}\\
~~~~$\rstar$ ($\rsun$)\dotfill      &  \hatcurISOrlongeccen{42}   &  \hatcurISOrlongeccen{43}\\
~~~~$\loggstar$ (cgs)\dotfill       &  \hatcurISOloggeccen{42}    &  \hatcurISOloggeccen{43}\\
~~~~$\lstar$ ($\lsun$)\dotfill      &  \hatcurISOlumeccen{42}     &  \hatcurISOlumeccen{43}\\
~~~~$M_V$ (mag)\dotfill             &  \hatcurISOmveccen{42}      &  \hatcurISOmveccen{43}\\
~~~~$M_K$ (mag,\hatcurjhkfilset{42})\dotfill &  \hatcurISOMKeccen{42} &  \hatcurISOMKeccen{43}\\
~~~~Age (Gyr)\dotfill               &  \hatcurISOageeccen{42}     &  \hatcurISOageeccen{43}\\
~~~~$A_{V}$ (mag)\tablefootmark{c}\dotfill           &  \hatcurXAveccen{42}  &  \hatcurXAveccen{43}\\
~~~~Distance (pc)\dotfill           &  \hatcurXdistredeccen{42}  &  \hatcurXdistredeccen{43}\\
\hline
\end{tabular}
\tablefoot{
Quantities and abbreviations are as in \reftabl{stellar}, which gives
our adopted values, determined assuming circular orbits.  We do not
list parameters that are independent of the eccentricity.
}
\end{table*}
%% --------------------------------------------------------------------

%% --------------------------------------------------------------------
%% Table of orbital and planetary parameters, assuming circular orbits 
%%
\begin{table*}
\caption{\label{tab:planetparam}Adopted orbital and planetary
  parameters for \hatcurb{42} and \hatcurb{43} assuming circular
  orbits.}  
\centering
\begin{tabular}{lcc}
\hline\hline
    &{\bf HAT-P-42}&{\bf HAT-P-43}\\
~~~~~~~~Parameter~~~~~~~~&Value&Value\\
\hline
\multicolumn{3}{l}{\em \Lc{} parameters}\\
~~~$P$ (days)             \dotfill    & $\hatcurLCP{42}$              & $\hatcurLCP{43}$              \\
~~~$T_c$ (${\rm BJD}$)    
      \tablefootmark{a}   \dotfill    & $\hatcurLCT{42}$              & $\hatcurLCT{43}$              \\
~~~$T_{14}$ (days)
      \tablefootmark{a}   \dotfill    & $\hatcurLCdur{42}$            & $\hatcurLCdur{43}$            \\
~~~$T_{12} = T_{34}$ (days)
      \tablefootmark{a}   \dotfill    & $\hatcurLCingdur{42}$         & $\hatcurLCingdur{43}$         \\
~~~$\arstar$              \dotfill    & $\hatcurPPar{42}$             & $\hatcurPPar{43}$             \\
~~~$\zrstar$\tablefootmark{b}              \dotfill    & $\hatcurLCzeta{42}$       & $\hatcurLCzeta{43}$       \\
~~~$\rpl/\rstar$          \dotfill    & $\hatcurLCrprstar{42}$        & $\hatcurLCrprstar{43}$        \\
~~~$b^2$                  \dotfill    & $\hatcurLCbsq{42}$            & $\hatcurLCbsq{43}$            \\
~~~$b \equiv a \cos i/\rstar$
                          \dotfill    & $\hatcurLCimp{42}$            & $\hatcurLCimp{43}$            \\
~~~$i$ (deg)              \dotfill    & $\hatcurPPi{42}$          & $\hatcurPPi{43}$          \\

\multicolumn{3}{l}{\em Limb-darkening coefficients \tablefootmark{c}}\\
~~~$c_1,i$ (linear term)  \dotfill    & $\hatcurLBii{42}$             & $\hatcurLBii{43}$             \\
~~~$c_2,i$ (quadratic term) \dotfill  & $\hatcurLBiii{42}$            & $\hatcurLBiii{43}$            \\

\multicolumn{3}{l}{\em RV parameters}\\
~~~$K$ (\ms)              \dotfill    & $\hatcurRVK{42}$      & $\hatcurRVK{43}$      \\
~~~$e$                    \dotfill    & $0$ (fixed)          & $0$ (fixed)          \\
\multicolumn{3}{l}{\em Planetary parameters}\\
~~~$\mpl$ ($\mjup$)       \dotfill    & $\hatcurPPmlong{42}$          & $\hatcurPPmlong{43}$          \\
~~~$\rpl$ ($\rjup$)       \dotfill    & $\hatcurPPrlong{42}$          & $\hatcurPPrlong{43}$          \\
~~~$C(\mpl,\rpl)$
    \tablefootmark{d}     \dotfill    & $\hatcurPPmrcorr{42}$         & $\hatcurPPmrcorr{43}$         \\
~~~$\rhopl$ (\gcmc)       \dotfill    & $\hatcurPPrho{42}$            & $\hatcurPPrho{43}$            \\
~~~$\log g_p$ (cgs)       \dotfill    & $\hatcurPPlogg{42}$           & $\hatcurPPlogg{43}$           \\
~~~$a$ (AU)               \dotfill    & $\hatcurPParel{42}$           & $\hatcurPParel{43}$           \\
~~~$T_{\rm eq}$ (K)\tablefootmark{e}        \dotfill   & $\hatcurPPteff{42}$           & $\hatcurPPteff{43}$           \\
~~~$\Theta$\tablefootmark{f} \dotfill & $\hatcurPPtheta{42}$          & $\hatcurPPtheta{43}$          \\
~~~$\langle F \rangle$ ($10^{8}$\ergscmsq) \tablefootmark{g}
                          \dotfill    & $\hatcurPPfluxavg{42}$        & $\hatcurPPfluxavg{43}$        \\
\hline
\end{tabular}
\tablefoot{
\tablefoottext{a}{%%
    Reported times are in Barycentric Julian Date calculated directly
    from UTC, {\em without} correction for leap seconds.
    \ensuremath{T_c}: Reference epoch of mid transit that
    minimizes the correlation with the orbital period.
    \ensuremath{T_{14}}: total transit duration, time
    between first to last contact;
    \ensuremath{T_{12}=T_{34}}: ingress/egress time, time between first
    and second, or third and fourth contact.
}
\tablefoottext{b}{
    Reciprocal of the half duration of the transit used as a jump
    parameter in our MCMC analysis in place of $\arstar$. It is
    related to $\arstar$ by the expression $\zrstar = \arstar
    (2\pi(1+e\sin \omega))/(P \sqrt{1 - b^{2}}\sqrt{1-e^{2}})$
    Bakos et al. (2010).
}
\tablefoottext{c}{
    Values for a quadratic law, adopted from the tabulations by
    Claret (2004) according to the spectroscopic (SPC) parameters
    listed in \reftabl{stellar}.
}
\tablefoottext{d}{
    Correlation coefficient between the planetary mass \mpl\ and radius
    \rpl.
}
\tablefoottext{e}{
    Planet equilibrium temperature averaged over the orbit, calculated
    assuming a Bond albedo of zero, and that flux is re-radiated from
    the full planet surface.
}
\tablefoottext{f}{
    The Safronov number is given by $\Theta = \frac{1}{2}(V_{\rm
    esc}/V_{\rm orb})^2 = (a/\rpl)(\mpl / \mstar )$
    (see Hansen \& Barman 2007).
}
\tablefoottext{g}{
    Incoming flux per unit surface area, averaged over the orbit.
}
}
\end{table*}

%% --------------------------------------------------------------------
%% Table of orbital and planetary parameters, assuming eccentric orbits 
%%
\begin{table*}
\caption{\label{tab:planetparameccen}Orbital and planetary
  parameters for \hatcurb{42} and \hatcurb{43} allowing eccentric orbits.}
\centering
\begin{tabular}{lcc}
\hline\hline
    &{\bf HAT-P-42}&{\bf HAT-P-43}\\
~~~~~~~~Parameter~~~~~~~~&Value&Value\\
\hline
\multicolumn{3}{l}{\em \Lc{} parameters}\\
~~~$\arstar$              \dotfill    & $\hatcurPPareccen{42}$             & $\hatcurPPareccen{43}$             \\
~~~$\zrstar$              \dotfill    & $\hatcurLCzetaeccen{42}$       & $\hatcurLCzetaeccen{43}$       \\
~~~$i$ (deg)              \dotfill    & $\hatcurPPieccen{42}$          & $\hatcurPPieccen{43}$          \\

\multicolumn{3}{l}{\em RV parameters}\\
~~~$K$ (\ms)              \dotfill    & $\hatcurRVKeccen{42}$      & $\hatcurRVKeccen{43}$      \\
~~~$\sqrt{e} \cos \omega$ 
                          \dotfill    & $\hatcurRVrkeccen{42}$          & $\hatcurRVrkeccen{43}$\\
~~~$\sqrt{e} \sin \omega$
                          \dotfill    & $\hatcurRVrheccen{42}$              & $\hatcurRVrheccen{43}$\\
~~~$e \cos \omega$ 
                          \dotfill    & $\hatcurRVkeccen{42}$          & $\hatcurRVkeccen{43}$\\
~~~$e \sin \omega$
                          \dotfill    & $\hatcurRVheccen{42}$              & $\hatcurRVheccen{43}$\\
~~~$e$                    \dotfill    & $\hatcurRVecceneccen{42}$          & $\hatcurRVecceneccen{43}$\\
~~~$\omega$ (deg)         \dotfill    & $\hatcurRVomegaeccen{42}$      & $\hatcurRVomegaeccen{43}$\\

\multicolumn{3}{l}{\em Secondary eclipse parameters}\\
~~~$T_s$ (BJD)            \dotfill    & $\hatcurXsecondaryeccen{42}$       & $\hatcurXsecondaryeccen{43}$\\
~~~$T_{s,14}$              \dotfill   & $\hatcurXsecdureccen{42}$          & $\hatcurXsecdureccen{43}$\\
~~~$T_{s,12}$              \dotfill   & $\hatcurXsecingdureccen{42}$       & $\hatcurXsecingdureccen{43}$\\

\multicolumn{3}{l}{\em Planetary parameters}\\
~~~$\mpl$ ($\mjup$)       \dotfill    & $\hatcurPPmlongeccen{42}$          & $\hatcurPPmlongeccen{43}$          \\
~~~$\rpl$ ($\rjup$)       \dotfill    & $\hatcurPPrlongeccen{42}$          & $\hatcurPPrlongeccen{43}$          \\
~~~$C(\mpl,\rpl)$
         \dotfill    & $\hatcurPPmrcorreccen{42}$         & $\hatcurPPmrcorreccen{43}$         \\
~~~$\rhopl$ (\gcmc)       \dotfill    & $\hatcurPPrhoeccen{42}$            & $\hatcurPPrhoeccen{43}$            \\
~~~$\log g_p$ (cgs)       \dotfill    & $\hatcurPPloggeccen{42}$           & $\hatcurPPloggeccen{43}$           \\
~~~$a$ (AU)               \dotfill    & $\hatcurPPareleccen{42}$           & $\hatcurPPareleccen{43}$           \\
~~~$T_{\rm eq}$ (K)        \dotfill   & $\hatcurPPteffeccen{42}$           & $\hatcurPPteffeccen{43}$           \\
~~~$\Theta$ \dotfill & $\hatcurPPthetaeccen{42}$          & $\hatcurPPthetaeccen{43}$          \\
~~~$\langle F \rangle$ ($10^{8}$\ergscmsq) 
                          \dotfill    & $\hatcurPPfluxavgeccen{42}$        & $\hatcurPPfluxavgeccen{43}$        \\
\hline
\end{tabular}
\tablefoot{
    Quantities and definitions are as in \reftabl{planetparam}, which
    gives our adopted values, determined assuming circular orbits. 
    Here we do not list parameters that are effectively independent of
    the eccentricity.
}
\end{table*}

%_________________________________________________________________
\section{Discussion}
\label{sec:conclusion}

We have presented the discovery, confirmation and characterization 
of two new transiting planets.

\hatcurb{42} and \hatcurb{43} are inflated hot Jupiters with
$P=\hatcurLCP{42}$ and $\hatcurLCP{43}$ days, $\mpl =
\hatcurPPmlong{42}$ and $\hatcurPPmlong{43}$ $\mjup$, and
$\rpl=\hatcurPPrlong{42}$ and $\hatcurPPrlong{43}$ $\rjup$, respectively. 
Assuming
zero albedo and full heat redistribution, and using the stellar and
planetary parameters determined by our analysis, 
their equilibrium temperatures are in the same regime,
i.e.~$T_{\rm eq}=\hatcurPPteff{42}$ and $\hatcurPPteff{43}$\,K,
respectively. 
\hatcurb{42} has larger uncertainties on its derived parameters as the
transit depth is shallower due to the larger radius of the slightly
more evolved host.  
In Fig.~\ref{fig:MassRadius} and
Fig.~\ref{fig:TeqRadius} we plot\footnote{We used
the up-to-date catalogue of the transiting planets TEPCat
\url{http://www.astro.keele.ac.uk/jkt/tepcat/}.}
these planets in the mass--radius and $T_{\rm
eq}$--radius diagram, showing that their positions are typical compared to the
 sample of well-characterized transiting exoplanets.  Within $1\sigma$,
\hatcurb{42} and \hatcurb{43} have the same radius and $T_{\rm eq}$. 
They mainly differ by their periods (semi-major axes) and by their masses
(densities).  \hatcurb{43} is very similar to HAT-P-4b
($\mpl=0.672\mjup$, $P=3.05$d, $\rpl=1.27\rjup$, Kovacs et al. 2007)
with HAT-P-4 being a little more massive, hotter and more evolved star, with
similar metallicity ($\feh=0.24$), and to OGLE-Tr-10b  ($\mpl=0.63\mjup$, $P=3.1$d,
$\rpl=1.25\rjup$, Torres et al. 2008). 
\hatcurb{42} is resembling to CoRoT-19b ($\mpl=1.107\mjup$, $P=3.9$d,
$\rpl=1.29\rjup$, Guenther et al. 2012). HAT-P-13b is the other closest analogous to \hatcurb{42} and \hatcurb{43} but in the intermediate mass-regime ($\mpl=0.851\mjup$,
$P=2.9$d, $\rpl=1.28\rjup$, Winn et al. 2010, Bakos et al. 2009).  
The stellar hosts of all the previously quoted planets are slow rotators,
with late F/early G spectral type, and with enhanced metallicity with
respect to the Sun (except for CoRoT-19, which is solar).

\hatcurb{42} and \hatcurb{43} are inflated hot Jupiters compared to models of coreless giant planets. The models of Fortney et al. (2007) predict a maximal radius of $\sim1.1$$\rjup$ for both planet (see their Table~4 and Figure~7) considering the system age, planet distance and mass (maximal in the meaning of a core free planet of pure H-He). Laughlin et al. (2011) derived a simple fitting relation from the Bodenheimer et al. (2003) models to infer the expected radius of a H-He composition planets for different masses and insolation. From the Laughlin et al. (2011) relations, we derived an predicted radius of 1.19 and 1.17$\rjup$, respectively for \hatcurb{42} and \hatcurb{43}, still at the 1-$\sigma$ and 3-$\sigma$ level from our observations.

Enoch et al. (2012) determined an empirical relation between $\rpl$,
$T_{\rm eq}$, and the semi-major axis $a$ (see their Eq.~9).  This leads
to an estimated radius of $1.325\pm0.054\,\rjup$ for \hatcurb{42}, matching 
the observed radius within the error bars.  The same relations predict 
a radius of $1.180^{+0.021}_{-0.023}\rjup$ for \hatcurb{43}, which is 3-$\sigma$ below our measured value.  Considering also
the results observed for the highly inflated hot Jupiters presented by
Hartman et al. (2012), it is possible that other parameters than $T_{\rm eq}$ and $a$ play a role in the planetary radius in the
mass domain where the equation was derived for $0.5<\mpl<2\mjup$.
Independently, Beky et al. (2011) derived a relation to determine the radius for the planets with $0.3<\mpl<0.8\mjup$ that we could apply to \hatcurb{43}. Their equation leads to an inferred radius of $1.133\pm0.044\,\rjup$.
As was already remarked for HAT-P-39b/40b/41b by Hartman et al. (2012), the predicted radius is smaller than the measured value.

Another caveat in using the above relations is the assumption of zero Bond
albedo and full redistribution of the heat when calculating $T_{\rm eq}$,
whereas in reality these assumptions are unlikely to be true for all of the
planets. 

Few measurements per orbit lead to poor constraints on a small eccentricity.  We followed the recommendation of 
Anderson et al. (2012), and adopted circular orbits, after a
significance test on the eccentricity measurement.  
We note that observing the secondary transit detection is a very efficient way to characterize
the shape of the orbit, that would need several tens of RV values to
obtain an equivalent precision (e.g. Boisse et al. 2009,
Husnoo et al. 2011).  Another clue to constrain a small eccentricity is the measurement of the RVs of planetary lines during the transit, but would currently need brighter hosts to be significantly detected (Snellen et al. 2010, Montalto et al. 2011).

The Rossiter-McLaughlin effect measured via spectroscopy during the transit
allows the measurement of the sky-plane projected angle between the stellar
spin-axis and the planetary orbital momentum vector. 
With
the hypothesis that the projected angle is close to zero, the expected
semi-amplitude of the effect is of $26$ and $34\,\ms$ for
\hatcur{42} and \hatcur{43}, respectively.  

We note that the transit of \hatcurb{42} is far from equatorial ($b = \hatcurLCimp{42}$),
which facilitates the measurement of the projected angle, since there is a
smaller correlation between the stellar $\vsini$ and the projected
spin-orbit angle.  According to Winn et al. (2010a), planets around 
stars with $\teffstar < 6250$\,K are preferentially aligned (except the systems with longest timescales for obliquity damping, e.g. HD80606). Additionally, the (old) age of the host stars as well as the (low) mass of
the planets also increase the chances that both planetary systems are
well-aligned (Triaud 2011, H\'ebrard et al. 2011).  We applied Eq.2 of
Albrech tet al. (2012) where they estimated the tidal timescale to
align a planet with the stellar equator from a calibration of binary studies. These results also favor 
alignment for both \hatcur{42} and \hatcur{43} (respectively $\tau_{\mathrm CE}=247$ and $558$ to be reported in their Fig.~24).

To conclude, hot gas giant planets are surprisingly diverse, and 
the physical reasons behind the diversity and the observed 
correlations are not well understood.  It is likely that the
physical properties of such planets depend on a
large number of parameters. Hence, there is a continued need to
broaden the sample of well-characterized transiting
planets in order to develop a comprehensive view of exoplanet
formation, structure and evolution.

\begin{acknowledgements}
The authors thank all the staff of Haute-Provence Observatory for their
contribution to the success of the ELODIE and SOPHIE projects and their
support at the 1.93-m telescope.  IB acknowledges the support of the
European Research Council/ European Community under the FP7 through a
Starting Grant, as well from Fundac{c}ao para a Ci\^encia e a
Tecnologia (FCT), Portugal, through SFRH/BPD/81084/2011 and the project
PTDC/CTE-AST/098528/2008.
We acknowledge partial funding for HATNet operations by
NASA grant NNX08AF23G,
and support for performing follow-up observations by NSF
grant NSFAST-1108686. 
G.K. acknowledges the support of the Hungarian Scientific Research Foundation (OTKA) through grant K-81373.

\end{acknowledgements}

% for the bibliography, at the end
%\bibliographystyle{aa} % style aa.bst
\bibliography{htrbib.bib} % your references Yourfile.bib

\end{document}